\documentclass[aps,prd,twocolumn,amsmath,amssymb,superscriptaddress,nofootinbib,longbibliography,10pt]{revtex4-2}
\usepackage{amssymb,amsmath}
\usepackage{mathrsfs}
\usepackage{mathtools}
\usepackage[normalem]{ulem}
\usepackage{dcolumn}
\usepackage{wasysym}
\usepackage{tikz}
\allowdisplaybreaks
\usepackage[colorlinks=true,citecolor=blue,urlcolor=blue]{hyperref}

\allowdisplaybreaks
\newcommand{\CGP}{\affiliation{Center for Gravitational Physics, University of Texas at Austin, Austin, TX 78712, USA}}
\newcommand{\FlatIron}{\affiliation{Center for Computational Astrophysics, Flatiron Institute, 162 5th Avenue, New York, NY 10010, USA}}

\newcommand{\Th}{\text{\Thorn}}

\begin{document}

\title{Computing spectral shifts for Johannsen-Psaltis black holes}

\author{David G. Wu}
    \email{david.wu@utexas.edu}
\CGP
\author{Asad Hussain}
\CGP\FlatIron
\author{Aaron Zimmerman}
\CGP

\begin{abstract}
    The growing number of gravitational wave (GW) detections and the increasing sensitivity of GW detectors have enabled precision tests of General Relativity (GR) in the strong-field regime. 
    The recent observation of multiple quasinormal modes (QNMs) in GW250114 marks a major advance for observational black hole spectroscopy.
    This clear signal, together with the growing number of GW detections, highlights the need for accurate predictions of QNM spectra in beyond-GR theories in order to carry out precision searches for new physics. 
    In this work, we continue to lay the foundation for such predictions using a modified Teukolsky formalism in conjunction with the eigenvalue perturbation method.
    We compute the spectral shifts of slowly rotating Johannsen–Psaltis black holes for \(2\leq \ell \leq 10\), all \(m\), and overtones \(n=0,1,2\), and confirm the large-\(\ell\) behavior of the modes by comparing with the WKB approximation. 
    We find that these black holes admit definite-parity modes but break the isospectrality between even- and odd-parity QNMs at all spins, and that the shifts depend linearly on \(m\) for slow spins. 
    We further derive a general parity condition that any beyond-GR modification to the metric must satisfy to support definite-parity modes, providing
    new insights into isospectrality breaking and parity structure in gravitational perturbations.
\end{abstract}

\maketitle

\section{Introduction}
Advanced ground-based laser interferometers~\cite{abbott_prospects_2020,LIGOScientific:2014pky,VIRGO:2014yos,10.1093/ptep/ptaa125} have now detected hundreds of binary black hole (BBH) mergers via their gravitational wave (GW) signatures~\cite{abac_gw250114_2025}. 
These observations open a window into the strong-field and dynamical regime of gravity which enable new tests of Einstein’s General Relativity (GR). 
These tests complement weak-field experiments such as those conducted within the Solar System~\cite{will_confrontation_2014} where GR has already been verified with high precision. 
The GW signals from these BBH coalescences can be broken down into three stages: inspiral, merger, and ringdown. 
Theoretical predictions for the GWs during each stage can be modified with free parameters, enabling theory-agnostic tests of GR, e.g.~\cite{Yunes:2009ke,Li:2011cg,Mehta:2022pcn,Yunes:2025xwp,the_ligo_scientific_collaboration_tests_2021}.
To date these tests have not revealed convincing signatures of deviations from GR.

While theory-agnostic tests are powerful tools for searching for the imprint of unknown new physics, the strongest possible constraints require theoretical predictions from specific beyond GR (bGR) theories.
For a given event, multiple free deviation parameters would be required to mimic the effect of a specific bGR theory, which means testing either a partial effect on the GW signal or constraining several parameters.
Theory-specific tests also allow for the easy combination of constraints from multiple GW events.
Each GW event tests the same set of bGR parameters, allowing multiple events to be combined to improve bounds \cite{Zimmerman:2019wzo,perkins_improved_2021,sennett_gravitational-wave_2020} and simplifying the required hierarchical analysis~\cite{Isi:2019asy}.
Theoretical predictions are also needed to interpret any potential future detections of deviations from GR.

Predicting GW signals in specific bGR theories is itself a daunting challenge.
For the long inspiral, high-order post-Newtonian computations must be carried out to match GR predictions, and many results are available at leading post-Newtonian orders (see e.g.~\cite{Yunes:2025xwp}).
Many bGR theories involve additional fields which couple to the curvature of spacetime~\cite{sotiriou_black_2014, alexander_chernsimons_2009, kanti_classical_1995}, adding new degrees of freedom, complicating both analytic and numerical approaches.
Nevertheless, numerical simulations of compact binary mergers in bGR theories have emerged in recent years, e.g.~\cite{okounkova_numerical_2020,tahura_parameterized_2018,julie_inspiral-merger-ringdown_2025,roy_black_2025}. 
By producing complete inspiral–merger–ringdown waveforms, these simulations open new avenues for testing GR. 
However, such simulations are computationally intensive and often limited to a small subset of bGR theories and parameter ranges. 
Consequently, analytic and perturbative approaches remain essential to build waveform models.

A promising avenue in terms of both observational constraints and bGR modeling is the ringdown phase following BBH merger.
During ringdown, the remnant black hole (BH) produces quasinormal modes (QNMs), which are rapidly decaying and oscillating GW modes labeled by the integers  \((\ell, m ,n)\) \cite{vishveshwara_scattering_1970,vishveshwara_stability_1970, berti_black_2025}. 
In GR, the properties of a BH are governed by the no hair theorem \cite{israel_event_1967, carter_axisymmetric_1971, robinson_uniqueness_1975}, stating that a BH spacetime is completely determined by only its mass and angular momentum.
These two properties completely determine the QNM spectrum. 
The measurement of a single QNM's frequency and decay provide a mass and spin measurement, and any additional modes allow for testing the properties of the BH, a process known as black hole spectroscopy~\cite{dreyer_black_2004, berti_gravitational-wave_2006, berti_extreme_2018,berti_black_2025}.
Recent observations of high signal-to-noise BBHs, particularly GW250114~\cite{abac_gw250114_2025,the_ligo_scientific_collaboration_black_2025}, have allowed for the clear measurement of multiple QNMs, advancing BH spectroscopy as an observational science.

Within BH spectroscopy, the simplest approach remains theory-agnostic tests, where deviations from the GR prediction of the spectrum are constrained.
However, the relative simplicity of predicting the QNM spectrum in GR~\cite{berti_black_2025} makes ringdown predictions in bGR theories a promising theoretical target.
Many methods and approximations have been applied to study QNMs in bGR theories.
The eikonal and Wentzel-Kramers-Brillouin (WKB) approximations offer important insights into bGR QNM spectra \cite{cardoso_geodesic_2009, konoplya_quasinormal_2011}, but are most accurate for high frequency (large-\(\ell\)) modes which are subdominant in GW signals. 
Alternatively, the QNMs of non-rotating or slowly spinning black holes in bGR theories can be calculated by applying extensions of the Regge-Wheeler-Zerilli formalism~\cite{regge_stability_1957,zerilli_gravitational_1970,martel_gravitational_2005}.
 
By separating the metric perturbation into even- and odd-parity components, the QNM spectrum for a number of bGR theories has been explored \cite{antoniou_gravitational_2025,blazquez-salcedo_perturbed_2016,cardoso_black_2018,guo_quasi-normal_2025}. 
Although these approaches are important for understanding key QNM features within each bGR theory, the majority of observed BBH mergers produce remnants with spins \(a\gtrsim 0.6\) \cite{the_ligo_scientific_collaboration_gwtc-40_2025}. 
Consequently, approaches to calculating bGR QNMs that are general in spin or are valid for very large spins are required to address GW data.

Driven by the need for methods, much effort has been made to more thoroughly calculate bGR QNMs for all modes and all spins. 
This effort has been aided in recent years by the development of the modified Teukolsky formalism \cite{hussain_approach_2022,li_perturbations_2023}, which can be applied in conjunction with the eigenvalue perturbation (EVP) method \cite{mark_quasinormal_2015} to compute the shifts to the QNM spectrum in bGR theories, in principle. 
This approach allows for the calculation of QNMs in a bGR theory perturbatively in a deviation parameter but general in spin. 
It can also help reveal promising signatures of bGR theories in ringdown, such as the breaking of isospectrality~\cite{li_isospectrality_2024}.

Using the modified Teukolsky formalism, a number of bGR theories have been studied to calculate QNMs. 
Dynamical Chern-Simons (dCS) gravity, which introduces parity-violating corrections via a scalar field \cite{jackiw_chern-simons_2003, alexander_chernsimons_2009}, has been explored for slowly rotating black holes \cite{wagle_perturbations_2023, li_perturbations_2025}. 
Higher-derivative theories, motivated by quantum and string-inspired corrections to GR, have also been analyzed within this framework \cite{cano_higher-derivative_2024, cano_universal_2023}. 
Parametrized bGR metrics, which aim to capture generic deviations from GR without specifying a particular underlying theory, have likewise been investigated for the non-rotating case \cite{weller_spectroscopy_2025}. 
Complementary approaches based on spectral methods to deal with metric perturbations have been employed to study QNMs in bGR metrics up to large spins, covering theories including scalar-Gauss-Bonnet (sGB), dCS, and axi-dilation gravity \cite{chung_quasi-normal_2024,chung_spectral_2024,chung_ringing_2024,lam_near-extremal_2025, chung_quasinormal_2025,lam_analytic_2025}. 
Collectively, these works illustrate the growing ability to compute bGR QNMs across a broad range of spins and modes.

In this work we continue to lay the foundation for bGR QNM calculation by applying the modified Teukolsky approach advanced in Ref.~\cite{hussain_approach_2022} to a specific bGR metric developed by Johannsen and Psaltis \cite{johannsen_metric_2011}. 
We refer to this as the JP metric, and it is a parametrized perturbation to the Kerr metric.
The JP metric was originally developed as a Kerr-like metric that could be used to test the no hair theorem using electromagnetic observations, e.g.~\cite{johannsen_inner_2013, wang_revisiting_2025,kong_constraints_2014,john_bondi_2019}.
However, it can also be applied to gravitational tests of the no hair theorem. 
The JP metric provides a relatively simple testbed for our methods, but computing the shifted QNMs requires the implementation of multiple key elements of the general formalism. 

In this work, we extend the formalism of~\cite{hussain_approach_2022} to clarify the connection between positive and negative frequency QNMs and even- and odd-parity perturbations.
We then apply the insights of Ref.~\cite{li_isospectrality_2024} to our formalism and develop a simple check for whether the bGR theory breaks isospectrality and whether it still admits definite-parity perturbations.
With this we show that the JP metric breaks isospectrality but retains definite-parity modes, with the even- and odd-parity QNMs receiving a Zeeman-like splitting.
Finally, we  compute the QNM shifts for \(2\leq\ell\leq 10\), all \(m\), and \(n=0,1,2\) of the slowly rotating JP metric.
We limit our mode computations to the slow rotation regime, linearizing in spin, due to computational constraints.
We empirically find the slowly rotating JP QNM shifts are linear in \(m\), and we compare our results to the scalar WKB approximation, showing that they follow the expected large \(\ell\) behavior.

The structure of this paper is as follows. 
In Sec.~\ref{sec:JPMetric}, we describe the JP metric as constructed by \cite{johannsen_metric_2011}. 
Then in Sec.~\ref{sec:ModTeukEqn} we describe our implementation of the modified Teukolsky formalism outlined in \cite{hussain_approach_2022,li_perturbations_2023}. Section \ref{sec:SpectrumProperties} details the isospectral and parity properties of the spectral shifts in our formalism. 
Finally, we present our results and discuss avenues for future work in Secs.~\ref{sec:Results} and~\ref{sec:Conclusion}.
Additional technical details and tabulated results are given in the Appendices.

\section{JP Metric}
\label{sec:JPMetric}
The Johannsen–Psaltis metric, first presented in \cite{johannsen_metric_2011}, generalizes the Kerr spacetime by introducing additional free parameters beyond mass and spin, while requiring that the spacetime remain regular outside the horizon. 
The metric, which we refer to as the JP metric, is constructed starting from a parametric deviation to the Schwarzschild metric, and then extended to spinning black holes using the Newman-Janis algorithm \cite{newman_note_1965}. 
The result is a Kerr-like metric that, up to a threshold value in spin, does not have singularities or regions with closed timelike curves outside the horizon. 
This metric was originally developed as a Kerr-like metric that could be used to test the no hair theorem using electromagnetic observations. 
For example, the deviations from the Kerr spacetime can significantly affect the location of the inner edges of an accretion disk, a potential observable through relativistically broadened iron lines \cite{johannsen_inner_2013}.
Additional electromagnetic observables include black hole shadows \cite{johannsen_inner_2013, wang_revisiting_2025,kong_constraints_2014} and inferred accretion rates \cite{john_bondi_2019}.
Due to its relatively simple form, it also provides an excellent testing ground for computing the shifts to the QNM spectrum in a bGR theory using the modified Teukolsky formalism and EVP method. 

The JP metric takes the form
\begin{align}
\label{eqn:JPMetric}
    ds^2=&-[1+h(r,\theta)]\left(1-\frac{2Mr}{\Sigma}\right)dt^2 
    \notag\\ &
    -\frac{4aMr \sin^2\theta}{\Sigma}[1+h(r,\theta)]dtd\phi
    \notag\\ &
    +\frac{\Sigma[1+h(r,\theta)]}{\Delta+a^2\sin^2\theta h(r,\theta)}dr^2+\Sigma d\theta^2
    \notag\\ &
    +\Bigg[\sin^2\theta\left(r^2+a^2+\frac{2a^2Mr\sin^2\theta}{\Sigma}\right)
    \notag\\ & \qquad
    +h(r,\theta)\frac{a^2(\Sigma+2Mr)\sin^4\theta}{\Sigma}\Bigg]d\phi^2\,,
\end{align}
where we use
\begin{align}
    &\Sigma= r^2+a^2 \cos^2\theta\,,\\
    &\Delta= r^2-2Mr+a^2\,,
\end{align}
and where
\begin{align}
    h(r,\theta)=\sum_{k=0}^{\infty}& \left(\epsilon_{2k}+ \epsilon_{2k+1}\frac{Mr}{\Sigma}\right)\left(\frac{M^2}{\Sigma}\right)^k 
\end{align}
describes the deviation away form Kerr. Enforcing asymptotic flatness requires that \(h(r)\sim 1/r^n\) with \(n\geq 2\), so \(\epsilon_0=\epsilon_1=0\). 
The next deviation parameter, \(\epsilon_2\), has observational constraints on weak-field deviations from GR in the parametrized post-Newtonian (PPN) framework~\cite{will_confrontation_2014}. 
The Lunar Laser Ranging experiment puts constraints on the \(\beta\) PPN parameter, giving \(\epsilon_2=(1.24\pm1.42)\times10^{-4}\) using the relation \(\epsilon_2=2(\beta-1)\)~\cite{biskupek_benefit_2021}. 
As such we take \(\epsilon_2=0\). 
For \(\epsilon_3\), X-ray observations of black hole accretion disks impose the relatively weaker bound \(\epsilon_3<5\) \cite{kong_constraints_2014}. 
This parameter encodes modifications to the quadrupole and higher multipolar moments of BHs as viewed from large distances.
A more recent analysis~\cite{santos_testing_2024} of several GWTC-3 events further constrains \(\epsilon_3\) by introducing parameterized post-Einsteinian~\cite{Yunes:2009ke} corrections into the IMRPhenomXPHM waveform model~\cite{Pratten:2020ceb}, accounting for the leading effects of the corresponding quadrupolar deviation on BBH inspiral.
Their analysis using \texttt{PyCBC} \cite{nitz_gwastropycbc_2024} and \texttt{Bilby} \cite{ashton_bilby_2019} gives constraints of \(|\epsilon_3|\lesssim O(1)\), with the 95\% credible interval for GW150914 being \(\epsilon_3=0.13^{+0.45}_{-0.27}\). 
Accordingly, we focus on the \(\epsilon_3\) deviation away from Kerr, giving
\begin{align}
    h(r,\theta)=\epsilon\frac{M^3 r}{\Sigma^2} \,,
\end{align}
where we have redefined \(\epsilon=\epsilon_3\) here and for the remainder of the paper. 
Now, \(\epsilon\) encodes the size of the bGR deviation, and the normal Kerr metric in Boyer-Lindquist coordinates is recovered when \(\epsilon\rightarrow0\).

While the JP metric is linear in the perturbation parameter \(\epsilon\), we also linearize in the spin \(a\) for this work, as explained in Sec.~\ref{sec:ContourIntegration}. 
This restricts us to slowly rotating JP black holes. 
With this approximation, the slowly rotating JP metric takes the form
\begin{align}
\label{eqn:SlowlyRotatingJPMetric}
    ds^2=&-\left(1-\frac{2M}{r}\right)[1+h(r)]dt^2
    \notag  -\frac{4Ma\sin^2\theta}{r}\\ &
    \times [1+h(r)]dtd\phi+\left(1-\frac{2M}{r}\right)^{-1}[1+h(r)]dr^2
    \notag \\ &
    +r^2d\theta^2+r^2\sin^2\theta d \phi^2 \,,
\end{align}
where
\begin{align}
    h(r)=\epsilon \frac{M^3}{r^3}\,.
\end{align}
In the interest of studying black holes described by this metric, we require a closed horizon, a condition that depends on the value of \(\epsilon\). 
When \(\epsilon\) is negative, the horizon is always closed. 
However, for positive values of \(\epsilon\), there is a threshold \(a_\text{crit}\) where the horizon is no longer closed for \(|a|>a_\text{crit}\) \cite{johannsen_metric_2011}. 
Given the constraints on \(\epsilon\), only high spins would result in a naked singularity. 
We avoid this regime in this work due to the constraints of the small spin expansion.

\section{Modified Teukolsky Equation}
\label{sec:ModTeukEqn}
To analyze the QNM spectrum of the slowly rotating JP metric, we use the modified Teukolsky formalism developed in \cite{li_perturbations_2023,hussain_approach_2022} to construct the leading order QNM shifts in \(\epsilon\). More specifically, we seek \(\omega^{(1)}\), where the QNM frequencies are expanded as 
\begin{align}
    \omega\approx\omega^{(0)}+\epsilon \, \omega^{(1)} \,.
\end{align}
Following \cite{hussain_approach_2022}, we adopt a two parameter expansion in \(\epsilon\) and \(\eta\).  In this expansion, \(\epsilon\) characterizes the deviation away from Kerr in the slowly rotating JP metric, and \(\eta\) characterizes the GW perturbation on the background spacetime. For the metric and the spin-weighted scalar solutions to the Teukolsky equation, the expansion is
\begin{align}
    g_{ab}&=g^{(0,0)}_{ab}+\epsilon g_{ab}^{(1,0)}+\eta h_{ab}^{(0,1)}\, ,\\
    \prescript{}{s}{\psi}=\prescript{}{s}{\psi}^{(0,0)}+&\epsilon \prescript{}{s}{\psi}^{(1,0)}+\eta \prescript{}{s}{\psi}^{(0,1)}+\epsilon \eta \prescript{}{s}{\psi}^{(1,1)} \,,
\end{align}
where terms of order \(O(\epsilon^n,\eta^m)\) are denoted with a \((n,m)\) superscript. Terms of order \(O(\epsilon^1,\eta^0)\) are driven by the bGR changes to the spacetime. Terms of order \(O(\epsilon^0,\eta^1)\) correspond to GWs in GR, while terms of order \(O(\epsilon^1,\eta^1)\) are the bGR GW perturbations which are of interest in this work. Consequently, we work entirely at order \(O(\eta^1)\) for the remainder of this work, and we exclude the \(\eta\) order counting notation and only retain the \(\epsilon\) order counting for simplicity.

Using this expansion, Ref.~\cite{hussain_approach_2022} finds a modified Teukolsky equation of the form
\begin{align}
\label{eqn:ModTeukOrig}
    \mathcal{O}[\prescript{}{s}{\psi}^{(0)}]+\epsilon\mathcal{O}[\prescript{}{s}{\psi}^{(1)}]+\epsilon\mathcal{V}[h^{(0)}]=0 \, ,\\
    \label{eqn:VOperator}
    \mathcal{V}[h^{(0)}]=\mathscr{S}_4^{ab}\left(2G^{(2)}_{ab}\left[h^{(0)},g^{(1)}\right]\right)\,,
\end{align}
where \(\mathcal{O}\) is the background Teukolsky operator and \(\mathscr{S}_4\) is the source (or decoupling) operator that constructs the sources for the Teukolsky equation from the stress-energy tensor \cite{pound_black_2022}. 
Using the Geroch, Held, and Penrose (GHP) formalism, the \(\mathscr{S}_4\) source operator is given by
\begin{align}
\label{eqn:S4}
\mathscr{S}_4^{ab} = 
  \tfrac12(\eth'-\bar{\tau}-4\tau') 
    \big[(\Th'-2\bar{\rho}')n^a\bar{m}^b -(\eth'-\bar{\tau})n^an^b  \big] \nonumber \\ 
  + \tfrac12(\Th'-4\rho'-\bar{\rho}')
    \big[(\eth'-2\bar{\tau}) n^a\overline{m}^b - (\Th'-\bar{\rho}')\bar{m}^a\bar{m}^b  \big] \, .
\end{align}
For a review of the GHP formalism and the GHP quantities used, refer to Ref.~\cite{pound_black_2022}. 
Finally, \(G^{(2)}_{ab}\) is the second order Einstein tensor operator. Normally, under some metric perturbation \(g_{ab}=g_{ab}^{(0)}+h_{ab}\), the Einstein tensor would expand as 
\begin{align}
    G_{ab}[g]\approx G_{ab}^{(0)}+G_{ab}^{(1)}[h]+G_{ab}^{(2)}[h,h]\,,
\end{align}
where \(G_{ab}^{(2)}[h,h]\) is quadratic in \(h_{ab}\). However, we have two different metric perturbations \(g_{ab}^{(1)}\) and \(h_{ab}^{(0)}\). 
To move from \(G_{ab}^{(2)}[h,h]\) to \(G_{ab}^{(2)}[h^{(0)},g^{(1)}]\), we take each quadratic term and replace one copy of \(h_{ab}\) with \(h_{ab}^{(0)}\), replace the other with \(g_{ab}^{(1)}\), and then symmetrize over both possible substitutions. 
With this construction, Eq.~\eqref{eqn:VOperator} can be thought of as an effective source term from the combination of the bGR and QNM sourcing metric perturbations.

While we have included the effects of the bGR geometry in Eq.~\eqref{eqn:ModTeukOrig} through Eq.~\eqref{eqn:VOperator}, one would ideally also include contributions of the underlying field equations of the JP metric. 
After all, the JP metric is a vacuum solution for an unknown set of field equations, or equivalently, a sourced solution to Einstein's equations.
Such contributions are encoded in additional contributions from the \(\mathcal C\) operator to the modified Teukolsky equation~\eqref{eqn:ModTeukOrig} in the formalism of Ref.~\cite{hussain_approach_2022}.
However, because these field equations are unknown, they cannot be directly incorporated into our analysis.
We must take a similar approach to the authors of \cite{glampedakis_post-kerr_2017}, who uses an eikonal approximation to calculate the QNMs of a JP black hole while neglecting the effect of the unknown fields. 

Neglecting the effect of the fields can be heuristically motivated by comparison to the Cowling approximation \cite{cowling_non-radial_1941}, originally developed for studying fluid oscillations in stars. 
In that context, one treats the background potential of a star as fixed when computing its spectrum of normal modes, although in reality the fluid oscillations are coupled to the potential.
For stellar perturbations this coupling has a small effect because, roughly speaking, the perturbations to the potential average out across the oscillation mode~\cite{Dalsgaard:2003book}.
By analogy, if the fields associated with the JP metric weakly couple to the geometry of the spacetime, one can invoke the spirit of the Cowling approximation and neglect their effects in an analogous manner.
A useful point of comparison is the Dudley–Finley approximation, which can be interpreted as a Cowling-like scheme that artificially decouples gravitational and electromagnetic perturbations of a Kerr-Newman black hole \cite{DudleyI,DudleyII}.
An EVP analysis for weakly charged Kerr-Newman~\cite{mark_quasinormal_2015} shows that the Dudley-Finley approximation can lead to appreciable errors in the correction to the Kerr frequencies (see also \cite{SahaDudleyFinley} for an investigation of this approximation for charged black holes).
More broadly, incorporating dynamical contributions has been important to accurately compute QNMs in sGB, dCS, and higher derivative theories \cite{cano_higher-derivative_2024,chung_quasi-normal_2024,li_perturbations_2025}.
Also, the Cowling approximation is worst for the low-order modes most relevant to current BH spectroscopic tests. 
As such, exploring the uncertainties due to this approximation would be important for a complete analysis.
Doing so, however, requires identifying an appropriate source for the JP solution, either in the form of matter or an underlying alternative theory.
Regardless, our results address the contributions arising from \(\mathcal V\) to \(\omega^{(1)}\).

\subsection{Metric Reconstruction}
\label{sec:MetricReconstruction}
In its current form, the modified Teukolsky equation in Eq.~\eqref{eqn:ModTeukOrig} involves the metric perturbation \(h_{ab}^{(0)}\). 
However, our goal is a formulation expressed solely in terms of differential operators acting on spin-weighted scalars that encode the curvature perturbations of the spacetime.
Of primary interest is the \(\Psi_4\) Weyl scalar, which describes the transverse GWs propagating along the outgoing null direction at asymptotic infinity.
This corresponds to a spin weight \(s=-2\) Newman-Penrose scalar~\cite{teukolsky_perturbations_1973}, and we take \(s=-2\) for the remainder of the paper unless noted otherwise, dropping the \(s\) subscripts to reduce notational clutter. 
For example, \(\prescript{}{-2}{\psi}^{(0)}=\psi^{(0)}\). 

Fortunately, \(h_{ab}^{(0)}\) can be reconstructed from \(\psi^{(0)}\) and its complex conjugate using the CCK-Ori procedure \cite{chrzanowski_vector_1975, kegeles_constructive_1979,ori_reconstruction_2003} which relies on radiation gauges. 
For our computation of \(\omega^{(1)}\), we require perturbations of Kerr that are regular on the future event horizon, and so it is most convenient to choose ingoing radiation gauge (IRG)~\cite{ori_reconstruction_2003}. 
To reconstruct the metric from \(\psi^{(0)}\), the CCK-Ori procedure makes use of a Hertz potential \(\Psi_H\). 
The Hertz potential and \(\psi^{(0)}\) are related by~\cite{keidl_finding_2007, keidl_gravitational_2010}
\begin{align}
\label{eqn:HertzPotentialRelation}
    \psi^{(0)}=\frac{1}{16}\left(\mathcal{L}^{\dagger4}\overline{\Psi_H}-12M \partial_t \Psi_H\right) \,,
\end{align}
where \(\mathcal{L}^{\dagger4}=\mathcal{L}^{\dagger}_{-1}\mathcal{L}^{\dagger}_{0}\mathcal{L}^{\dagger}_{1}\mathcal{L}^{\dagger}_{2}\), and 
\begin{align}
    \mathcal{L}^\dagger_n=-(\partial_\theta+n\cot\theta-i\csc\theta\partial_\phi)+i a \sin\theta\partial_r\,.
\end{align}
The reconstructed metric is then given by
\begin{align}
    h^{(0)}_{ab}=\mathscr{S}_0^\dagger[\Psi_H]_{ab}+ \overline{\mathscr{S}_0^\dagger[\Psi_H]_{ab}} \,,
\end{align}
where \(\mathscr{S}_0^\dagger\) is the metric reconstruction operator, and the complex conjugate term ensures that the metric perturbation remains real. 
This metric reconstruction operator is expressed in the GHP formalism as
\begin{align}
\label{eqn:MetricReconstructionOp}
    \mathscr{S}_0^\dagger[\Psi]_{ab}=&
    -\tfrac{1}{2}l_a l_b (\eth-\tau)(\eth+3\tau)\Psi
    \notag\\ &
    -\tfrac{1}{2}m_am_b(\Th-\rho)(\Th+3\rho)\Psi
    \notag\\ &
    +\tfrac{1}{2}l_{(a}m_{b)}\Big[(\Th-\rho+\overline{\rho})(\eth+3\tau)
    \notag\\ & \qquad
    +(\eth-\tau+\overline{\tau}')(\Th+3\rho)\Big]\Psi \,.
\end{align}
This still must be recast in terms of \(\psi^{(0)}\) rather than the Hertz potential by inverting the relationship in Eq.~\eqref{eqn:HertzPotentialRelation}. 

In Kerr, QNM frequencies with a positive real part are paired with a mode of the same imaginary part and real part of the opposite sign, a property related to the isospectrality between even and odd perturbations~\cite{chrzanowski_vector_1975,nichols_visualizing_2012}. 
The QNM frequencies are consequently split into plus \((+)\) and minus \((-)\) modes according to the sign of their real part, leading to the relationship 
\begin{align}
\label{eqn:MirrorModes}
    \omega_{\ell mn}^+=-\overline{\omega}_{\ell -m n}^-\,.
\end{align}
Henceforth, \((\ell m n)\) will always be associated with the plus mode and \((\ell -m n)\) with the minus mode, allowing us to drop the \((\ell m n)\) in favor of the notation
\begin{subequations}
    \begin{align}
        \omega_+=\omega_{\ell mn}^+\,,\\
        \omega_-=\omega_{\ell -m n}^-\,.
    \end{align}
\end{subequations}
To invert Eq.~\eqref{eqn:HertzPotentialRelation}, consider a spin-weighted scalar \(\psi^{(0)}\) composed of one plus mode and one minus mode with the form
\begin{subequations}
    \begin{align}
    \label{eqn:HarmonicDecomp}
    \psi^{(0)}&=\psi_+^{(0)}+\overline{\gamma}\psi_-^{(0)}\,,\\
    \psi_+^{(0)}&=R_+(r)S_+(\theta)e^{-i\omega^{}_+ t+im \phi}\,,\\
    \psi_-^{(0)}&=R_-(r)S_-(\theta)e^{-i \omega^{}_- t - i m \phi}\,,
\end{align}
\end{subequations}
where \(\overline{\gamma}\) is a complex constant encoding the mixing of the two modes, and is defined as such for later convenience. 
The radial and angular parts of the plus or minus QNM wavefunctions are \(R_\pm(r)\) and \(S_\pm(\theta)\). 
We also define a mode-dependent version of the \(\mathcal{L}^{\dagger4}\) operator which we denote with a tilde,
\begin{align}
    e^{-i \omega_{\pm}t \pm i m \phi}\tilde{\mathcal{L}}^{\dagger4}_{\pm}[\tilde{\psi}(r,\theta)]=\mathcal{L}^{\dagger 4}[\tilde{\psi}(r,\theta)e^{-i \omega_{\pm}t \pm i m \phi}]\, .
\end{align}
Due to the symmetries of the Teukolsky equation, the radial functions have the property 
\begin{align}
\label{eqn:RadialPMRelation}
    R_+(r)\propto \overline{R_-}(r)\, .
\end{align}
The proportionality constant relating the plus and minus radial modes encodes the normalization of the radial functions. 
We choose this proportionality constant to be unity for convenience. 
Additionally, as shown in \cite{chandrasekhar_mathematical_1983}, the angular solution obeys
\begin{align}
    \tilde{\mathcal{L}}^{\dagger 4}_\pm \overline{S_\mp}(\theta)=D_\pm S_\pm(\theta)\,,
\end{align}
with
\begin{align}
    D_\pm^2 =& \lambda_\pm^2(\lambda_\pm+2)^2+8\lambda_\pm(5\lambda_\pm+6)(a^2\omega_\pm^2-a m \omega_\pm)
    \notag \\&
    +96\lambda_\pm a^2 \omega_\pm^2+144(a^2 \omega_\pm^2-a m \omega_\pm)^2\,, \\
     \lambda_\pm=&\mathcal{A}_\pm+a^2 \omega_\pm^2-2a m \omega_\pm \,,
\end{align}
and \(\mathcal{A}_\pm\) is the angular separation constant in the Teukolsky equation. 
Using the Teukolsky-Starobinsky constant \(\mathfrak{C}\)~\cite{teukolsky_perturbations_1974}, 
\(D_\pm\) can also be written as 
\begin{align}
    D_\pm^2=\mathfrak{C}_\pm^2-144M^2 \omega_\pm^2\,.
\end{align}
With \(\psi^{(0)}\) composed of a plus mode and minus mode along with the above properties, inverting Eq.~\eqref{eqn:HertzPotentialRelation} gives 
\begin{align}
    &\Psi_H  =A\psi_+^{(0)}+\overline{B}\psi_-^{(0)} \,,\\
    &A =\dfrac{16}{\mathfrak{C}^2_+}(D_+ \gamma-12 i M \omega_+) \,,\\
    &B =\dfrac{16}{\mathfrak{C}_+^2}\left(D_+ - 12 i M \omega_+ \gamma\right)\,.
\end{align}

This inversion of Eq.~\eqref{eqn:HertzPotentialRelation} is closely related to that of Ref.~\cite{cano_universal_2023}, with the added advantage that our formulation is explicitly independent of the radial Teukolsky--Starobinsky constants.
These constants, denoted \(C_{\pm 2}\) in that work with \(C_{+2}C_{-2} = 1/\mathfrak{C}^2\), appear analytically in the modified Teukolsky equation of~\cite{cano_universal_2023}, although the equation is found to be numerically independent of them. 
This independence is necessary because the radial Teukolsky--Starobinsky constants depend on the normalization of the radial functions, whereas the modified Teukolsky equation itself cannot depend on such a choice.
Our approach can be viewed as choosing a convenient normalization of the radial functions from the outset, thereby eliminating any analytic dependence on \(C_{\pm 2}\).

Since the metric reconstruction operator \(\mathscr{S}_0^\dagger\) is a linear differential operator, the \(A\) and \(B\) constants can be pulled outside the operators, and the reconstructed metric can be expressed as
\begin{align}
\label{eqn:ReconstructedMetric}
h_{ab}^{(0)}=& A\,\mathscr{S}_0^\dagger\Big[\psi_+^{(0)}\Big]_{ab}
+B\,\overline{\mathscr{S}_0^\dagger\Big[\psi_-^{(0)}\Big]}_{ab}
\notag \\ &
+\overline{A}\,\overline{\mathscr{S}_0^\dagger\Big[\psi_+^{(0)}\Big]}_{ab}
+\overline{B}\,\mathscr{S}_0^\dagger\Big[\psi_-^{(0)}\Big]_{ab}\,,
\end{align}
eliminating the Hertz potential. 
Because the Hertz potential is composed of both \((+)\) and \((-)\) modes, so too is the reconstructed metric. This means that for a reconstructed metric with a temporal and azimuthal dependence of \(e^{-i\omega_+ t+i m \phi}\), both \(\psi_+^{(0)}\) and \(\psi_-^{(0)}\) are needed. 
The exact form of the operators needed for metric reconstruction derived here in Eq.~\eqref{eqn:ReconstructedMetric} extends the work of \cite{hussain_approach_2022}, which implicitly defined the operators needed for metric reconstruction. 
Additionally, this approach provides a computational advantage to metric reconstruction in similar approaches like Ref.~\cite{li_isospectrality_2024,li_perturbations_2025} which leave the inversion of  Eq.~\eqref{eqn:HertzPotentialRelation} in terms of fourth order derivatives stemming from the Teukolsky-Starobinsky identities \cite{teukolsky_perturbations_1973,starobinskii_amplification_1973, starobinskil_amplification_1974,chandrasekhar_mathematical_1983}. 
By analytically evaluating these derivatives and finding the constants \(A\) and \(B\) similar to \cite{Berens_2024}, we trade these derivatives for quantities depending only on the angular separation constant and the background QNM frequencies, which are more easily calculated. 

\subsection{Gravitational QNM Shifts}
Now that the reconstructed metric is a function of the spin-weighted scalars, we can substitute Eq.~\eqref{eqn:ReconstructedMetric} into Eq.~\eqref{eqn:ModTeukOrig} for a modified Teukolsky equation of the form 
\begin{multline}
\label{eqn:TeukEqnTempSplit}
    \mathcal{O}[\psi_+]
    +\epsilon\left( A\mathcal{H}\Big[\psi_+^{(0)}\Big]+B\mathcal{I}\Big[\overline{\psi_-^{(0)}}\Big] 
    \right)\\
    +\overline{\gamma}\mathcal{O}[\psi_-]
    +\epsilon\left(
    \overline{B}\mathcal{H}\Big[\psi_-^{(0)}\Big]+\overline{A}\mathcal{I}\Big[\overline{\psi_+^{(0)}}\Big]
    \right)\\
    =0 \,,
\end{multline}
where we have defined
\begin{align}
\label{eqn:TeukOpRMSub}
    \mathcal{H}=\mathcal{V}\mathscr{S}_0^\dagger\,,\\
    \mathcal{I}=\mathcal{V}\overline{\mathscr{S}_0^{\dagger}}\,.
\end{align}
Note that the first line of Eq.~\eqref{eqn:TeukEqnTempSplit} has a temporal and azimuthal dependence of \(e^{-i\omega_+ t+i m \phi}\) while the second line has a dependence of \(e^{-i \omega_- t - i m \phi}\). Each of these lines is then describing a distinct mode, so both lines must vanish independently. This gives two independent equations where the temporal and azimuthal dependence can be divided out. To do so, we will once again define mode-dependent operators denoted with a tilde, for example
\begin{align}
 e^{-i \omega^{}_\pm t\pm i m \phi}   \tilde{\mathcal{H}}_\pm[\tilde{\psi}(r,\theta)]=\mathcal{H}[\tilde{\psi}(r,\theta) e^{-i \omega^{}_\pm t\pm i m \phi}]\,.
\end{align}
The two resulting equations, one for the plus mode and one for the minus mode, are
\begin{subequations}
\label{eqn:SeparatedModTeukEqn}
\begin{align}
\label{eqn:SeparatedModTeukEqn_a}
     \tilde{\mathcal{O}}_+[\tilde{\psi}_+]+\epsilon(A\tilde{\mathcal{H}}_+[\tilde{\psi}_+^{(0)}]+B\tilde{\mathcal{I}}[\overline{\tilde{\psi}_-^{(0)}}])=0\,,\\
\label{eqn:SeparatedModTeukEqn_b}
    \overline{\gamma}\tilde{\mathcal{O}}_-[\tilde{\psi}_-]+\epsilon(\overline{B}\tilde{\mathcal{H}}_-[\tilde{\psi}_-^{(0)}]+\overline{A}\tilde{\mathcal{I}}_-[\overline{\tilde{\psi}_+^{(0)}}])=0\,.
\end{align}
\end{subequations}
Using the expansion 
\begin{multline}
    \psi_\pm \approx\left(\tilde{\psi}_\pm^{(0)}(r,\theta)+\epsilon \tilde{\psi}_\pm^{(1)}(r,\theta)\right)e^{-i\left(\omega_\pm^{(0)}+\epsilon\omega_\pm^{(1)}\right)t\pm im\phi}\,,
\end{multline}
the mode-dependent Teukolsky operator behaves as
\begin{align}
    \tilde{\mathcal{O}}_\pm[\tilde{\psi}_\pm]\approx\epsilon\left(\omega_\pm^{(1)}(\partial_\omega\tilde{\mathcal{O}})_\pm[\tilde{\psi}_+^{(0)}]+\tilde{\mathcal{O}}_\pm[\tilde{\psi}_\pm^{(1)}]\right)\,.
\end{align}
With this expansion substituted back into Eqs.~\eqref{eqn:SeparatedModTeukEqn}, we arrive at two independent equations at order \(\epsilon\),
\begin{subequations}
\label{eqn:HarmonicModTeuk}
    \begin{align}
     \omega_+^{(1)}(\partial_\omega\tilde{\mathcal{O}})_+[\tilde{\psi}_+^{(0)}]+A\tilde{\mathcal{H}}_+[\tilde{\psi}_+^{(0)}]+B\tilde{\mathcal{I}}_+[\overline{\tilde{\psi}_-^{(0)}}] \nonumber\\
     +\tilde{\mathcal{O}}_+[\tilde{\psi}_+^{(1)}]=0\,,
\end{align}
\begin{align}
\label{eqn:HarmonicModTeuk_b}
   \overline{\gamma} \omega_-^{(1)}(\partial_\omega\tilde{\mathcal{O}})_-[\tilde{\psi}_-^{(0)}]+\overline{B}\tilde{\mathcal{H}}_-[\tilde{\psi}_-^{(0)}]+\overline{A}\tilde{\mathcal{I}}_-[\overline{\tilde{\psi}_+^{(0)}}] \nonumber\\
     +\overline{\gamma}\tilde{\mathcal{O}}_-[\tilde{\psi}_-^{(1)}]=0\,.
\end{align}
\end{subequations}
However, \(\tilde{\psi}_\pm^{(1)}\) appears within these equations, which is unknown. 

\subsection{EVP Method and Contour Integration}
\label{sec:ContourIntegration}
To isolate \(\omega^{(1)}\) and eliminate \(\tilde{\psi}_\pm^{(1)}\), we use the EVP method developed in \cite{mark_quasinormal_2015,Zimmerman:2014aha}, analogous to degenerate perturbation theory in quantum mechanics. 
In this analogy, \(\omega^{(1)}\) replaces the first-order energy shift, and the combination of \(\tilde{\mathcal{H}}\) and \(\tilde{\mathcal{I}}\) take the role of the first-order Hamiltonian perturbation. 
In degenerate perturbation theory, left multiplying by the zeroth-order state allows for the elimination of terms with the first-order state utilizing the fact that the zeroth-order Hamiltonian is self-adjoint. We can do the same to our modified Teukolsky equation in Eqs.~\eqref{eqn:HarmonicModTeuk}
by left multiplying with a zeroth-order wavefunction and then take a specially defined scalar product, \(\langle |\rangle\). 
The scalar product is defined such that \(\tilde{\mathcal{O}}_\pm\) is self-adjoint. 
This is done by the addition of a weight function, \(w(r,\theta)\), within the scalar product. 
By making the Teukolsky operator self-adjoint, the left multiplication by the applicable zeroth-order \(\psi^{(0)}\) annihilates the terms with \(\psi^{(1)}\). 
Then there are only terms containing \(\psi^{(0)}\), which we separate into its radial and angular components. 

The radial solution of \(\psi^{(0)}\) can be written using a series expansion of confluent Heun functions, and the full form is given in Appendix \ref{sec:RadContInteg}. 
The angular solutions are the spin-weighted spheroidal harmonics which can be expanded in terms of spin-weighted spherical harmonics,
\begin{align}
    _sS_{\ell m}(x;c)=\sum_{\ell'=\ell_\text{min}}^\infty C_{\ell' \ell m}(c) _sS_{\ell' m}(x;0)\,,
\end{align}
with \(x=\cos\theta, c=a\omega\). The coefficients \( C_{\ell' \ell m}(c)\) can be obtained with recurrence relations and the angular separation constant of the Teukolsky equation, as detailed in \cite{cook_gravitational_2014}. Importantly, the infinite series in both the radial and angular solution can be reasonably truncated with small numerical error. 

\begin{figure}
\includegraphics[width=1.0\columnwidth]{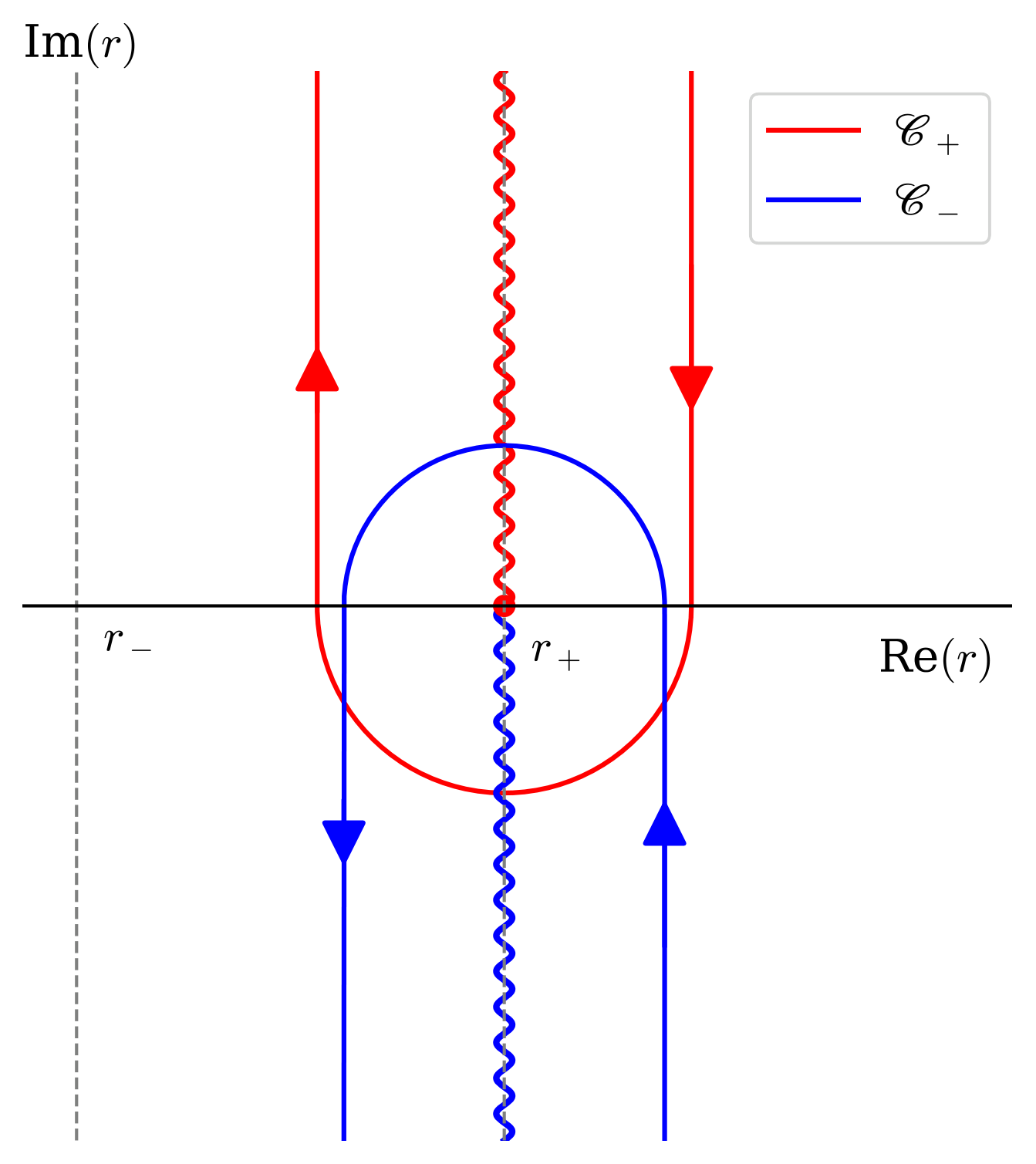} 
\caption{The radial integration contours in the complex plane. The radial portions of the spin-weighted scalars have a branch cut at the outer horizon $r_+$, which the contour wraps around. \(\mathscr{C}_+\) wraps around the upwards-pointing branch cut, and \(\mathscr{C}_-\) wraps around the downwards-pointing branch cut.}
\label{fig:Contour}
\end{figure}

While it is easy to integrate over the angular portion, the spin-weighted scalars are not square-integrable in \(r\) because they diverge at the inner horizon \(r_-\), the outer horizon \(r_+\), and at infinity. 
To deal with this issue, we promote \(r\) to be complex and then integrate around a contour, a trick introduced by Leaver \cite{leaver_spectral_1986}. 
However, under analytic continuation, the radial functions have branch points, and the branch cut extending from the outer horizon is of particular interest.
For \(\psi_+\), we take the branch cut to point upwards in the complex \(r\)-plane because \(\psi_+\) decays exponentially towards positive imaginary \(r\) values.
Conversely, we take the \(\psi_-\) branch cut to point downwards in the complex plane because \(\psi_-\) decays exponentially towards negative imaginary \(r\) values. 
Note that complex conjugation of these wavefunctions produces a reflection over the real \(r\)-axis, flipping the direction of the branch cuts and the exponential decay. 
The exponential decay of the wavefunctions means we can define contours in the complex \(r\)-plane that wrap around the branches and decay exponentially at the boundaries, allowing for numerical evaluation of the radial integral. 
The form of these contours are shown in Fig.~\ref{fig:Contour}. 
We refer to the upwards facing contour as \(\mathscr{C}_+\) and the downwards facing contour as \(\mathscr{C}_-\). 
The process of choosing our phase conventions to rotate the branch cuts to the desired position is described in Appendix~\ref{sec:RadContInteg}.

With the branch cuts pointing in the proper directions, we can define the scalar product that will allow us to calculate the gravitational QNM shifts. 
Using the GHP form of the Teukolsky operator from~\cite{pound_black_2022}, we define the scalar product
\begin{subequations}
\label{eqn:ScalarProduct}
    \begin{align}
    \label{eqn:ScalarProduct_a}
    \langle f(r,\theta)& |g(r,\theta)\rangle_\pm  \equiv \int_{\mathscr{C}_\pm}\int_{0}^\pi f(r,\theta)g(r,\theta)w(r,\theta)d\theta dr\,,
    \end{align}
    \begin{align}
    \label{eqn:WeightFunction}
    & w(r,\theta)=\frac{\Sigma (r-ia\cos\theta)^4}{\Delta^2} \,,
    \end{align}
\end{subequations}
which makes the Teukolsky operator self-adjoint. 
To get a consistent solution for the QNM shift, we require
\begin{align}
\label{eqn:PertPlusMinusRelation}
    \omega_+^{(1)}=-\overline{\omega}_-^{(1)} \,,
\end{align}
which allows us to solve for \(\gamma\).
When we left multiply Eqs.~\eqref{eqn:HarmonicModTeuk} by the proper zeroth-order wavefunction and use the scalar product~\eqref{eqn:ScalarProduct}, the terms with \(\psi^{(1)}\) vanish because the Teukolsky operator is self-adjoint. 
There are then a number of scalar products to evaluate which we define as 
\begin{align}
\label{eqn:InnerProd1}
\langle \delta \mathcal O_+ \rangle & = 
\langle \tilde{\psi}^{(0)}_+ | (\partial_\omega \tilde{\mathcal O})_{+}[\tilde{\psi}^{(0)}_+]\rangle_+ \,, \\
\label{eqn:InnerProd2}
\langle \mathcal H_+ \rangle & = 
\langle \tilde \psi^{(0)}_{+} |\tilde {\mathcal H}_{+} [\tilde \psi^{(0)}_{+}]\rangle_+ \,,\\
\label{eqn:InnerProd3}
\langle \mathcal I_+ \rangle & =  \langle \tilde{\psi}^{(0)}_{+} | \tilde{\mathcal I}_{+}[\overline{\tilde{\psi}^{(0)}_{-}{}}] \rangle_+ \,,
\\
\label{eqn:InnerProd4}
\langle \delta \mathcal O_- \rangle & = 
\overline{\langle \tilde{\psi}^{(0)}_{-} | (\partial_\omega \tilde{\mathcal O})_{-}[\tilde{\psi}^{(0)}_{-}] \rangle_-} \,, \\
\label{eqn:InnerProd5}
\langle \mathcal H_- \rangle & = 
\overline{\langle \tilde \psi^{(0)}_{-} | \tilde {\mathcal H}_{-} [\tilde \psi^{(0)}_{-}] \rangle_-}\,,  \\
\label{eqn:InnerProd6}
\langle \mathcal I_- \rangle & = \overline{\langle  \tilde{\psi}^{(0)}_{-} | \tilde{\mathcal I}_{-} [\overline{\tilde \psi^{(0)}_{+}{}}] \rangle_-} \,.
\end{align}
In this notation, the system of equations for the gravitational QNM shift becomes
\begin{align}
\label{eqn:SolveForOmega_a}
    \omega_+^{(1)}\langle\delta\mathcal{O}_+\rangle+A\langle \mathcal{H}_+\rangle+B\langle\mathcal{I}_+\rangle &=0\,,\\
    \label{eqn:SolveForOmega_b}
    -\omega_+^{(1)}\gamma\langle\delta\mathcal{O}_-\rangle+B\langle \mathcal{H}_-\rangle+A\langle\mathcal{I}_-\rangle &=0\,.
\end{align}
Note when we take the scalar product of Eq.~\eqref{eqn:HarmonicModTeuk_b}, we have also taken the complex conjugate, allowing us to use Eq.~\eqref{eqn:PertPlusMinusRelation} to get an equation containing \(\omega_+^{(1)}\). This is also the reason we use \(\overline{\gamma}\) in Eq.~\eqref{eqn:HarmonicDecomp}. For a consistent solution, Eqs.~\eqref{eqn:SolveForOmega_a} and \eqref{eqn:SolveForOmega_b} must give the same \(\omega_+^{(1)}\), meaning \(\gamma\) must satisfy the quadratic equation given by
\begin{align}
\label{eqn:GammaQuad}
    -\frac{A\langle\mathcal{H}_+\rangle+B\langle\mathcal{I}_+\rangle}{\langle\delta\mathcal{O}_+\rangle}=-\frac{A\langle\mathcal{I}_-\rangle+B\langle\mathcal{H}_-\rangle}{\gamma \langle\delta\mathcal{O}_-\rangle}\,,
\end{align}
recalling that both \(A\) and \(B\) linearly depend on \(\gamma\).
Again, borrowing from the language of degenerate perturbation theory, \(\gamma\) encodes the linear combinations of \(\psi_+\) and \(\psi_-\) that form the ``good" states which diagonalize the perturbation in the degenerate subspace. 
Given the appropriate values of \(\gamma\) that satisfy the quadratic, the final shift to the gravitational QNMs is
\begin{align}
    \omega_+^{(1)}= -\frac{A\langle\mathcal{H}_+\rangle+B\langle\mathcal{I}_+\rangle}{\langle\delta\mathcal{O}_+\rangle}\,.
\end{align}

Although the methodology above holds for a general spin \(a\) and the full JP metric, in this work we limit our investigation to a linear expansion in the spin parameter \(a\) because of computational limitations.
In particular, this limitation is driven by the second-order Einstein tensor, \(G^{(2)}\), whose full form is given by Eq.~(A10) of \cite{hussain_approach_2022}. 
When expanding the required sums for general values of \(a\), the number of terms becomes too large for our \texttt{Mathematica}~\cite{wolfram_research_inc_mathematica_nodate} implementation to handle effectively, especially due to the complicated form of covariant derivatives acting on the reconstructed metric.
The issue is only exacerbated when further applying the source operator \(\mathscr{S}_4\) to this result for use in our scalar products.
However, our methods are able to handle these expressions when the background Kerr metric and the JP metric perturbation are linearized in spin. 
The result is that our gravitational QNM calculations linearized in spin as well. 

\section{Properties of the Shifted Spectrum}
\label{sec:SpectrumProperties}
With the modified Teukolsky formalism and EVP method in hand, we can now explore what they reveal about the structure of the QNM spectrum itself. 
In particular, this methodology makes it easy to discuss the isospectrality and parity properties of the spectrum. 
In this section, we first discuss the WKB method, which allows for a verification of the large-\(\ell\) behavior of our calculations. 
We next investigate what types of metric perturbations produce isospectral QNMs and which produce modes of definite parity from a general viewpoint before applying it to the JP metric.

\subsection{Scalar WKB Approximation}
As a check of our work, we perform a first-order WKB approximation~\cite{schutz_black_1985,iyer_black-hole_1987,seidel_black-hole_1990,yang_quasinormal-mode_2012} of the scalar QNM modes for the slowly rotating JP metric. 
The WKB approximation is a semi-analytic approach that casts the wave equation into a Schr{\"o}dinger-like equation involving an effective potential \(V\),
\begin{align}
    \frac{d^2\psi}{dr_*^2}+V(r,\omega)\psi=0\,.
\end{align}
Here $r_*$ is the standard Kerr tortoise coordinate~\cite{teukolsky_perturbations_1973}, which can be also used for the slowly rotating JP metric. 
The real part of the QNM frequency is implicitly given by the conditions~\cite{iyer_black-hole_1987,yang_quasinormal-mode_2012}
\begin{align}
    \label{eqn:WKBconditions}
    V(r_0,\omega_R)=\frac{\partial V}{\partial r}\biggr|_{(r_0,\omega_R)}=0 \,,
\end{align}
where \(r_0\) is the location of the peak of the effective potential.
For Kerr, \(\omega_R\) can be solved for as a function of \(r_0\), the latter of which is given by the root of a polynomial equation that can be found numerically.
The imaginary part is then given by 
\begin{align}
    \omega_I=(n+1/2)\frac{\sqrt{2 \, d^2 V/dr_*^2|_{r_0,\omega_R}}}{\partial V/\partial\omega|_{r_0,\omega_R}}\,.
\end{align}
For the JP metric, the expression for \(V\) is straightforward to derive, but is lengthy and not illuminating.
In this work we numerically solve Eqs.~\eqref{eqn:WKBconditions} for \(\omega_R\) and \(r_0\) jointly.
From these we can derive \(\omega_I\).  
We then use finite differencing in small \(\epsilon\) and \(\chi\) to identify the real and imaginary QNM shifts for slowly spinning JP at leading order in \(\epsilon\).
Valid for large \(L=\ell+1/2\), this first-order scalar WKB approximation converges to the gravitational modes as \(\sim 1/L\). 
With this first-order WKB method, we also find good agreement with the higher order WKB analysis of \cite{konoplya_radiation_2013} for non-rotating JP black holes. 
Consequently, we can use this approximation to verify that the QNMs calculated from the modified Teukolsky formalism have the proper limiting behavior.

\subsection{Isospectrality Breaking}
\label{sec:Isospectrality}
For non-rotating BHs in GR, spherical symmetry can be used to separate the angular and radial sectors of the metric perturbations directly, as well as into axial (odd parity) and polar (even parity) sectors.
In each sector, the metric components are directly related to a master scalar variable, and these obey wave equations involving the Regge-Wheeler \cite{regge_stability_1957} and Zerilli-Moncrief \cite{zerilli_gravitational_1970, moncrief_gravitational_1974} potentials respectively. 
These equations can be solved to find the QNM spectrum for the axial and polar modes. 
Despite being governed by different potentials, both sectors have the same QNM frequencies.
This degeneracy in the spectrum is referred to as isospectrality \cite{chandrasekhar_mathematical_1983}. 

In Kerr, the Teukolsky formalism is required due to the lack of spherical symmetry \cite{teukolsky_perturbations_1973,press_perturbations_1973,teukolsky_perturbations_1974}. 
Rather than starting from the metric perturbations directly, $h_{ab}$ must be   reconstructed via the Weyl scalars. 
Using the Teukolsky formalism, definite parity metric perturbations are produced from a combination of even and odd parity modes, as first shown in \cite{chrzanowski_vector_1975}. 
Metric perturbations of both even and odd parity again have the same QNM frequencies, such that Kerr is isospectral like Schwarzschild.

For modified theories of gravity, the properties of Schwarzschild and Kerr that result in isospectrality are not generally present. 
By applying the EVP method to a modified Teukolsky equation, we can derive the specific conditions a modified theory must satisfy for isospectrality, and then apply them to the slowly rotating JP metric perturbation. 
For this we follow the approach of Ref.~\cite{li_isospectrality_2024}, adapting the details to our formalism.
Generally, the EVP method admits two different shifts to the QNM spectrum since Eq.~\eqref{eqn:GammaQuad} is quadratic in \(\gamma\). 
If isospectrality is to be retained, \(\gamma\) can only have one unique solution, whereas two solutions would break isospectrality through a Zeeman-like splitting of the QNM frequencies.

By expanding Eq.~\eqref{eqn:GammaQuad} into the form \(b\gamma^2+c\gamma+d=0\), we see \(\gamma\) has one unique solution if either \(b=0\) or \(c^2-4bd=0\). In terms of our scalar products and other variables, these conditions are
\begin{subequations}
\label{eqn:QuadConditions}
    \begin{align}
    \label{eqn:FirstQuadCondition}
        D_+\langle\mathcal{H}_+\rangle-12i M \omega_+^{(0)} \langle\mathcal{I}_+\rangle=0\,,
    \end{align}
    \begin{align}
    \label{eqn:SecondQuadCondition}
       &4(D_+ \langle\mathcal{H}_-\rangle-12 i M \omega_+^{(0)} \langle\mathcal{I}_-\rangle)(D_+ \langle\mathcal{H}_+\rangle-12 i M \omega_+^{(0)} \langle\mathcal{I}_+\rangle) \nonumber\\
       &+\left(D_+ (\langle\mathcal{I}_-\rangle-\langle\mathcal{I}_+\rangle)-12 i M \omega_+^{(0)} (\langle\mathcal{H}_-\rangle-\langle\mathcal{H}_+\rangle)\right)^2=0\,.
    \end{align}
\end{subequations}
While these conditions may not give a good physical intuition as to what types of perturbations will preserve isospectrality, it is clear that these conditions are quite strict and not easily satisfied. 
Only a perturbation that has just the right structure can have isospectral QNMs. 
For example, we can treat a slowly rotating Kerr black hole as a linear perturbation away from Schwarzschild with the spin \(a\) as our perturbative parameter. This type of perturbation satisfies Eq.~\eqref{eqn:FirstQuadCondition}, leading to an isospectral shift in the QNM spectrum, exactly as we would expect from a Kerr black hole in GR. 
Conversely, despite the simplicity of the slowly rotating JP metric perturbation, neither of the conditions in Eqs.~\eqref{eqn:QuadConditions} are satisfied. 
This means the JP metric perturbation breaks the isospectrality of Kerr, discussed further in Sec.~\ref{sec:Results}.

\subsection{Definite Parity}
\label{sec:DefiniteParity}
As discussed above, perturbations of Schwarschild and Kerr BHs separate into independent sectors of definite parity.
To extend this concept to bGR theories, assume that the background spacetime admits Boyer–Lindquist–like coordinates \((t,r,\theta,\phi)\), as is the case for Kerr and the JP metric.
To analyze definite parity modes, let us begin by defining the conjugate-parity operator \(\hat{\mathcal{P}}\)~\cite{li_isospectrality_2024},
\begin{align}
\hat{\mathcal{P}}f(t,r,\theta,\phi)=\hat{C}\hat{P}f(t,r,\theta,\phi)=\overline{f}(t,r,\pi-\theta,\phi+\pi)\,,
\end{align}
where \(\hat{C}\) and \(\hat{P}\) are the complex conjugation and the parity transformation respectively. 
Under the conjugate-parity operator, modes with definite even (E) and odd (O) parity and angular quantum number \(\ell\) behave as
\begin{align}
    \label{eqn:CPTransform}
    \hat{\mathcal{P}}[\psi^{\text{E,O}}]=\pm(-1)^\ell \psi^{\text{E,O}}\,.
\end{align}
Work from \cite{li_isospectrality_2024,nichols_visualizing_2012} showed that with a Kerr background, the reconstructed metric \(h_{ab}^{(0)}\) has definite parity when \(\psi^{(0)}\) also has definite parity, so that \(h_{ab}^{(0)}\) transforms in the manner of Eq.~\eqref{eqn:CPTransform}. 
With Kerr as the background, the spin-weighted scalar combinations with definite parity are  
\begin{subequations}
\label{eqn:DefParityModes}
    \begin{align}
    \psi^{(0)}_\text{E}&=\psi_+^{(0)}+\psi_-^{(0)}\,,\\
    \psi^{(0)}_\text{O}&=\psi_+^{(0)}-\psi_-^{(0)}\,.
\end{align}
\end{subequations}
Comparing to Eq.~\eqref{eqn:HarmonicDecomp}, this means definite parity metric perturbations correspond to \(\gamma=\pm1\). 
Reference \cite{li_isospectrality_2024} proved that the modified Teukolsky equation admits solutions of \(\gamma=\pm1\) if and only if the operators perturbing the Teukolsky equation are conjugate-parity invariant.
In our case, where we take the Cowling approximation and so there are no new fields coupling to the spacetime curvature, this reduces to the condition
\begin{align}
\label{eqn:InvariantV}
    \hat{\mathcal{P}}[\mathcal{V}]=\mathcal{V}\,,
\end{align}
where we recall that \(\mathcal{V}\) is a linear operator acting on the reconstructed metric, or equivalently on the spin-weighted scalar corresponding to \(\Psi_4\). 
Equation~\eqref{eqn:InvariantV} states that the required condition is that \(\mathcal{V}\) be parity preserving.
This makes intuitive sense if we look back at the modified Teukolsky equation in Eq.~\eqref{eqn:ModTeukOrig}. 
The normal Teukolsky operator \(\mathcal{O}\) is parity preserving, so to admit definite parity modes, the additional piece in the modified Teukolsky equation must also be parity preserving. 
Our goal now is to derive the conditions satisfied by \(g^{(1)}_{ab}\) such that Eq.~\eqref{eqn:InvariantV} holds.
The \(\mathcal{V}\) operator is composed of the source operator \(\mathscr{S}_4\) and the second order Einstein tensor \(G^{(2)}_{ab}\) which is a function of the metric perturbation \(g^{(1)}_{ab}\). 
To see when \(\mathcal{V}\) is parity-preserving, we must dive deeper into the properties of \(\mathscr{S}_4\) and \(G^{(2)}_{ab}\).

\subsection{Parity of Operators and Tetrad}
To start disentangling the behavior of these operators, it is helpful to first explore the behavior of the GHP quantities under the conjugate-parity transformation. 
We label background quantities that remain unchanged under the conjugate-parity transformation \(\hat{\mathcal{P}}\) as \((+)\) parity, while those that gain an overall minus sign are \((-)\) parity. 
We also want to classify tensors where its components may not all have the same parity. 
Due to the parity operation that takes \(\theta\rightarrow\pi-\theta\), many tensors gain an overall minus sign in their \(\theta\)-components. 
A tensor \(T_{a_1,a_2,...}^{b_1,b_2,...}\) has \(\uparrow\) (upwards) parity if components of \(T\) with an even number of \(\theta\) indices are of \((+)\) parity and components with an odd number of \(\theta\) indices are of \((-)\) parity. 
Conversely, \(T_{a_1,a_2,...}^{b_1,b_2,...}\) has \(\downarrow\) (downwards) parity if components of \(T\) with an even number of \(\theta\) indices are of \((-)\) parity and components with an odd number of \(\theta\) indices are of \((+)\) parity. 
As an example, take the arbitrary tensor \(A^\uparrow_{ab}\) with \(\uparrow\) parity and \(B^\downarrow_{ab}\) which has \(\downarrow\) parity. 
Schematically, the parity of their components looks like
\begin{align}
A^\uparrow_{ab}=
    \begin{pmatrix}
    + & + & - & + \\
    + & + & - & + \\
    - & - & + & - \\
    + & + & - & + 
\end{pmatrix} \qquad 
B^\downarrow_{ab}=
    \begin{pmatrix}
    - & - & + & - \\
    - & - & + & - \\
    + & + & - & + \\
    - & - & + & - 
\end{pmatrix}
\end{align}
for coordinates \((t,r,\theta,\phi)\). 
Importantly, multiplying something of \(\uparrow\)/\(\downarrow\) parity by something of \((-)\) parity results in \(\downarrow\)/\(\uparrow\) parity. 
Conversely, multiplying by something of \((+)\) parity leaves \(\uparrow\)/\(\downarrow\) parity objects unchanged. 
Contractions work similarly. 
Contracting objects of the same \(\uparrow\)/\(\downarrow\) parity on one index results in a \(\uparrow\) parity object. 
Contracting objects of opposite \(\uparrow\)/\(\downarrow\) parities on one index results in a \(\downarrow\) parity object. 
Since the background Kerr metric has \(\uparrow\) parity, raising or lowering the indices of a \(\uparrow\)/\(\downarrow\) parity object does not change its \(\uparrow\)/\(\downarrow\) parity.

Now that the different types of parities have been defined, we can look specifically at the parities for GHP spin coefficients \(\rho,\tau,\beta, \epsilon\) as well as the GHP derivative operators \(\Th, \eth\) following the conventions of \cite{pound_black_2022}. 
The parity of each of these objects is summarized in Table \ref{tab:Parity}. Each of these also has a primed and complex-conjugated variant, but the parity remains unchanged under these operations. Using the parity of the Newman-Penrose quantities from \cite{li_isospectrality_2024}, we find that \(\Th,\rho,\epsilon\) have \((+)\) parity while \(\eth,\tau,\beta\) have \((-)\) parity.

The null tetrad also appears within the \(\mathscr{S}_4\) operator so it is important to understand its behavior as well. Using the Kinnersley tetrad \cite{kinnersley_type_1969}, \(l^\mu\) and \(n^\mu\) both remain unchanged under the conjugate-parity transformation. However, note that \(l^\theta=0\) and \(n^\theta=0\) meaning they can be considered as having \(\uparrow\) parity. Conversely, \(m^\mu\) gains an overall minus sign except for in the \(m^\theta\) component under conjugate-parity, meaning it has \(\downarrow\) parity. 

Finally, we must consider the covariant derivative \(\nabla_\mu\). Assuming a Kerr background, \(\nabla_\mu\) is a \(\uparrow\)/\(\downarrow\) parity-preserving operator. If it acts on a tensor with \(\uparrow\) parity, the resulting tensor will also have \(\uparrow\) parity and vice-versa. This happens because \(\partial_\theta\) gains a minus sign under conjugate parity, and the Kerr Christoffel symbols behave as if they have \(\uparrow\) parity. When acting on a tensor with definite \(\uparrow\)/\(\downarrow\) parity, this becomes a \(\uparrow\)/\(\downarrow\) parity-preserving effect. This preserving effect means we can think of the covariant derivative having \(\uparrow\) parity.

\begin{table}[tb]
\centering
\renewcommand{\arraystretch}{1} % spacing
\setlength{\tabcolsep}{12pt}
\begin{tabular}{cccc}
\hline\hline
$(+)$ & $(-)$ & $\uparrow$ & $\downarrow$ \\
\hline
\rule{0pt}{3ex} $\Th, \rho,\epsilon$ 
 & $\eth,\tau,\beta$ 
 & $l^\mu, n^\mu, \nabla_\mu$ 
 & $m^\mu, \overline{m}^\mu$  \\
\hline\hline
\end{tabular}
\caption{Parity properties of GHP quantities, the null tetrad, and the covariant derivative.}
\label{tab:Parity}
\end{table}

\subsubsection{Parity of \texorpdfstring{$\mathscr{S}_4$}{S4}}
With the parity behavior of these quantities now understood, we can look at the parity behavior of the \(\mathscr{S}_4\) operator as a whole. Looking at the form of \(\mathscr{S}_4\) in Eq.~\eqref{eqn:S4}, we can see that each set of parentheses contains GHP objects of the same parity. Furthermore, the pairs of tetrad legs combine to form tensors of definite \(\uparrow\)/\(\downarrow\) parity. 
Using \((+)\)/\((-)\) and \(\uparrow\)/\(\downarrow\), \(\mathscr{S}_4\) schematically can be written as
\begin{align}
\label{eqn:SchematicSourceOp}
\mathscr{S}_4^{ab}  \sim &
  \tfrac12(-) 
    \big[(+)(\downarrow)^{ab}  -(-)(\uparrow)^{ab}  \big] \nonumber\\
  &+ \tfrac12(+)
    \big[(-) (\downarrow)^{ab} - (+)(\uparrow)^{ab}  \big]\,.
\end{align}
From this schematic form, it becomes evident that overall, \(\mathscr{S}_4\) has \(\uparrow\) parity. 
This realization is important for our analysis because it restricts what \(\mathscr{S}_4\) can act on in order to satisfy Eq.~\eqref{eqn:InvariantV}, as it must result in something of \((+)\) parity.
Say \(\mathscr{S}_4\) acts on a tensor \(T_{ab}\) without definite \(\uparrow\)/\(\downarrow\) parity. 
We can always break this tensor up into components of \(\uparrow\) and \(\downarrow\) parity and see how \(\mathscr{S}_4\) acts on these components:
\begin{align}
    T_{ab}=T^\uparrow_{ab}+T^\downarrow_{ab}\,, \\
    \mathscr{S}_4^{ab}T^\uparrow_{ab}\sim(+)\,, \\
    \mathscr{S}_4^{ab}T^\downarrow_{ab}\sim(-)\,.
\end{align}
However, \(\mathscr{S}_4^{ab}T_{ab}\) itself must result in a scalar of even parity, so 
\begin{align}
    \mathscr{S}_4^{ab}T^\downarrow_{ab}=0
\end{align}
is required for there to be definite parity modes. 

\subsubsection{Parity of \texorpdfstring{$G^{(2)}_{ab}$}{G{(2)}{ab}}}

In practice, \(\mathscr{S}_4\) acts on the second order Einstein tensor \(G^{(2)}_{ab}\), which takes in the reconstructed metric \(h^{(0)}_{ab}\) and depends on the \(g^{(1)}_{ab}\) metric perturbation. 
It is now helpful to think of the \(g^{(1)}_{ab}\) dependence as part of the operator, and the operator takes in \(h^{(0)}_{ab}\). 
The \(G^{(2)}_{ab}\) operator is the product of the background metric \(g^{(0)}_{ab}\), the \(g^{(1)}_{ab}\) metric perturbation, and covariant derivatives, all acting on the \(h^{(0)}_{ab}\) metric perturbation. 
Still assuming a Kerr background, \(g^{(0)}_{ab}\) and the covariant derivatives have \(\uparrow\) parity. 
Then \(g^{(1)}_{ab}\) can be broken into its \(\uparrow\) and \(\downarrow\) parity components,
\begin{align}
    g^{(1)}_{ab}=g^{(1)\uparrow}_{ab}+g^{(1)\downarrow}_{ab}\,.
\end{align}
Based on the structure of \(G^{(2)}_{ab}\), when it depends on the upwards-parity component \(g^{(1)\uparrow}_{ab}\), \(G^{(2)}_{ab}\) also has upwards-parity when viewed as an operator that acts on  \(h^{(0)}_{ab}\). 
Likewise, when it depends on the downwards-parity component \(g^{(1)\downarrow}_{ab}\), \(G^{(2)}_{ab}\) will also have downwards-parity:
\begin{align}
    G^{(2)}_{ab}(g^{(1)}_\uparrow)\sim G^{(2)\uparrow}_{ab}\, \\
    G^{(2)}_{ab}(g^{(1)}_\downarrow)\sim G^{(2)\downarrow}_{ab}\,.
\end{align}
From this behavior, a metric perturbation \(g^{(1)}_{ab}\) of completely \(\uparrow\) parity will result in definite parity modes. 
In such a case, \(\mathscr{S}_4^{ab} G^{(2)}_{ab}\) would have \((+)\) parity, meaning it is invariant under \(\hat{\mathcal{P}}\) as required. 
Alternately, we can place restrictions on the completely \(\downarrow\) parity portion of the metric perturbation and still have \(\mathcal V\) be parity-preserving. 
Namely, the restriction is that
\begin{align}
    \mathscr{S}_4^{ab}\left[G^{(2)}_{ab}[g^{(1)}_\downarrow,h^{(0)}]\right]=0\,.
\end{align}
This implies that, in some sense, the \(\downarrow\) parity portion of the metric perturbation cannot couple to gravitational radiation. 

From these restrictions on \(g^{(1)}_{ab}\), it becomes clear that slowly rotating JP metric perturbations separates into definite parity modes. 
The slowly rotating JP metric perturbation in Eq.~\eqref{eqn:SlowlyRotatingJPMetric} has completely \(\uparrow\) parity with no \(\downarrow\) parity. 
This means that \(\mathcal{V}\) obeys Eq.~\eqref{eqn:InvariantV}, guaranteeing that the modes are of definite parity. 
While our work primarily focuses on the slowly rotating JP metric, note that this can be easily extended to the full JP metric. 
From Eq.~\eqref{eqn:JPMetric}, we can see that the full JP metric perturbation has \(\uparrow\) parity, meaning the full JP metric has produce definite parity modes as well.

\section{Results}
\label{sec:Results}

\begin{figure*}[t]
\includegraphics[width=\textwidth]{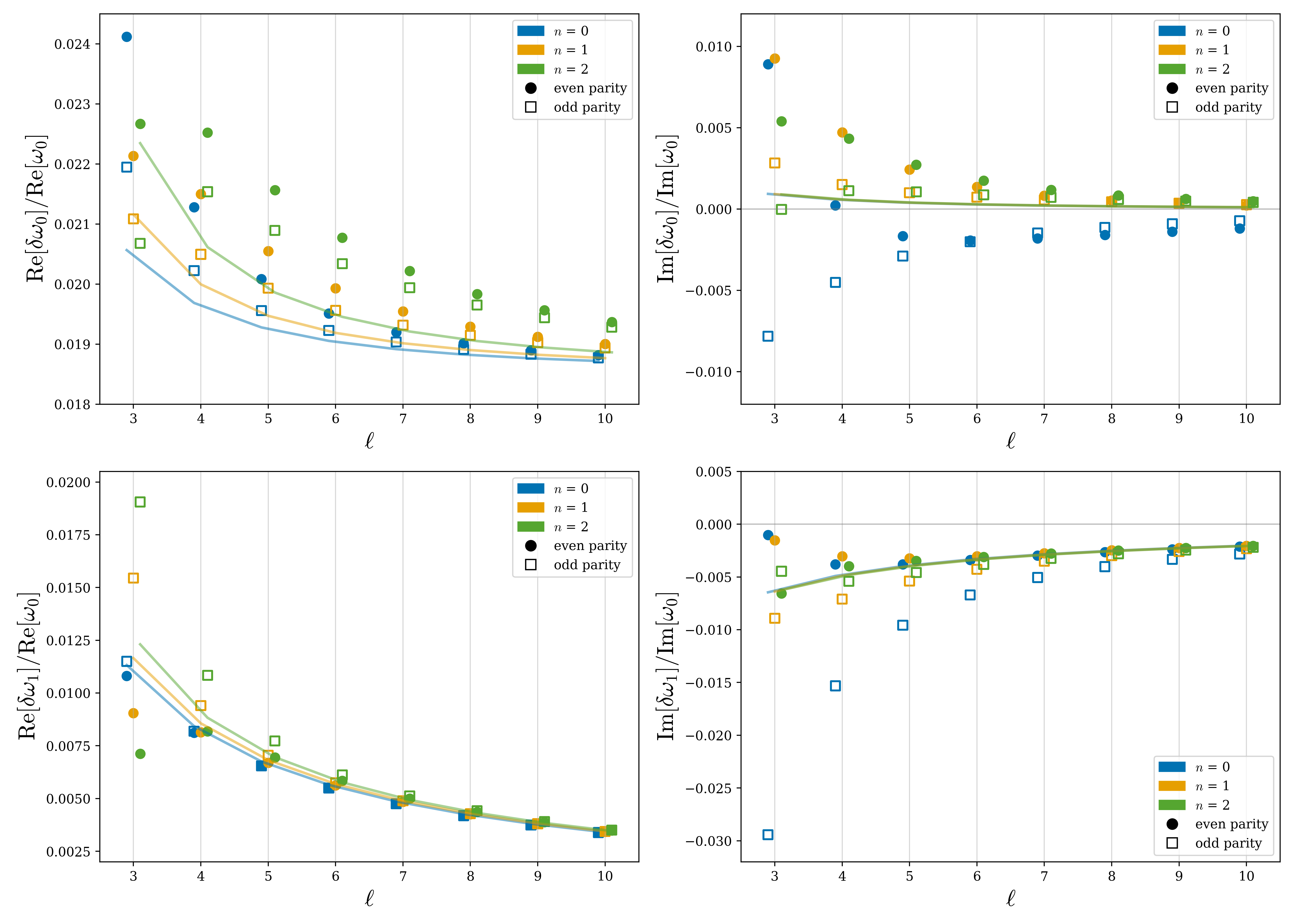}
\caption{The real (left column) and imaginary (right column) QNM shifts of \((\delta\omega_0)_{\ell n}\) (top row) and \((\delta\omega_1)_{\ell n}\) (bottom row) for modes \(3\leq\ell \leq 10\) and \(n=0,1,2\). For each value of \(\ell\) and \(n\), both even and odd parity shifts are shown. All shifts are normalized by \(\omega_0\), the corresponding Schwarzschild QNM frequency for each mode. The scalar WKB approximation for each overtone normalized by \(\omega_0\) is given by the line of the corresponding color. Note that although the real part of the WKB approximation has no \(n\) dependence, the normalization by \(\omega_0\) causes a splitting in between the overtones. A small horizontal jitter has also been added to improve the visibility of overlapping points.}
\label{fig:Shifts}
\end{figure*}

Using the modified Teukolsky formalism in conjunction with the EVP method, we calculate the shifts in the QNM spectrum for the slowly rotating JP metric for \(2\leq \ell \leq 10\), all \(m\), and \(n=0,1,2\). 
In our numerical integration procedure, all QNMs are computed with a relative tolerance of \(O(10^{-4})\), ensuring that systematic \(O(a^2)\) error quickly becomes dominant for any appreciable spin. 
Interestingly, the slowly rotating JP QNM shifts behave similarly to those of the slowly rotating Kerr QNMs computed on a Schwarzschild background. 
When treating slowly rotating Kerr BHs as a perturbation to Schwarzschild with the dimensionless spin \(\chi=a/M\) as the perturbative parameter, the QNMs can be written as
\begin{align}
    \omega^\text{Kerr}_{\ell m n} = \omega^\text{Schw}_{\ell n}+\chi m \, \delta\omega^{}_{\ell n} +O(\chi^2)\,.
\end{align}
Due to the linearization in spin, \(m\) can be pulled out of the QNM shift. A proof of this using the EVP method is presented in Appendix \ref{sec:LinKerrShifts} and hinges on the fact that the Schwarzschild background has spherical symmetry. 
Despite having a slowly rotating background, the slowly rotating JP QNMs behave in a similar way, where the QNM shifts can be decomposed as 
\begin{align}
\label{eqn:JPQNMDecomp}
    \omega^{(1)}_{\ell m n} \approx (\delta\omega^{}_{0})_{\ell n}+\chi m \, (\delta\omega^{}_{1})_{\ell n}\,,
\end{align}
where \(\delta\omega_0\) is the spin-independent shift, and \(\delta\omega_1\) is the spin-dependent shift.
We find this result empirically through our numerical calculations, and the analytic origin of this simplifying features is unclear.
Deriving the reason for this simplification is a goal for future work.

\begin{figure*}[t]
\includegraphics[width=\textwidth]{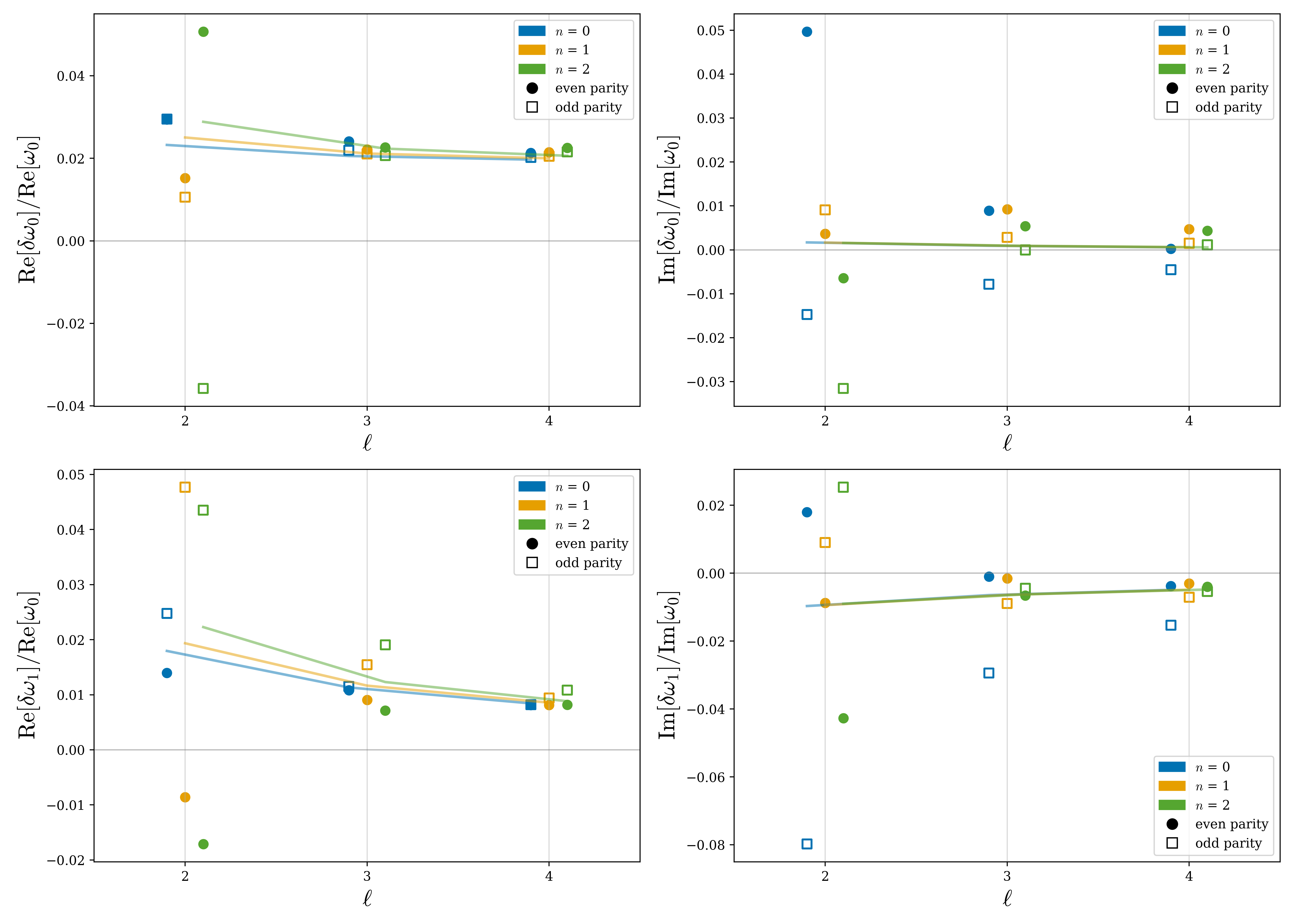}
\caption{The real (left column) and imaginary (right column) QNM shifts of \((\delta\omega_{0})_{\ell n}\) (top row) and \((\delta\omega_{1})_{\ell n}\) (bottom row) for modes \(2\leq\ell \leq 4\) and \(n=0,1,2\). For each value of \(\ell\) and \(n\), both even and odd parity shifts are shown. The scalar WKB approximation for each overtone normalized by \(\omega_0\) is given by the line of the corresponding color. Note that although the real part of the WKB approximation has no \(n\) dependence, the normalization by \(\omega_0\) causes a splitting in between the overtones. All shifts are normalized by \(\omega_0\), the corresponding Schwarzschild QNM frequency for each mode. A small horizontal jitter has also been added to improve the visibility of overlapping points.}
\label{fig:l2Shifts}
\end{figure*}

Accordingly, the real and imaginary parts of \(\delta\omega_0\) and \(\delta\omega_1\), normalized by the corresponding Schwarzschild frequency \(\omega_0\), are shown in Fig.~\ref{fig:Shifts} and Fig.~\ref{fig:l2Shifts} for \(n=0,1,2\) and \(2\leq \ell \leq 10\). 
We also plot the scalar WKB approximation for each mode. 
All computed shifts are given in Table \ref{tab:AllShifts} of Appendix \ref{sec:TabulatedQNMs}, and are publicly available \cite{wu2026beyondgrqnms}. 
As discussed in Sec.~\ref{sec:Isospectrality}, we see that the slowly rotating JP metric breaks isospectrality, leading to a split in the QNM frequency that lifts the even/odd parity perturbation degeneracy. 
Furthermore, we confirm numerically our proof from Sec.~\ref{sec:Isospectrality} that the modes have definite parity. 
When solving for the ``good" states of the degenerate subspace, we find that \(\gamma=\pm 1\) as expected for modes of definite parity \cite{li_isospectrality_2024}. 
Even parity metric perturbations correspond to \(\gamma=+1\) and odd parity metric perturbations correspond to \(\gamma=-1\). 
Consequently, we refer to the associated QNM shifts as the even and odd parity modes. 

From Fig.~\ref{fig:Shifts}, we see that as \(\ell\) increases, the even and odd parity modes begin to converge. This is expected because at large \(\ell\), the QNMs are predominantly determined by the geometry of the spacetime. In particular, we approach the eikonal limit where the QNMs are governed by the dynamics of unstable null orbits on the photon sphere.
These dynamics are agnostic of the parity of the perturbation \cite{ferrari_new_1984}. 
It should be noted that this behavior does not generally hold for bGR theories.
In GR, the eikonal limit corresponds to the geometric optics limit, and all GWs travel along null geodesics regardless of their parity, a direct result from the linearized Einstein equations in this limit.
However, dynamical bGR effects and coupling to new bGR fields can modify the linearized Einstein equations such that this is longer the case.
The propagation speed of a GW can then depend on its parity, meaning even and odd parity waves will travel at different speeds. 
Consequently, GWs of opposite parity travel along different paths, which breaks isospectrality, e.g.~\cite{bryantEikonalQuasinormalModes2021,canoIsospectralityEffectiveField2024}. 
If dynamical effects (see Sec.~\ref{sec:ModTeukEqn}) were incorporated in our investigation of the JP QNMs, it is thus possible that the even and odd parity modes would not converge in the eikonal limit.

Interestingly, there is not any hierarchy between even and odd parity modes. 
This is particularly evident in the \(\text{Im}[(\delta\omega_0)_{\ell 0}]\) shifts. 
For large \(\ell\), the odd parity shifts lie above the even parity shifts, but this is reversed for smaller \(\ell\).
Additionally, it is important to look at the relative size of these QNM shifts. 
Since we have normalized by the corresponding Schwarzschild frequency, Fig.~\ref{fig:Shifts} and Fig.~\ref{fig:l2Shifts} show the percent-level shift caused by the slowly rotating JP perturbation. 
The largest \(\delta\omega_0\) shifts are \(\sim 5\%\). 
This is a small shift, and has an additional prefactor of \(\epsilon\lesssim O(1)\), decreasing the shift even further. 
Similarly, the largest \(\delta\omega_1\) shift is \(\sim 8\%\). 
This spin-dependent portion of the shift has the same prefactor of \(\epsilon\lesssim O(1)\) and an additional factor of \(\chi\) which is small due to our slow rotation expansion. 
Therefore, the total slowly rotating JP QNM shifts are few- to sub-percent-level. 
The small size of these shifts means that ringdown-based constraints on the JP metric may not the best avenue to search for such deviations from GR. 
QNM shifts of \(\sim 1\%\) require an SNR of at least \(\sim 150\) \cite{pacilio_identifying_2023}, almost twice as loud as the current loudest GW event \cite{abac_gw250114_2025}.

While the slowly rotating JP QNMs follow clear trends for \(3\leq \ell \leq 10\), we find interesting behavior in the \(\ell=2\) modes, highlighted in Fig.~\ref{fig:l2Shifts}. 
First, the \(\ell=2\) modes highlight the limits of the WKB approximation. 
Expected to hold at large \(\ell\), we see that many calculated QNM shifts for \(\ell=2\) deviate from the WKB approximation. 
Furthermore, many \(\ell=2\) modes lie off the trends apparent at larger \(\ell\). 
This is exemplified by the \(n=1\) overtone. 
The \(\text{Re}[(\delta\omega_0)_{\ell 1}]\) and \(\text{Re}[(\delta\omega_1)_{\ell 1}]\) shifts of both parities monotonically decrease for \(\ell\geq 3\). 
However, this trend is broken at \(\ell=2\) by \(\text{Re}[(\delta\omega_0)_{2 1}]\) for both parities, and for the even-parity \(\text{Re}[(\delta\omega_1)_{\ell 1}]\) shift. 
Similar monotonic trends are broken by the even-parity \(\text{Im}[(\delta\omega_0)_{2 1}]\) and \(\text{Re}[(\delta\omega_1)_{2 1}]\) shifts. 
Additionally, at the \(n=2\) overtone, the breaking of large-\(\ell\) monotonic behavior begins at \(\ell=3\). 
In particular, the even-parity \(\text{Re}[(\delta\omega_1)_{3 2}]\) and odd-parity \(\text{Im}[(\delta\omega_1)_{3 2}]\) deviate from their corresponding monotonic trends. 
In the even-parity \(\text{Re}[(\delta\omega_0)_{\ell 2}]\) shifts, the \(\text{Re}[(\delta\omega_0)_{3 2}]\) shift appears to start moving away from the corresponding monotonic trend, which is then clearly broken by the \(\text{Re}[(\delta\omega_0)_{2 2}]\) shift. 
It would be interesting to uncover the physical origin of the difference between lower-\(\ell\) shifts and the large--\(\ell\) trends, and our results suggest that the effect is overtone-dependent: the effect is absent for \(n=0\) modes for \(\ell \geq 2\), begins at \(\ell=2\) for the \(n=1\) modes, and moves further out to \(\ell=3\) for some \(n=2\) modes.

\begin{figure*}[t]
\includegraphics[width=\textwidth]{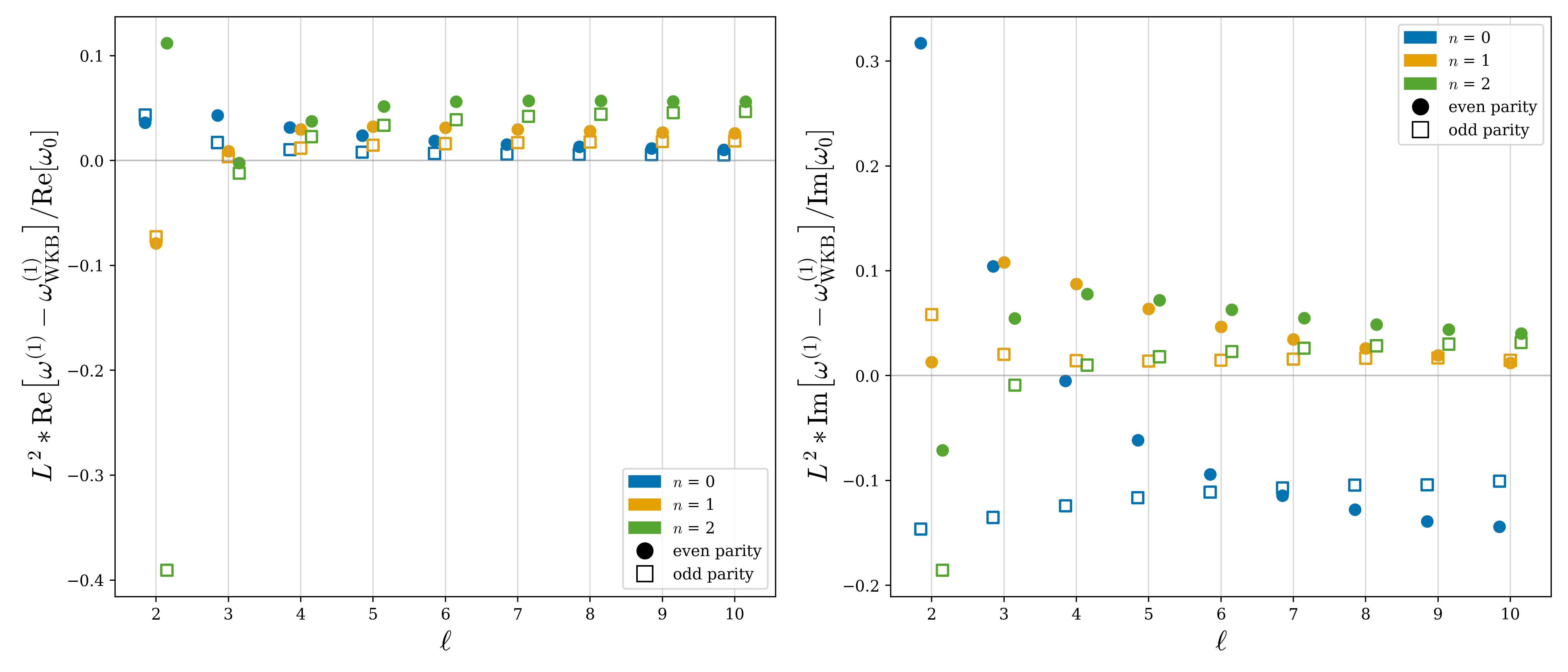} 
\caption{Normalized differences between the real (left) and imaginary (right) parts of gravitational modes from the EVP method and the scalar modes from the first-order WKB method for \(2\leq \ell\leq  10\), \(m=1\), \(n=0,1,2\), and \(\chi=0.1\). The differences are normalized by the respective Schwarzschild gravitational frequencies, \(\omega_0\). The normalized differences are further scaled by \(L^2\).}
\label{fig:WKB}
\end{figure*}

To verify the large-\(L\) behavior, we use our scalar WKB approximation. If we take the relative difference between the slowly rotating JP QNMs and those from the WKB approximation, we expect a convergence of \(1/L^2\) at large \(L\).
This convergence is shown in Fig.~\ref{fig:WKB} for a spin of \(\chi=0.1\), where we scale the normalized difference by \(L^2\). 
The normalized differences follow the expected convergence as all modes asymptote to some constant value for large \(\ell\) for both parities and all overtones. 
From this asymptotic behavior, we see that our slowly rotating QNMs have the proper limiting behavior for large \(\ell\). 
We recall that in general we would not expect this convergence if the bGR dynamical terms are included because these terms break the wave-nature of the linearized Einstein equations in the eikonal limit.

Due to the linearization in spin for this work, we are limited to small spins. Higher order expansions in spin lead to the same proliferation of terms described in Section \ref{sec:ContourIntegration} so it is difficult to ascertain the size of the \(O(a^2)\) error. To estimate the effect of the higher order spin corrections, we employ the JP eikonal approximation from \cite{glampedakis_post-kerr_2017} which is general in spin and valid for \(\ell\gg1\). Specializing to \(|m|=\ell\), the eikonal approximation takes the form
\begin{align}
    \omega_{\ell n}\approx (\ell+\tfrac{1}{2})\Omega - i (n+\tfrac{1}{2})\gamma_\text{L}\,,
\end{align}
where \(\Omega\) is the Keplerian orbital frequency of the equatorial light ring, and \(\gamma_\text{L}\) is the Lyapunov exponent. By linearizing their eikonal approximation in \(\chi\), we find that the linearized version remains within \(\sim1\%\) of the generalized version for \(\chi\lesssim0.2\). As such, we expect a similar regime of validity for our EVP method. For comparison, the slowly rotating Kerr QNMs remain within \(\lesssim 1\%\) error for \(\chi\lesssim 0.27\). 

Finally, we note that the convergence of our EVP results to the eikonal approximation is only clear at sufficiently large \(L\) values.
For example, the imaginary part of the even-parity \(n=0\) modes asymptote to a constant value slowly, as seen in Fig.~\ref{fig:WKB}.
This is interesting in the context of recent results for the QNMs of bumpy BHs~\cite{collins_towards_2004,weller_spectroscopy_2025}.
Recently Ref.~\cite{weller_spectroscopy_2025} found indications that the eikonal approximation may not be valid, but only studied modes with the \(\ell\leq 5\). 
Our results hint that the eikonal limit may still be valid for bumpy BHs, but this must be studied at even higher \(L\) values.

\section{Conclusion}
\label{sec:Conclusion}
In this work, we have used the modified Teukolsky formalism and EVP method to calculate the gravitational QNM shifts of the slowly rotating JP metric for \(2\leq\ell\leq10\), all \(m\), and overtones \(n=0,1,2\). 
Applying the EVP method to the modified Teukolsky equation requires solving a problem in degenerate perturbation theory, since the Kerr spectrum is isospectral.
Expanding on previous work~\cite{hussain_approach_2022}, we present a formulation of metric reconstruction that allows for convenient handling of the degeneracy.
Our approach makes it easy to assess if our bGR metric breaks isospectrality and if it retains definite parity sectors, properties governed by a mixing parameter \(\gamma\).
By introducing the notion of \(\uparrow\)/\(\downarrow\) parity and building on past results~\cite{li_isospectrality_2024}, we further derive conditions a bGR metric must satisfy in order for the shifted spectrum to admit definite parity modes, finding that metric perturbations of \(\uparrow\) parity like the JP metric do admit definite parity modes. 

These conditions allow for the simple identification of other theories which admit definite parity modes by examining the form of the bGR metric perturbation. 
Requirements for isospectrality, the property that the even- and odd-parity spectra are the same, naturally falls out of our formalism as well, showing the stringent conditions required.
The results derived here neglect the important cases where new fields couple to the curvature through bGR interactions, effects encoded in the additional \(\mathcal C\) operator of~\cite{hussain_approach_2022}.
In such cases our conditions are almost certainly still necessary conditions, but not sufficient to identify cases where definite parity modes persist.
Our results can be generalized to include such effects in future work.

We find that the slowly rotating JP QNM shifts can be written as \(\omega^{(1)}_{\ell m n}=(\delta\omega^{}_{0})_{\ell n}+\chi m \, (\delta\omega^{}_{1})_{\ell n}\), with a linear dependence on \(m\). 
Uncovering the analytical reason for this fact, which we find numerically, is the goal of future work. 
Identifying the general conditions under which such a decomposition holds would allow for the convenient calculation of each \(\ell,n\) mode shift at a single \(m\) to construct the QNM shifts at all \(m\) values. 
Our implementation also verifies that the slowly rotating JP spectrum breaks isospectrality while admitting definite parity modes, as predicted by the metric's \(\uparrow\) parity. 

While lower values of \(\ell\) dominate the ringdown of observed GW signals, this work is, to our knowledge, the first to apply the modified Teukolsky formalism to \(\ell>5\). 
Consequently, we are able to observe that even- and odd-parity modes converge to the eikonal limit with increasing \(\ell\) as a result of the QNMs being governed by the geometric optics limit. 
Moreover, we verify the large-\(\ell\) behavior of our calculations through comparison with a first-order scalar WKB approximation, ensuring the proper convergence. 
This convergence is not expected to occur in all bGR theories, but holds for the JP metric with our approximations, as discussed in Sec.~\ref{sec:Results}.
The QNM shifts for the \(n=1,2\) overtones have interesting behavior as well. 
While at large \(\ell\), the QNM shifts tend to lie along simple monotonic trends, this is not true for the higher overtones at \(\ell=2\). 
For the \(n=2\) overtone, the breaking of the trend appears to occur at \(\ell=3\), possibly suggesting an overtone dependence for this behavior. Although the physical reasoning behind this behavior is unknown, it nonetheless opens the door to explore low \(\ell\) QNM behavior in other bGR theories.

From the JP eikonal approximation of~\cite{glampedakis_post-kerr_2017} we estimate that our slowly rotating JP QNMs have an error \(\lesssim 1\%\) for \(\chi\lesssim 0.2\). 
However, with BH remnants from detected signals usually having spins of \(a\gtrsim 0.6\), the primary goal of future work is to extend our calculations to generic spin. 
Beyond the slow-rotation regime, the required operators in the modified Teukolsky formalism have so far proven cumbersome. 
One possible strategy is order reduction by application of the Teukolsky equation to remove higher derivatives.
While naively this may increase the number of terms in the operators, the existence of properties like the Teukolsky-Starobinsky identities and other possible simplifications mean that this approach may prove fruitful. 
This strategy could also be key to analytically proving why the slowly rotating JP QNMs shifts are linear in \(m\). 
Although computing QNM shifts for generic spins $0\leq \chi \lesssim 1$ has yet to be accomplished for any bGR theory using the modified Teukolsky formalism, the simplicity of the JP metric means it would serve as a good validation framework for the modified Teukolsky formalism general in spin for other bGR theories. 

In our implementation, we also neglect terms involving the underlying field equations of the JP metric. 
While the JP metric is not a solution to Einstein's equations but rather some alternative theory, the form of an appropriate alternative theory is unknown. 
Consequently, terms including the unknown underlying field equations simply cannot be included. 
If known, these terms can be included in the modified Teukolsky formalism as detailed by \cite{hussain_approach_2022}.
Neglecting these dynamical terms can be heuristically motivated by the Cowling approximation where we treat the background as fixed. 
However, these terms have made significant contributions to QNMs in other bGR theories \cite{cano_higher-derivative_2024,chung_quasi-normal_2024,li_perturbations_2025}, so they may be required for a reliable analysis.

Having established the behavior of the JP QNM shifts, we consider their potential observational relevance. 
This is limited by current GW detector sensitivity. The size of the QNM shifts for the slowly-rotating JP metric is small given the slow rotation and current constraints on \(\epsilon\) \cite{santos_testing_2024}. 
The largest shifts are low-percent level shifts, deviations which would not be detectable given the sensitivity of current detectors \cite{pacilio_identifying_2023}. 
The slow rotation approximation also limits the ability to utilize these QNM shifts in real GW signals as the spins of the remnant of most observed mergers is \(\chi\gtrsim 0.6\) \cite{the_ligo_scientific_collaboration_gwtc-40_2025}. 

Looking ahead, a natural next step is to explore how these theoretical predictions translate into measurable effects in GW signals. Although the slow-rotation regime restricts direct comparison with most observed mergers, injecting synthetic slowly rotating JP ringdown signals into ringdown parameter estimation codes, e.g.~\cite{Carullo:2019flw,isi_analyzing_2021,Capano:2021etf,Ma:2023cwe} would allow for further prototyping of data analysis pipelines for bGR QNM inference. 
Because the JP spectrum breaks isospectrality like many bGR theories, understanding how these signatures modify waveform morphology and bias parameter recovery will be essential for interpreting future high-precision observations. 
While current detectors are not sensitive enough to resolve these small frequency shifts, the methods developed here lay the foundation for systematically incorporating JP-like deviations into next-generation tests of GR in the strong-field regime.

\acknowledgments

We thank D.~Dzingeleski and Y.~Wang for valuable conversations, and Y.~Chen, D.~Li, P.~Wagle, and N.~Yunes for previous collaboration on topics related to the modified Teukolsky formalism.
DGW and AZ are supported by NSF Grant PHY-2308833.
The Flatiron Institute is a division of the Simons Foundation.

\appendix
\section{Radial Contour Integration}
\label{sec:RadContInteg}
To utilize the EVP formalism, we need to numerically evaluate the scalar products in Eqs.~\eqref{eqn:InnerProd1}--\eqref{eqn:InnerProd6}. First, we need an analytic form of the radial portion of \(\psi^{(0)}\). Using a series representation of Confluent Heun functions, Refs. \cite{leaver_analytic_1985, cook_gravitational_2014} represent the radial solution as 
\begin{multline}
\label{eqn:GenRadialFunc}
    R(r)=(r_+-r_-)^\alpha (r-r_-)^{\eta-\alpha}(r-r_+)^\eta \\\times e^{\zeta r} \sum_{n=0}^\infty a_n  \left(\frac{r-r_+}{r-r_-}\right)^n\,,
\end{multline}
where 
\begin{align}
    \alpha & = 1 + s + \xi + \eta - 2\zeta + \frac{i s\omega^{(0)}}{\zeta}\,,\\
    \zeta & =i \omega^{(0)}\,,\\
    \xi &=\frac{-s-(s+2i\sigma_+)}{2}\,,\\
    \eta &=\frac{-s-(s-2i\sigma_-)}{2}\,,\\
    \sigma_\pm &=\frac{2\omega^{(0)} M r_\pm -m a}{r_+-r_-}\,,
\end{align}
and \(a_n\) can be found via a recurrence relation.
As noted in Section \ref{sec:ContourIntegration}, the radial function diverges at the inner horizon \(r_-\), the outer horizon \(r_+\), and at infinity. The angular functions, however, are the spin-weighted spheroidal harmonics and can easily be integrated over. To get around the divergence of the radial functions, we perform a contour integral in the complex \(r\) plane. Using Eq.~\eqref{eqn:GenRadialFunc} for the radial solution, promoting \(r\) to be complex creates a branch cut extending leftwards on the real-\(r\) axis from the outer horizon \(r_+\). This branch cut is due to the complex logarithm and the standard definition of its principal branch which has a branch cut along the negative real axis. However, as described in Section \ref{sec:ContourIntegration}, we want the branch cut to extend in the direction of the exponential decay of the radial function. To rotate the branch cuts, we can add factors of \(\pm i\) in the asymptotic terms that multiply the series representation of the confluent Heun function:
\begin{multline}
    R_+(r)=(r_+-r_-)^\alpha (i(r-r_-))^{\eta-\alpha}(i(r-r_+))^\eta \\\times e^{\zeta r} \sum_{n=0}^\infty a_n  \left(\frac{r-r_+}{r-r_-}\right)^n\,,
\end{multline}
\begin{multline}
    R_-(r)=(r_+-r_-)^\alpha (-i(r-r_-))^{\eta-\alpha}(-i(r-r_+))^\eta \\\times e^{\zeta r} \sum_{n=0}^\infty a_n  \left(\frac{r-r_+}{r-r_-}\right)^n\,.
\end{multline}
This points the \(R_+(r)\) branch cut upwards in the complex plane and the \(R_-(r)\) branch cut downwards, the same direction as their respective exponential decays. The addition of these factors of \(i\) also maintain the relationship in Eq.~\eqref{eqn:RadialPMRelation} with a proportionality constant of \(1\). In effect, adding these factors of \(i\) amounts to adding a complex normalization constant while rotating the branch cuts in the desired direction. This allows us to deform the integration into a contour integration that wraps around the branch cut at \(r_+\), shown in Fig.~\ref{fig:Contour}. Because of the exponential decay in the direction of the branch cut, numerical integration along the contour can be truncated to the desired tolerance.

\begin{table*}[]
    \centering
    \setlength{\tabcolsep}{7pt}
    \begin{tabular}{cccccc}
        \hline\hline
        \rule{0pt}{2.5ex}
        $\ell$ & Overtone & $\delta\omega_0^\text{E}$ & $\delta\omega_1^\text{E}$ & $\delta\omega_0^\text{O}$ & $\delta\omega_1^\text{O}$ \\
        \hline
         & $n=0$ & $0.010993-0.004418i$ & $0.005213-0.001602i$ & $0.011031+0.001308i$ & $0.009256+0.007096i$ \\
        $\ell=2$ & $n=1$ & $0.005273-0.001006i$ & $-0.002978+0.002402i$ & $0.003665-0.002496i$ & $0.016534-0.002478i$ \\
         & $n=2$ & $0.015258+0.003087i$ & $-0.005149+0.020426i$ & $-0.010773+0.015097i$ & $0.013097-0.012111i$ \\
        \hline
         & $n=0$ & $0.014459-0.000826i$ & $0.006481+0.000094i$ & $0.013157+0.000725i$ & $0.006895+0.002728i$ \\
        $\ell=3$ & $n=1$ & $0.012896-0.002604i$ & $0.005271+0.000429i$ & $0.012286-0.000798i$ & $0.009001+0.002506i$ \\
         & $n=2$ & $0.012507-0.002582i$ & $0.003928+0.003144i$ & $0.011409+0.000008i$ & $0.010512+0.002136i$ \\
        \hline
         & $n=0$ & $0.017219-0.000021i$ & $0.006569+0.000358i$ & $0.016365+0.000425i$ & $0.006628+0.001442i$ \\
        $\ell=4$ & $n=1$ & $0.017127-0.001340i$ & $0.006486+0.000860i$ & $0.016327-0.000428i$ & $0.007491+0.002016i$ \\
         & $n=2$ & $0.017403-0.002076i$ & $0.006323+0.001907i$ & $0.016642-0.000541i$ & $0.008377+0.002597i$ \\
        \hline
         & $n=0$ & $0.020333+0.000157i$ & $0.006620+0.000361i$ & $0.019801+0.000274i$ & $0.006626+0.000908i$ \\
        $\ell=5$ & $n=1$ & $0.020596-0.000695i$ & $0.006711+0.000925i$ & $0.019975-0.000286i$ & $0.007067+0.001540i$ \\
         & $n=2$ & $0.021192-0.001310i$ & $0.006828+0.001667i$ & $0.020533-0.000506i$ & $0.007596+0.002201i$ \\
        \hline
         & $n=0$ & $0.023649+0.000184i$ & $0.006663+0.000323i$ & $0.023305+0.000191i$ & $0.006662+0.000638i$ \\
        $\ell=6$ & $n=1$ & $0.023990-0.000389i$ & $0.006768+0.000869i$ & $0.023544-0.000209i$ & $0.006921+0.001223i$ \\
         & $n=2$ & $0.024658-0.000842i$ & $0.006930+0.001490i$ & $0.024146-0.000421i$ & $0.007268+0.001833i$ \\
        \hline
         & $n=0$ & $0.027069+0.000173i$ & $0.006700+0.000285i$ & $0.026837+0.000141i$ & $0.006697+0.000483i$ \\
        $\ell=7$ & $n=1$ & $0.027411-0.000235i$ & $0.006791+0.000788i$ & $0.027091-0.000160i$ & $0.006866+0.001008i$ \\
         & $n=2$ & $0.028067-0.000569i$ & $0.006941+0.001331i$ & $0.027681-0.000345i$ & $0.007109+0.001553i$ \\
        \hline
         & $n=0$ & $0.030544+0.000152i$ & $0.006728+0.000252i$ & $0.030380+0.000108i$ & $0.006725+0.000385i$ \\
        $\ell=8$ & $n=1$ & $0.030866-0.000151i$ & $0.006803+0.000711i$ & $0.030633-0.000127i$ & $0.006844+0.000856i$ \\
         & $n=2$ & $0.031483-0.000404i$ & $0.006932+0.001194i$ & $0.031192-0.000283i$ & $0.007024+0.001343i$ \\
        \hline
         & $n=0$ & $0.034050+0.000133i$ & $0.006754+0.000227i$ & $0.033931+0.000087i$ & $0.006748+0.000319i$ \\
        $\ell=9$ & $n=1$ & $0.034348-0.000100i$ & $0.006817+0.000649i$ & $0.034173-0.000102i$ & $0.006833+0.000743i$ \\
         & $n=2$ & $0.034921-0.000298i$ & $0.006919+0.001076i$ & $0.034699-0.000235i$ & $0.006977+0.001184i$ \\
        \hline
         & $n=0$ & $0.037574+0.000114i$ & $0.006767+0.000203i$ & $0.037484+0.000069i$ & $0.006766+0.000271i$ \\
        $\ell=10$ & $n=1$ & $0.037849-0.000064i$ & $0.006825+0.000593i$ & $0.037716-0.000078i$ & $0.006841+0.000667i$ \\
         & $n=2$ & $0.038379-0.000229i$ & $0.006911+0.000984i$ & $0.038207-0.000198i$ & $0.006943+0.001051i$ \\
        \hline
        \hline\hline
    \end{tabular}
    \caption{Gravitational QNM shifts of the slowly rotating JP metric of the form \(\omega^{(1)}_{\ell m n}=(\delta\omega_0)_{\ell n}+\chi m\, (\delta\omega_1)_{\ell n}\) for even (E) and odd (O) parities with \(2\leq \ell \leq 10\) and \(n=0,1,2\).}
    \label{tab:AllShifts}
\end{table*}

\section{Tabulated QNM Shifts}
\label{sec:TabulatedQNMs}
Table \ref{tab:AllShifts} gives the slowly rotating gravitational JP QNM shifts in the form of Eq.~\eqref{eqn:JPQNMDecomp} for \(2\leq \ell \leq 10\), \(n=0,1,2\), also publicly available in \cite{wu2026beyondgrqnms}. Both even and odd parity modes are presented. All modes are calculated to a relative tolerance of \(O(10^{-4})\).

\section{Slowly rotating Kerr QNM Shifts}
\label{sec:LinKerrShifts}
In this appendix, we demonstrate using the EVP method that the slowly rotating Kerr QNMs can be written as
\begin{align}
    \omega^\text{Kerr}_{\ell mn}\approx \omega^{(0)}_{\ell n}+\chi m \, \delta\omega_{\ell n}\,.
\end{align}
Treating Schwarzschild as the background and the spin \(a\) as the perturbative parameter, the modified Teukolsky equation for a slowly rotating Kerr BH can be written as 
\begin{align}
    \mathcal{O} [\psi^{(1)}]+\delta\mathcal{O}[\psi^{(0)}]+\omega^{(1)}\partial_\omega \mathcal{O}[\psi^{(0)}]=0\,,
\end{align}
where \(\delta\mathcal{O}=\partial\mathcal{O}/\partial a\) and all operators are evaluated at \(a=0,\omega=\omega^{(0)}\). 
Using the EVP formalism to define a self-adjoint scalar product along the contour that wraps around the branch cut of the radial function, \(\omega^{(1)}\) can be calculated by 
\begin{align}
    \omega^{(1)}=\frac{\langle\tilde{\psi}^{(0)}_{\ell m n}|\delta\mathcal{O}_{\ell mn}[\tilde{\psi}^{(0)}_{\ell m n}]\rangle}{\langle\tilde{\psi}^{(0)}_{\ell m n}|\partial_\omega\mathcal{O}_{\ell mn}[\tilde{\psi}^{(0)}_{\ell m n}]\rangle}\,.
\end{align}
Now, our goal is to show that we can write \(\omega^{(1)}= m \, \delta\omega_{\ell n}\). 
Since the form of the full Teukolsky operator is known, we can show the operators are defined as
\begin{align}
    \delta\mathcal{O}_{\ell mn}=-\frac{4 m \omega^{(0)}}{r^2(r-2M)}\,,
\end{align}
\begin{align}
    \partial_\omega \mathcal{O}_{\ell m n}=\frac{2\omega^{(0)} r}{r-2M}\,.
\end{align}
Note that \(\omega^{(0)}\) is the Schwarzschild QNM frequency which is actually independent of \(m\), so \(\partial_\omega \mathcal{O}_{\ell m n}\) is independent of \(m\).
Then by factoring out the linear \(m\) dependence of \(\delta\mathcal{O}_{\ell mn}\), we find
\begin{align}
    \omega^{(1)}=m\frac{\langle\tilde{\psi}^{(0)}_{\ell m n}|\delta\mathcal{O}_{\ell n}[\tilde{\psi}^{(0)}_{\ell m n}]\rangle}{\langle\tilde{\psi}^{(0)}_{\ell m n}|\partial_\omega\mathcal{O}_{\ell n}[\tilde{\psi}^{(0)}_{\ell m n}]\rangle}\,.
\end{align}

The weight function \(w(r,\theta)\) we defined to make the background Teukolsky operator self-adjoint becomes independent of \(\theta\) for a Schwarzschild background, \(w(r,\theta)=w(r)\).
We can then separate out the radial and angular integrals because the operators and the weight function has no angular dependence.
The angular term simplifies to \(1\), giving
\begin{multline}
    \omega^{(1)}=m\frac{\int_\mathscr{C} R_{\ell m n}(r)w(r)\delta\mathcal{O}_{\ell n}[R_{\ell m n}(r)]dr}{\int_\mathscr{C} R_{\ell m n}(r)w(r)\partial_\omega\mathcal{O}_{\ell n}[R_{\ell m n}(r)]dr}\,.
\end{multline}
In hindsight, this makes sense at the integral is evaluated on the background, which if Schwarzschild, should have no angular dependence.
In the radial integral, the only \(m\) dependence comes from the radial functions.
However, with a Schwarzschild background, the radial functions are independent of \(m\).
We are left with
\begin{align}
     \omega^{(1)}=m\frac{\int_\mathscr{C} R_{\ell n}(r)w(r)\delta\mathcal{O}_{\ell n}[R_{\ell n}(r)]dr}{\int_\mathscr{C} R_{\ell n}(r)w(r)\partial_\omega\mathcal{O}_{\ell n}[R_{\ell n}(r)]dr}\,,
\end{align}
which is linear in \(m\) as desired. Therefore the small spin corrections of Kerr as a perturbation to Schwarzschild can be written as \(\omega^{(1)}_{\ell mn}=m\, \delta \omega_{\ell n}\). 

\bibliography{references}

\begin{thebibliography}{111}%
\makeatletter
\providecommand \@ifxundefined [1]{%
 \@ifx{#1\undefined}
}%
\providecommand \@ifnum [1]{%
 \ifnum #1\expandafter \@firstoftwo
 \else \expandafter \@secondoftwo
 \fi
}%
\providecommand \@ifx [1]{%
 \ifx #1\expandafter \@firstoftwo
 \else \expandafter \@secondoftwo
 \fi
}%
\providecommand \natexlab [1]{#1}%
\providecommand \enquote  [1]{``#1''}%
\providecommand \bibnamefont  [1]{#1}%
\providecommand \bibfnamefont [1]{#1}%
\providecommand \citenamefont [1]{#1}%
\providecommand \href@noop [0]{\@secondoftwo}%
\providecommand \href [0]{\begingroup \@sanitize@url \@href}%
\providecommand \@href[1]{\@@startlink{#1}\@@href}%
\providecommand \@@href[1]{\endgroup#1\@@endlink}%
\providecommand \@sanitize@url [0]{\catcode `\\12\catcode `\$12\catcode `\&12\catcode `\#12\catcode `\^12\catcode `\_12\catcode `\%12\relax}%
\providecommand \@@startlink[1]{}%
\providecommand \@@endlink[0]{}%
\providecommand \url  [0]{\begingroup\@sanitize@url \@url }%
\providecommand \@url [1]{\endgroup\@href {#1}{\urlprefix }}%
\providecommand \urlprefix  [0]{URL }%
\providecommand \Eprint [0]{\href }%
\providecommand \doibase [0]{https://doi.org/}%
\providecommand \selectlanguage [0]{\@gobble}%
\providecommand \bibinfo  [0]{\@secondoftwo}%
\providecommand \bibfield  [0]{\@secondoftwo}%
\providecommand \translation [1]{[#1]}%
\providecommand \BibitemOpen [0]{}%
\providecommand \bibitemStop [0]{}%
\providecommand \bibitemNoStop [0]{.\EOS\space}%
\providecommand \EOS [0]{\spacefactor3000\relax}%
\providecommand \BibitemShut  [1]{\csname bibitem#1\endcsname}%
\let\auto@bib@innerbib\@empty
%</preamble>
\bibitem [{\citenamefont {Abbott}\ \emph {et~al.}(2020)\citenamefont {Abbott} \emph {et~al.}}]{abbott_prospects_2020}%
  \BibitemOpen
  \bibfield  {author} {\bibinfo {author} {\bibfnamefont {B.~P.}\ \bibnamefont {Abbott}} \emph {et~al.},\ }\bibfield  {title} {\bibinfo {title} {Prospects for observing and localizing gravitational-wave transients with {Advanced} {LIGO}, {Advanced} {Virgo} and {KAGRA}},\ }\bibfield  {journal} {\bibinfo  {journal} {Living Reviews in Relativity}\ }\textbf {\bibinfo {volume} {23}},\ \href {https://doi.org/10.1007/s41114-020-00026-9} {10.1007/s41114-020-00026-9} (\bibinfo {year} {2020})\BibitemShut {NoStop}%
\bibitem [{\citenamefont {Aasi}\ \emph {et~al.}(2015)\citenamefont {Aasi} \emph {et~al.}}]{LIGOScientific:2014pky}%
  \BibitemOpen
  \bibfield  {author} {\bibinfo {author} {\bibfnamefont {J.}~\bibnamefont {Aasi}} \emph {et~al.} (\bibinfo {collaboration} {LIGO Scientific}),\ }\bibfield  {title} {\bibinfo {title} {{Advanced LIGO}},\ }\href {https://doi.org/10.1088/0264-9381/32/7/074001} {\bibfield  {journal} {\bibinfo  {journal} {Class. Quant. Grav.}\ }\textbf {\bibinfo {volume} {32}},\ \bibinfo {pages} {074001} (\bibinfo {year} {2015})},\ \Eprint {https://arxiv.org/abs/1411.4547} {arXiv:1411.4547 [gr-qc]} \BibitemShut {NoStop}%
\bibitem [{\citenamefont {Acernese}\ \emph {et~al.}(2015)\citenamefont {Acernese} \emph {et~al.}}]{VIRGO:2014yos}%
  \BibitemOpen
  \bibfield  {author} {\bibinfo {author} {\bibfnamefont {F.}~\bibnamefont {Acernese}} \emph {et~al.} (\bibinfo {collaboration} {VIRGO}),\ }\bibfield  {title} {\bibinfo {title} {{Advanced Virgo: a second-generation interferometric gravitational wave detector}},\ }\href {https://doi.org/10.1088/0264-9381/32/2/024001} {\bibfield  {journal} {\bibinfo  {journal} {Class. Quant. Grav.}\ }\textbf {\bibinfo {volume} {32}},\ \bibinfo {pages} {024001} (\bibinfo {year} {2015})},\ \Eprint {https://arxiv.org/abs/1408.3978} {arXiv:1408.3978 [gr-qc]} \BibitemShut {NoStop}%
\bibitem [{\citenamefont {Akutsu}\ \emph {et~al.}(2020)\citenamefont {Akutsu} \emph {et~al.}}]{10.1093/ptep/ptaa125}%
  \BibitemOpen
  \bibfield  {author} {\bibinfo {author} {\bibfnamefont {T.}~\bibnamefont {Akutsu}} \emph {et~al.},\ }\bibfield  {title} {\bibinfo {title} {Overview of {KAGRA}: Detector design and construction history},\ }\href {https://doi.org/10.1093/ptep/ptaa125} {\bibfield  {journal} {\bibinfo  {journal} {Progress of Theoretical and Experimental Physics}\ }\textbf {\bibinfo {volume} {2021}},\ \bibinfo {pages} {05A101} (\bibinfo {year} {2020})},\ \Eprint {https://arxiv.org/abs/https://academic.oup.com/ptep/article-pdf/2021/5/05A101/37974994/ptaa125.pdf} {https://academic.oup.com/ptep/article-pdf/2021/5/05A101/37974994/ptaa125.pdf} \BibitemShut {NoStop}%
\bibitem [{\citenamefont {Abac}\ \emph {et~al.}(2025{\natexlab{a}})\citenamefont {Abac} \emph {et~al.}}]{abac_gw250114_2025}%
  \BibitemOpen
  \bibfield  {author} {\bibinfo {author} {\bibfnamefont {A.~G.}\ \bibnamefont {Abac}} \emph {et~al.} (\bibinfo {collaboration} {LIGO Scientific, Virgo, and KAGRA Collaborations}),\ }\bibfield  {title} {\bibinfo {title} {{GW250114}: Testing {Hawking's} area law and the {Kerr} nature of black holes},\ }\href {https://doi.org/10.1103/kw5g-d732} {\bibfield  {journal} {\bibinfo  {journal} {Phys. Rev. Lett.}\ }\textbf {\bibinfo {volume} {135}},\ \bibinfo {pages} {111403} (\bibinfo {year} {2025}{\natexlab{a}})}\BibitemShut {NoStop}%
\bibitem [{\citenamefont {Will}(2014)}]{will_confrontation_2014}%
  \BibitemOpen
  \bibfield  {author} {\bibinfo {author} {\bibfnamefont {C.~M.}\ \bibnamefont {Will}},\ }\bibfield  {title} {\bibinfo {title} {The {Confrontation} between {General} {Relativity} and {Experiment}},\ }\href {https://doi.org/10.12942/lrr-2014-4} {\bibfield  {journal} {\bibinfo  {journal} {Living Reviews in Relativity}\ }\textbf {\bibinfo {volume} {17}},\ \bibinfo {pages} {4} (\bibinfo {year} {2014})}\BibitemShut {NoStop}%
\bibitem [{\citenamefont {Yunes}\ and\ \citenamefont {Pretorius}(2009)}]{Yunes:2009ke}%
  \BibitemOpen
  \bibfield  {author} {\bibinfo {author} {\bibfnamefont {N.}~\bibnamefont {Yunes}}\ and\ \bibinfo {author} {\bibfnamefont {F.}~\bibnamefont {Pretorius}},\ }\bibfield  {title} {\bibinfo {title} {{Fundamental Theoretical Bias in Gravitational Wave Astrophysics and the Parameterized Post-Einsteinian Framework}},\ }\href {https://doi.org/10.1103/PhysRevD.80.122003} {\bibfield  {journal} {\bibinfo  {journal} {Phys. Rev. D}\ }\textbf {\bibinfo {volume} {80}},\ \bibinfo {pages} {122003} (\bibinfo {year} {2009})},\ \Eprint {https://arxiv.org/abs/0909.3328} {arXiv:0909.3328 [gr-qc]} \BibitemShut {NoStop}%
\bibitem [{\citenamefont {Li}\ \emph {et~al.}(2012)\citenamefont {Li}, \citenamefont {Del~Pozzo}, \citenamefont {Vitale}, \citenamefont {Van Den~Broeck}, \citenamefont {Agathos}, \citenamefont {Veitch}, \citenamefont {Grover}, \citenamefont {Sidery}, \citenamefont {Sturani},\ and\ \citenamefont {Vecchio}}]{Li:2011cg}%
  \BibitemOpen
  \bibfield  {author} {\bibinfo {author} {\bibfnamefont {T.~G.~F.}\ \bibnamefont {Li}}, \bibinfo {author} {\bibfnamefont {W.}~\bibnamefont {Del~Pozzo}}, \bibinfo {author} {\bibfnamefont {S.}~\bibnamefont {Vitale}}, \bibinfo {author} {\bibfnamefont {C.}~\bibnamefont {Van Den~Broeck}}, \bibinfo {author} {\bibfnamefont {M.}~\bibnamefont {Agathos}}, \bibinfo {author} {\bibfnamefont {J.}~\bibnamefont {Veitch}}, \bibinfo {author} {\bibfnamefont {K.}~\bibnamefont {Grover}}, \bibinfo {author} {\bibfnamefont {T.}~\bibnamefont {Sidery}}, \bibinfo {author} {\bibfnamefont {R.}~\bibnamefont {Sturani}},\ and\ \bibinfo {author} {\bibfnamefont {A.}~\bibnamefont {Vecchio}},\ }\bibfield  {title} {\bibinfo {title} {{Towards a generic test of the strong field dynamics of general relativity using compact binary coalescence}},\ }\href {https://doi.org/10.1103/PhysRevD.85.082003} {\bibfield  {journal} {\bibinfo  {journal} {Phys. Rev. D}\ }\textbf {\bibinfo {volume} {85}},\ \bibinfo {pages} {082003} (\bibinfo {year} {2012})},\
  \Eprint {https://arxiv.org/abs/1110.0530} {arXiv:1110.0530 [gr-qc]} \BibitemShut {NoStop}%
\bibitem [{\citenamefont {Mehta}\ \emph {et~al.}(2023)\citenamefont {Mehta}, \citenamefont {Buonanno}, \citenamefont {Cotesta}, \citenamefont {Ghosh}, \citenamefont {Sennett},\ and\ \citenamefont {Steinhoff}}]{Mehta:2022pcn}%
  \BibitemOpen
  \bibfield  {author} {\bibinfo {author} {\bibfnamefont {A.~K.}\ \bibnamefont {Mehta}}, \bibinfo {author} {\bibfnamefont {A.}~\bibnamefont {Buonanno}}, \bibinfo {author} {\bibfnamefont {R.}~\bibnamefont {Cotesta}}, \bibinfo {author} {\bibfnamefont {A.}~\bibnamefont {Ghosh}}, \bibinfo {author} {\bibfnamefont {N.}~\bibnamefont {Sennett}},\ and\ \bibinfo {author} {\bibfnamefont {J.}~\bibnamefont {Steinhoff}},\ }\bibfield  {title} {\bibinfo {title} {{Tests of general relativity with gravitational-wave observations using a flexible theory-independent method}},\ }\href {https://doi.org/10.1103/PhysRevD.107.044020} {\bibfield  {journal} {\bibinfo  {journal} {Phys. Rev. D}\ }\textbf {\bibinfo {volume} {107}},\ \bibinfo {pages} {044020} (\bibinfo {year} {2023})},\ \Eprint {https://arxiv.org/abs/2203.13937} {arXiv:2203.13937 [gr-qc]} \BibitemShut {NoStop}%
\bibitem [{\citenamefont {Yunes}\ \emph {et~al.}(2025)\citenamefont {Yunes}, \citenamefont {Siemens},\ and\ \citenamefont {Yagi}}]{Yunes:2025xwp}%
  \BibitemOpen
  \bibfield  {author} {\bibinfo {author} {\bibfnamefont {N.}~\bibnamefont {Yunes}}, \bibinfo {author} {\bibfnamefont {X.}~\bibnamefont {Siemens}},\ and\ \bibinfo {author} {\bibfnamefont {K.}~\bibnamefont {Yagi}},\ }\bibfield  {title} {\bibinfo {title} {{Gravitational-wave tests of general relativity with ground-based detectors and pulsar-timing arrays}},\ }\href {https://doi.org/10.1007/s41114-024-00054-9} {\bibfield  {journal} {\bibinfo  {journal} {Living Rev. Rel.}\ }\textbf {\bibinfo {volume} {28}},\ \bibinfo {pages} {3} (\bibinfo {year} {2025})}\BibitemShut {NoStop}%
\bibitem [{\citenamefont {Abbott}\ \emph {et~al.}(2025)\citenamefont {Abbott} \emph {et~al.}}]{the_ligo_scientific_collaboration_tests_2021}%
  \BibitemOpen
  \bibfield  {author} {\bibinfo {author} {\bibfnamefont {R.}~\bibnamefont {Abbott}} \emph {et~al.} (\bibinfo {collaboration} {The LIGO Scientific Collaboration, the Virgo Collaboration, and the KAGRA Collaboration}),\ }\bibfield  {title} {\bibinfo {title} {Tests of general relativity with gwtc-3},\ }\href {https://doi.org/10.1103/PhysRevD.112.084080} {\bibfield  {journal} {\bibinfo  {journal} {Phys. Rev. D}\ }\textbf {\bibinfo {volume} {112}},\ \bibinfo {pages} {084080} (\bibinfo {year} {2025})},\ \bibinfo {note} {arXiv: 2112.06861 [gr-qc]}\BibitemShut {NoStop}%
\bibitem [{\citenamefont {Zimmerman}\ \emph {et~al.}(2019)\citenamefont {Zimmerman}, \citenamefont {Haster},\ and\ \citenamefont {Chatziioannou}}]{Zimmerman:2019wzo}%
  \BibitemOpen
  \bibfield  {author} {\bibinfo {author} {\bibfnamefont {A.}~\bibnamefont {Zimmerman}}, \bibinfo {author} {\bibfnamefont {C.-J.}\ \bibnamefont {Haster}},\ and\ \bibinfo {author} {\bibfnamefont {K.}~\bibnamefont {Chatziioannou}},\ }\bibfield  {title} {\bibinfo {title} {{On combining information from multiple gravitational wave sources}},\ }\href {https://doi.org/10.1103/PhysRevD.99.124044} {\bibfield  {journal} {\bibinfo  {journal} {Phys. Rev. D}\ }\textbf {\bibinfo {volume} {99}},\ \bibinfo {pages} {124044} (\bibinfo {year} {2019})},\ \Eprint {https://arxiv.org/abs/1903.11008} {arXiv:1903.11008 [astro-ph.IM]} \BibitemShut {NoStop}%
\bibitem [{\citenamefont {Perkins}\ \emph {et~al.}(2021)\citenamefont {Perkins}, \citenamefont {Nair}, \citenamefont {Silva},\ and\ \citenamefont {Yunes}}]{perkins_improved_2021}%
  \BibitemOpen
  \bibfield  {author} {\bibinfo {author} {\bibfnamefont {S.~E.}\ \bibnamefont {Perkins}}, \bibinfo {author} {\bibfnamefont {R.}~\bibnamefont {Nair}}, \bibinfo {author} {\bibfnamefont {H.~O.}\ \bibnamefont {Silva}},\ and\ \bibinfo {author} {\bibfnamefont {N.}~\bibnamefont {Yunes}},\ }\bibfield  {title} {\bibinfo {title} {Improved gravitational-wave constraints on higher-order curvature theories of gravity},\ }\href {https://doi.org/10.1103/PhysRevD.104.024060} {\bibfield  {journal} {\bibinfo  {journal} {Physical Review D}\ }\textbf {\bibinfo {volume} {104}},\ \bibinfo {pages} {024060} (\bibinfo {year} {2021})},\ \bibinfo {note} {arXiv:2104.11189 [gr-qc]}\BibitemShut {NoStop}%
\bibitem [{\citenamefont {Sennett}\ \emph {et~al.}(2020)\citenamefont {Sennett}, \citenamefont {Brito}, \citenamefont {Buonanno}, \citenamefont {Gorbenko},\ and\ \citenamefont {Senatore}}]{sennett_gravitational-wave_2020}%
  \BibitemOpen
  \bibfield  {author} {\bibinfo {author} {\bibfnamefont {N.}~\bibnamefont {Sennett}}, \bibinfo {author} {\bibfnamefont {R.}~\bibnamefont {Brito}}, \bibinfo {author} {\bibfnamefont {A.}~\bibnamefont {Buonanno}}, \bibinfo {author} {\bibfnamefont {V.}~\bibnamefont {Gorbenko}},\ and\ \bibinfo {author} {\bibfnamefont {L.}~\bibnamefont {Senatore}},\ }\bibfield  {title} {\bibinfo {title} {Gravitational-wave constraints on an effective-field-theory extension of general relativity},\ }\href {https://doi.org/10.1103/PhysRevD.102.044056} {\bibfield  {journal} {\bibinfo  {journal} {Physical Review D}\ }\textbf {\bibinfo {volume} {102}},\ \bibinfo {pages} {044056} (\bibinfo {year} {2020})}\BibitemShut {NoStop}%
\bibitem [{\citenamefont {Isi}\ \emph {et~al.}(2019)\citenamefont {Isi}, \citenamefont {Chatziioannou},\ and\ \citenamefont {Farr}}]{Isi:2019asy}%
  \BibitemOpen
  \bibfield  {author} {\bibinfo {author} {\bibfnamefont {M.}~\bibnamefont {Isi}}, \bibinfo {author} {\bibfnamefont {K.}~\bibnamefont {Chatziioannou}},\ and\ \bibinfo {author} {\bibfnamefont {W.~M.}\ \bibnamefont {Farr}},\ }\bibfield  {title} {\bibinfo {title} {{Hierarchical test of general relativity with gravitational waves}},\ }\href {https://doi.org/10.1103/PhysRevLett.123.121101} {\bibfield  {journal} {\bibinfo  {journal} {Phys. Rev. Lett.}\ }\textbf {\bibinfo {volume} {123}},\ \bibinfo {pages} {121101} (\bibinfo {year} {2019})},\ \Eprint {https://arxiv.org/abs/1904.08011} {arXiv:1904.08011 [gr-qc]} \BibitemShut {NoStop}%
\bibitem [{\citenamefont {Sotiriou}\ and\ \citenamefont {Zhou}(2014)}]{sotiriou_black_2014}%
  \BibitemOpen
  \bibfield  {author} {\bibinfo {author} {\bibfnamefont {T.~P.}\ \bibnamefont {Sotiriou}}\ and\ \bibinfo {author} {\bibfnamefont {S.-Y.}\ \bibnamefont {Zhou}},\ }\bibfield  {title} {\bibinfo {title} {Black {Hole} {Hair} in {Generalized} {Scalar}-{Tensor} {Gravity}},\ }\href {https://doi.org/10.1103/PhysRevLett.112.251102} {\bibfield  {journal} {\bibinfo  {journal} {Physical Review Letters}\ }\textbf {\bibinfo {volume} {112}},\ \bibinfo {pages} {251102} (\bibinfo {year} {2014})}\BibitemShut {NoStop}%
\bibitem [{\citenamefont {Alexander}\ and\ \citenamefont {Yunes}(2009)}]{alexander_chernsimons_2009}%
  \BibitemOpen
  \bibfield  {author} {\bibinfo {author} {\bibfnamefont {S.}~\bibnamefont {Alexander}}\ and\ \bibinfo {author} {\bibfnamefont {N.}~\bibnamefont {Yunes}},\ }\bibfield  {title} {\bibinfo {title} {Chern–{Simons} modified general relativity},\ }\href {https://doi.org/10.1016/j.physrep.2009.07.002} {\bibfield  {journal} {\bibinfo  {journal} {Physics Reports}\ }\textbf {\bibinfo {volume} {480}},\ \bibinfo {pages} {1} (\bibinfo {year} {2009})}\BibitemShut {NoStop}%
\bibitem [{\citenamefont {Kanti}\ and\ \citenamefont {Tamvakis}(1995)}]{kanti_classical_1995}%
  \BibitemOpen
  \bibfield  {author} {\bibinfo {author} {\bibfnamefont {P.}~\bibnamefont {Kanti}}\ and\ \bibinfo {author} {\bibfnamefont {K.}~\bibnamefont {Tamvakis}},\ }\href {https://doi.org/10.48550/arXiv.hep-th/9502093} {\bibinfo {title} {Classical moduli hair for {Kerr} black holes in {String} {Gravity}}} (\bibinfo {year} {1995}),\ \bibinfo {note} {arXiv:hep-th/9502093}\BibitemShut {NoStop}%
\bibitem [{\citenamefont {Okounkova}\ \emph {et~al.}(2020)\citenamefont {Okounkova}, \citenamefont {Stein}, \citenamefont {Moxon}, \citenamefont {Scheel},\ and\ \citenamefont {Teukolsky}}]{okounkova_numerical_2020}%
  \BibitemOpen
  \bibfield  {author} {\bibinfo {author} {\bibfnamefont {M.}~\bibnamefont {Okounkova}}, \bibinfo {author} {\bibfnamefont {L.~C.}\ \bibnamefont {Stein}}, \bibinfo {author} {\bibfnamefont {J.}~\bibnamefont {Moxon}}, \bibinfo {author} {\bibfnamefont {M.~A.}\ \bibnamefont {Scheel}},\ and\ \bibinfo {author} {\bibfnamefont {S.~A.}\ \bibnamefont {Teukolsky}},\ }\bibfield  {title} {\bibinfo {title} {Numerical relativity simulation of {GW150914} beyond general relativity},\ }\href {https://doi.org/10.1103/PhysRevD.101.104016} {\bibfield  {journal} {\bibinfo  {journal} {Physical Review D}\ }\textbf {\bibinfo {volume} {101}},\ \bibinfo {pages} {104016} (\bibinfo {year} {2020})},\ \bibinfo {note} {arXiv:1911.02588 [gr-qc]}\BibitemShut {NoStop}%
\bibitem [{\citenamefont {Tahura}\ and\ \citenamefont {Yagi}(2018)}]{tahura_parameterized_2018}%
  \BibitemOpen
  \bibfield  {author} {\bibinfo {author} {\bibfnamefont {S.}~\bibnamefont {Tahura}}\ and\ \bibinfo {author} {\bibfnamefont {K.}~\bibnamefont {Yagi}},\ }\bibfield  {title} {\bibinfo {title} {Parameterized {Post}-{Einsteinian} {Gravitational} {Waveforms} in {Various} {Modified} {Theories} of {Gravity}},\ }\href {https://doi.org/10.1103/PhysRevD.98.084042} {\bibfield  {journal} {\bibinfo  {journal} {Physical Review D}\ }\textbf {\bibinfo {volume} {98}},\ \bibinfo {pages} {084042} (\bibinfo {year} {2018})},\ \bibinfo {note} {arXiv:1809.00259 [gr-qc]}\BibitemShut {NoStop}%
\bibitem [{\citenamefont {Julié}\ \emph {et~al.}(2025)\citenamefont {Julié}, \citenamefont {Pompili},\ and\ \citenamefont {Buonanno}}]{julie_inspiral-merger-ringdown_2025}%
  \BibitemOpen
  \bibfield  {author} {\bibinfo {author} {\bibfnamefont {F.-L.}\ \bibnamefont {Julié}}, \bibinfo {author} {\bibfnamefont {L.}~\bibnamefont {Pompili}},\ and\ \bibinfo {author} {\bibfnamefont {A.}~\bibnamefont {Buonanno}},\ }\bibfield  {title} {\bibinfo {title} {Inspiral-merger-ringdown waveforms in {Einstein}-scalar-{Gauss}-{Bonnet} gravity within the effective-one-body formalism},\ }\href {https://doi.org/10.1103/PhysRevD.111.024016} {\bibfield  {journal} {\bibinfo  {journal} {Physical Review D}\ }\textbf {\bibinfo {volume} {111}},\ \bibinfo {pages} {024016} (\bibinfo {year} {2025})},\ \bibinfo {note} {arXiv:2406.13654 [gr-qc]}\BibitemShut {NoStop}%
\bibitem [{\citenamefont {Roy}\ \emph {et~al.}(2025)\citenamefont {Roy}, \citenamefont {Küchler}, \citenamefont {Pound},\ and\ \citenamefont {Macedo}}]{roy_black_2025}%
  \BibitemOpen
  \bibfield  {author} {\bibinfo {author} {\bibfnamefont {A.}~\bibnamefont {Roy}}, \bibinfo {author} {\bibfnamefont {L.}~\bibnamefont {Küchler}}, \bibinfo {author} {\bibfnamefont {A.}~\bibnamefont {Pound}},\ and\ \bibinfo {author} {\bibfnamefont {R.~P.}\ \bibnamefont {Macedo}},\ }\href {https://doi.org/10.48550/arXiv.2510.11793} {\bibinfo {title} {Black hole mergers beyond general relativity: a self-force approach}} (\bibinfo {year} {2025}),\ \bibinfo {note} {arXiv:2510.11793 [gr-qc]}\BibitemShut {NoStop}%
\bibitem [{\citenamefont {Vishveshwara}(1970{\natexlab{a}})}]{vishveshwara_scattering_1970}%
  \BibitemOpen
  \bibfield  {author} {\bibinfo {author} {\bibfnamefont {C.~V.}\ \bibnamefont {Vishveshwara}},\ }\bibfield  {title} {\bibinfo {title} {Scattering of {Gravitational} {Radiation} by a {Schwarzschild} {Black}-hole},\ }\href {https://doi.org/10.1038/227936a0} {\bibfield  {journal} {\bibinfo  {journal} {Nature}\ }\textbf {\bibinfo {volume} {227}},\ \bibinfo {pages} {936} (\bibinfo {year} {1970}{\natexlab{a}})}\BibitemShut {NoStop}%
\bibitem [{\citenamefont {Vishveshwara}(1970{\natexlab{b}})}]{vishveshwara_stability_1970}%
  \BibitemOpen
  \bibfield  {author} {\bibinfo {author} {\bibfnamefont {C.~V.}\ \bibnamefont {Vishveshwara}},\ }\bibfield  {title} {\bibinfo {title} {Stability of the {Schwarzschild} {Metric}},\ }\href {https://doi.org/10.1103/PhysRevD.1.2870} {\bibfield  {journal} {\bibinfo  {journal} {Physical Review D}\ }\textbf {\bibinfo {volume} {1}},\ \bibinfo {pages} {2870} (\bibinfo {year} {1970}{\natexlab{b}})}\BibitemShut {NoStop}%
\bibitem [{\citenamefont {Berti}\ \emph {et~al.}(2025)\citenamefont {Berti}, \citenamefont {Cardoso}, \citenamefont {Carullo} \emph {et~al.}}]{berti_black_2025}%
  \BibitemOpen
  \bibfield  {author} {\bibinfo {author} {\bibfnamefont {E.}~\bibnamefont {Berti}}, \bibinfo {author} {\bibfnamefont {V.}~\bibnamefont {Cardoso}}, \bibinfo {author} {\bibfnamefont {G.}~\bibnamefont {Carullo}}, \emph {et~al.},\ }\href {https://doi.org/10.48550/arXiv.2505.23895} {\bibinfo {title} {Black hole spectroscopy: from theory to experiment}} (\bibinfo {year} {2025}),\ \bibinfo {note} {arXiv:2505.23895 [gr-qc]}\BibitemShut {NoStop}%
\bibitem [{\citenamefont {Israel}(1967)}]{israel_event_1967}%
  \BibitemOpen
  \bibfield  {author} {\bibinfo {author} {\bibfnamefont {W.}~\bibnamefont {Israel}},\ }\bibfield  {title} {\bibinfo {title} {Event {Horizons} in {Static} {Vacuum} {Space}-{Times}},\ }\href {https://doi.org/10.1103/PhysRev.164.1776} {\bibfield  {journal} {\bibinfo  {journal} {Physical Review}\ }\textbf {\bibinfo {volume} {164}},\ \bibinfo {pages} {1776} (\bibinfo {year} {1967})}\BibitemShut {NoStop}%
\bibitem [{\citenamefont {Carter}(1971)}]{carter_axisymmetric_1971}%
  \BibitemOpen
  \bibfield  {author} {\bibinfo {author} {\bibfnamefont {B.}~\bibnamefont {Carter}},\ }\bibfield  {title} {\bibinfo {title} {Axisymmetric {Black} {Hole} {Has} {Only} {Two} {Degrees} of {Freedom}},\ }\href {https://doi.org/10.1103/PhysRevLett.26.331} {\bibfield  {journal} {\bibinfo  {journal} {Physical Review Letters}\ }\textbf {\bibinfo {volume} {26}},\ \bibinfo {pages} {331} (\bibinfo {year} {1971})}\BibitemShut {NoStop}%
\bibitem [{\citenamefont {Robinson}(1975)}]{robinson_uniqueness_1975}%
  \BibitemOpen
  \bibfield  {author} {\bibinfo {author} {\bibfnamefont {D.~C.}\ \bibnamefont {Robinson}},\ }\bibfield  {title} {\bibinfo {title} {Uniqueness of the {Kerr} {Black} {Hole}},\ }\href {https://doi.org/10.1103/PhysRevLett.34.905} {\bibfield  {journal} {\bibinfo  {journal} {Physical Review Letters}\ }\textbf {\bibinfo {volume} {34}},\ \bibinfo {pages} {905} (\bibinfo {year} {1975})}\BibitemShut {NoStop}%
\bibitem [{\citenamefont {Dreyer}\ \emph {et~al.}(2004)\citenamefont {Dreyer}, \citenamefont {Kelly}, \citenamefont {Krishnan}, \citenamefont {Finn}, \citenamefont {Garrison},\ and\ \citenamefont {Lopez-Aleman}}]{dreyer_black_2004}%
  \BibitemOpen
  \bibfield  {author} {\bibinfo {author} {\bibfnamefont {O.}~\bibnamefont {Dreyer}}, \bibinfo {author} {\bibfnamefont {B.}~\bibnamefont {Kelly}}, \bibinfo {author} {\bibfnamefont {B.}~\bibnamefont {Krishnan}}, \bibinfo {author} {\bibfnamefont {L.~S.}\ \bibnamefont {Finn}}, \bibinfo {author} {\bibfnamefont {D.}~\bibnamefont {Garrison}},\ and\ \bibinfo {author} {\bibfnamefont {R.}~\bibnamefont {Lopez-Aleman}},\ }\bibfield  {title} {\bibinfo {title} {Black {Hole} {Spectroscopy}: {Testing} {General} {Relativity} through {Gravitational} {Wave} {Observations}},\ }\href {https://doi.org/10.1088/0264-9381/21/4/003} {\bibfield  {journal} {\bibinfo  {journal} {Classical and Quantum Gravity}\ }\textbf {\bibinfo {volume} {21}},\ \bibinfo {pages} {787} (\bibinfo {year} {2004})},\ \bibinfo {note} {arXiv:gr-qc/0309007}\BibitemShut {NoStop}%
\bibitem [{\citenamefont {Berti}\ \emph {et~al.}(2006)\citenamefont {Berti}, \citenamefont {Cardoso},\ and\ \citenamefont {Will}}]{berti_gravitational-wave_2006}%
  \BibitemOpen
  \bibfield  {author} {\bibinfo {author} {\bibfnamefont {E.}~\bibnamefont {Berti}}, \bibinfo {author} {\bibfnamefont {V.}~\bibnamefont {Cardoso}},\ and\ \bibinfo {author} {\bibfnamefont {C.~M.}\ \bibnamefont {Will}},\ }\bibfield  {title} {\bibinfo {title} {Gravitational-wave spectroscopy of massive black holes with the space interferometer {LISA}},\ }\href {https://doi.org/10.1103/PhysRevD.73.064030} {\bibfield  {journal} {\bibinfo  {journal} {Physical Review D}\ }\textbf {\bibinfo {volume} {73}},\ \bibinfo {pages} {064030} (\bibinfo {year} {2006})}\BibitemShut {NoStop}%
\bibitem [{\citenamefont {Berti}\ \emph {et~al.}(2018)\citenamefont {Berti}, \citenamefont {Yagi}, \citenamefont {Yang},\ and\ \citenamefont {Yunes}}]{berti_extreme_2018}%
  \BibitemOpen
  \bibfield  {author} {\bibinfo {author} {\bibfnamefont {E.}~\bibnamefont {Berti}}, \bibinfo {author} {\bibfnamefont {K.}~\bibnamefont {Yagi}}, \bibinfo {author} {\bibfnamefont {H.}~\bibnamefont {Yang}},\ and\ \bibinfo {author} {\bibfnamefont {N.}~\bibnamefont {Yunes}},\ }\bibfield  {title} {\bibinfo {title} {Extreme {Gravity} {Tests} with {Gravitational} {Waves} from {Compact} {Binary} {Coalescences}: ({II}) {Ringdown}},\ }\href {https://doi.org/10.1007/s10714-018-2372-6} {\bibfield  {journal} {\bibinfo  {journal} {General Relativity and Gravitation}\ }\textbf {\bibinfo {volume} {50}},\ \bibinfo {pages} {49} (\bibinfo {year} {2018})},\ \bibinfo {note} {arXiv:1801.03587 [gr-qc]}\BibitemShut {NoStop}%
\bibitem [{\citenamefont {{The LIGO Scientific Collaboration}}\ \emph {et~al.}(2025)\citenamefont {{The LIGO Scientific Collaboration}}, \citenamefont {{the Virgo Collaboration}},\ and\ \citenamefont {{the KAGRA Collaboration}}}]{the_ligo_scientific_collaboration_black_2025}%
  \BibitemOpen
  \bibfield  {author} {\bibinfo {author} {\bibnamefont {{The LIGO Scientific Collaboration}}}, \bibinfo {author} {\bibnamefont {{the Virgo Collaboration}}},\ and\ \bibinfo {author} {\bibnamefont {{the KAGRA Collaboration}}},\ }\href {https://arxiv.org/abs/2509.08099} {\bibinfo {title} {Black {Hole} {Spectroscopy} and {Tests} of {General} {Relativity} with {GW250114}}} (\bibinfo {year} {2025}),\ \bibinfo {note} {arXiv: 2509.08099 [gr-qc]}\BibitemShut {NoStop}%
\bibitem [{\citenamefont {Cardoso}\ \emph {et~al.}(2009)\citenamefont {Cardoso}, \citenamefont {Miranda}, \citenamefont {Berti}, \citenamefont {Witek},\ and\ \citenamefont {Zanchin}}]{cardoso_geodesic_2009}%
  \BibitemOpen
  \bibfield  {author} {\bibinfo {author} {\bibfnamefont {V.}~\bibnamefont {Cardoso}}, \bibinfo {author} {\bibfnamefont {A.~S.}\ \bibnamefont {Miranda}}, \bibinfo {author} {\bibfnamefont {E.}~\bibnamefont {Berti}}, \bibinfo {author} {\bibfnamefont {H.}~\bibnamefont {Witek}},\ and\ \bibinfo {author} {\bibfnamefont {V.~T.}\ \bibnamefont {Zanchin}},\ }\bibfield  {title} {\bibinfo {title} {Geodesic stability, {Lyapunov} exponents and quasinormal modes},\ }\href {https://doi.org/10.1103/PhysRevD.79.064016} {\bibfield  {journal} {\bibinfo  {journal} {Physical Review D}\ }\textbf {\bibinfo {volume} {79}},\ \bibinfo {pages} {064016} (\bibinfo {year} {2009})},\ \bibinfo {note} {arXiv:0812.1806 [hep-th]}\BibitemShut {NoStop}%
\bibitem [{\citenamefont {Konoplya}\ and\ \citenamefont {Zhidenko}(2011)}]{konoplya_quasinormal_2011}%
  \BibitemOpen
  \bibfield  {author} {\bibinfo {author} {\bibfnamefont {R.~A.}\ \bibnamefont {Konoplya}}\ and\ \bibinfo {author} {\bibfnamefont {A.}~\bibnamefont {Zhidenko}},\ }\bibfield  {title} {\bibinfo {title} {Quasinormal modes of black holes: from astrophysics to string theory},\ }\href {https://doi.org/10.1103/RevModPhys.83.793} {\bibfield  {journal} {\bibinfo  {journal} {Reviews of Modern Physics}\ }\textbf {\bibinfo {volume} {83}},\ \bibinfo {pages} {793} (\bibinfo {year} {2011})},\ \bibinfo {note} {arXiv:1102.4014 [gr-qc]}\BibitemShut {NoStop}%
\bibitem [{\citenamefont {Regge}\ and\ \citenamefont {Wheeler}(1957)}]{regge_stability_1957}%
  \BibitemOpen
  \bibfield  {author} {\bibinfo {author} {\bibfnamefont {T.}~\bibnamefont {Regge}}\ and\ \bibinfo {author} {\bibfnamefont {J.~A.}\ \bibnamefont {Wheeler}},\ }\bibfield  {title} {\bibinfo {title} {Stability of a {Schwarzschild} {Singularity}},\ }\href {https://doi.org/10.1103/PhysRev.108.1063} {\bibfield  {journal} {\bibinfo  {journal} {Physical Review}\ }\textbf {\bibinfo {volume} {108}},\ \bibinfo {pages} {1063} (\bibinfo {year} {1957})}\BibitemShut {NoStop}%
\bibitem [{\citenamefont {Zerilli}(1970)}]{zerilli_gravitational_1970}%
  \BibitemOpen
  \bibfield  {author} {\bibinfo {author} {\bibfnamefont {F.~J.}\ \bibnamefont {Zerilli}},\ }\bibfield  {title} {\bibinfo {title} {Gravitational {Field} of a {Particle} {Falling} in a {Schwarzschild} {Geometry} {Analyzed} in {Tensor} {Harmonics}},\ }\href {https://doi.org/10.1103/PhysRevD.2.2141} {\bibfield  {journal} {\bibinfo  {journal} {Physical Review D}\ }\textbf {\bibinfo {volume} {2}},\ \bibinfo {pages} {2141} (\bibinfo {year} {1970})}\BibitemShut {NoStop}%
\bibitem [{\citenamefont {Martel}\ and\ \citenamefont {Poisson}(2005)}]{martel_gravitational_2005}%
  \BibitemOpen
  \bibfield  {author} {\bibinfo {author} {\bibfnamefont {K.}~\bibnamefont {Martel}}\ and\ \bibinfo {author} {\bibfnamefont {E.}~\bibnamefont {Poisson}},\ }\bibfield  {title} {\bibinfo {title} {Gravitational perturbations of the schwarzschild spacetime: A practical covariant and gauge-invariant formalism},\ }\href {https://doi.org/10.1103/PhysRevD.71.104003} {\bibfield  {journal} {\bibinfo  {journal} {Phys. Rev. D}\ }\textbf {\bibinfo {volume} {71}},\ \bibinfo {pages} {104003} (\bibinfo {year} {2005})}\BibitemShut {NoStop}%
\bibitem [{\citenamefont {Antoniou}\ \emph {et~al.}(2025)\citenamefont {Antoniou}, \citenamefont {Gualtieri},\ and\ \citenamefont {Pani}}]{antoniou_gravitational_2025}%
  \BibitemOpen
  \bibfield  {author} {\bibinfo {author} {\bibfnamefont {G.}~\bibnamefont {Antoniou}}, \bibinfo {author} {\bibfnamefont {L.}~\bibnamefont {Gualtieri}},\ and\ \bibinfo {author} {\bibfnamefont {P.}~\bibnamefont {Pani}},\ }\bibfield  {title} {\bibinfo {title} {Gravitational quasinormal modes of black holes in quadratic gravity},\ }\href {https://doi.org/10.1103/PhysRevD.111.064059} {\bibfield  {journal} {\bibinfo  {journal} {Physical Review D}\ }\textbf {\bibinfo {volume} {111}},\ \bibinfo {pages} {064059} (\bibinfo {year} {2025})}\BibitemShut {NoStop}%
\bibitem [{\citenamefont {Blázquez-Salcedo}\ \emph {et~al.}(2016)\citenamefont {Blázquez-Salcedo}, \citenamefont {Macedo}, \citenamefont {Cardoso}, \citenamefont {Ferrari}, \citenamefont {Gualtieri}, \citenamefont {Khoo}, \citenamefont {Kunz},\ and\ \citenamefont {Pani}}]{blazquez-salcedo_perturbed_2016}%
  \BibitemOpen
  \bibfield  {author} {\bibinfo {author} {\bibfnamefont {J.~L.}\ \bibnamefont {Blázquez-Salcedo}}, \bibinfo {author} {\bibfnamefont {C.~F.~B.}\ \bibnamefont {Macedo}}, \bibinfo {author} {\bibfnamefont {V.}~\bibnamefont {Cardoso}}, \bibinfo {author} {\bibfnamefont {V.}~\bibnamefont {Ferrari}}, \bibinfo {author} {\bibfnamefont {L.}~\bibnamefont {Gualtieri}}, \bibinfo {author} {\bibfnamefont {F.~S.}\ \bibnamefont {Khoo}}, \bibinfo {author} {\bibfnamefont {J.}~\bibnamefont {Kunz}},\ and\ \bibinfo {author} {\bibfnamefont {P.}~\bibnamefont {Pani}},\ }\bibfield  {title} {\bibinfo {title} {Perturbed black holes in {Einstein}-dilaton-{Gauss}-{Bonnet} gravity: {Stability}, ringdown, and gravitational-wave emission},\ }\href {https://doi.org/10.1103/PhysRevD.94.104024} {\bibfield  {journal} {\bibinfo  {journal} {Physical Review D}\ }\textbf {\bibinfo {volume} {94}},\ \bibinfo {pages} {104024} (\bibinfo {year} {2016})},\ \bibinfo {note} {arXiv:1609.01286 [gr-qc]}\BibitemShut {NoStop}%
\bibitem [{\citenamefont {Cardoso}\ \emph {et~al.}(2018)\citenamefont {Cardoso}, \citenamefont {Kimura}, \citenamefont {Maselli},\ and\ \citenamefont {Senatore}}]{cardoso_black_2018}%
  \BibitemOpen
  \bibfield  {author} {\bibinfo {author} {\bibfnamefont {V.}~\bibnamefont {Cardoso}}, \bibinfo {author} {\bibfnamefont {M.}~\bibnamefont {Kimura}}, \bibinfo {author} {\bibfnamefont {A.}~\bibnamefont {Maselli}},\ and\ \bibinfo {author} {\bibfnamefont {L.}~\bibnamefont {Senatore}},\ }\bibfield  {title} {\bibinfo {title} {Black holes in an {Effective} {Field} {Theory} extension of {GR}},\ }\href {https://doi.org/10.1103/PhysRevLett.121.251105} {\bibfield  {journal} {\bibinfo  {journal} {Physical Review Letters}\ }\textbf {\bibinfo {volume} {121}},\ \bibinfo {pages} {251105} (\bibinfo {year} {2018})},\ \bibinfo {note} {arXiv:1808.08962 [gr-qc]}\BibitemShut {NoStop}%
\bibitem [{\citenamefont {Guo}\ \emph {et~al.}(2025)\citenamefont {Guo}, \citenamefont {Shashank},\ and\ \citenamefont {Bambi}}]{guo_quasi-normal_2025}%
  \BibitemOpen
  \bibfield  {author} {\bibinfo {author} {\bibfnamefont {Y.}~\bibnamefont {Guo}}, \bibinfo {author} {\bibfnamefont {S.}~\bibnamefont {Shashank}},\ and\ \bibinfo {author} {\bibfnamefont {C.}~\bibnamefont {Bambi}},\ }\bibfield  {title} {\bibinfo {title} {Quasi-normal modes of slowly-rotating {Johannsen} black holes},\ }\href {https://doi.org/10.1140/epjc/s10052-025-14118-9} {\bibfield  {journal} {\bibinfo  {journal} {The European Physical Journal C}\ }\textbf {\bibinfo {volume} {85}},\ \bibinfo {pages} {425} (\bibinfo {year} {2025})},\ \bibinfo {note} {arXiv:2412.08205 [gr-qc]}\BibitemShut {NoStop}%
\bibitem [{\citenamefont {Abac}\ \emph {et~al.}(2025{\natexlab{b}})\citenamefont {Abac} \emph {et~al.}}]{the_ligo_scientific_collaboration_gwtc-40_2025}%
  \BibitemOpen
  \bibfield  {author} {\bibinfo {author} {\bibfnamefont {A.~G.}\ \bibnamefont {Abac}} \emph {et~al.} (\bibinfo {collaboration} {The LIGO Scientific Collaboration, the Virgo Collaboration, and the KAGRA Collaboration}),\ }\href {https://arxiv.org/abs/2508.18082} {\bibinfo {title} {{GWTC}-4.0: {Updating} the {Gravitational}-{Wave} {Transient} {Catalog} with {Observations} from the {First} {Part} of the {Fourth} {LIGO}-{Virgo}-{KAGRA} {Observing} {Run}}} (\bibinfo {year} {2025}{\natexlab{b}}),\ \bibinfo {note} {arXiv: 2508.18082 [gr-qc]}\BibitemShut {NoStop}%
\bibitem [{\citenamefont {Hussain}\ and\ \citenamefont {Zimmerman}(2022)}]{hussain_approach_2022}%
  \BibitemOpen
  \bibfield  {author} {\bibinfo {author} {\bibfnamefont {A.}~\bibnamefont {Hussain}}\ and\ \bibinfo {author} {\bibfnamefont {A.}~\bibnamefont {Zimmerman}},\ }\bibfield  {title} {\bibinfo {title} {An approach to computing spectral shifts for black holes beyond {Kerr}},\ }\href {https://doi.org/10.1103/PhysRevD.106.104018} {\bibfield  {journal} {\bibinfo  {journal} {Physical Review D}\ }\textbf {\bibinfo {volume} {106}},\ \bibinfo {pages} {104018} (\bibinfo {year} {2022})},\ \bibinfo {note} {arXiv:2206.10653 [gr-qc]}\BibitemShut {NoStop}%
\bibitem [{\citenamefont {Li}\ \emph {et~al.}(2023)\citenamefont {Li}, \citenamefont {Wagle}, \citenamefont {Chen},\ and\ \citenamefont {Yunes}}]{li_perturbations_2023}%
  \BibitemOpen
  \bibfield  {author} {\bibinfo {author} {\bibfnamefont {D.}~\bibnamefont {Li}}, \bibinfo {author} {\bibfnamefont {P.}~\bibnamefont {Wagle}}, \bibinfo {author} {\bibfnamefont {Y.}~\bibnamefont {Chen}},\ and\ \bibinfo {author} {\bibfnamefont {N.}~\bibnamefont {Yunes}},\ }\bibfield  {title} {\bibinfo {title} {Perturbations of spinning black holes beyond {General} {Relativity}: {Modified} {Teukolsky} equation},\ }\href {https://doi.org/10.1103/PhysRevX.13.021029} {\bibfield  {journal} {\bibinfo  {journal} {Physical Review X}\ }\textbf {\bibinfo {volume} {13}},\ \bibinfo {pages} {021029} (\bibinfo {year} {2023})},\ \bibinfo {note} {arXiv:2206.10652 [gr-qc]}\BibitemShut {NoStop}%
\bibitem [{\citenamefont {Mark}\ \emph {et~al.}(2015)\citenamefont {Mark}, \citenamefont {Yang}, \citenamefont {Zimmerman},\ and\ \citenamefont {Chen}}]{mark_quasinormal_2015}%
  \BibitemOpen
  \bibfield  {author} {\bibinfo {author} {\bibfnamefont {Z.}~\bibnamefont {Mark}}, \bibinfo {author} {\bibfnamefont {H.}~\bibnamefont {Yang}}, \bibinfo {author} {\bibfnamefont {A.}~\bibnamefont {Zimmerman}},\ and\ \bibinfo {author} {\bibfnamefont {Y.}~\bibnamefont {Chen}},\ }\bibfield  {title} {\bibinfo {title} {Quasinormal modes of weakly charged {Kerr}-{Newman} spacetimes},\ }\href {https://doi.org/10.1103/PhysRevD.91.044025} {\bibfield  {journal} {\bibinfo  {journal} {Physical Review D}\ }\textbf {\bibinfo {volume} {91}},\ \bibinfo {pages} {044025} (\bibinfo {year} {2015})}\BibitemShut {NoStop}%
\bibitem [{\citenamefont {Li}\ \emph {et~al.}(2024)\citenamefont {Li}, \citenamefont {Hussain}, \citenamefont {Wagle}, \citenamefont {Chen}, \citenamefont {Yunes},\ and\ \citenamefont {Zimmerman}}]{li_isospectrality_2024}%
  \BibitemOpen
  \bibfield  {author} {\bibinfo {author} {\bibfnamefont {D.}~\bibnamefont {Li}}, \bibinfo {author} {\bibfnamefont {A.}~\bibnamefont {Hussain}}, \bibinfo {author} {\bibfnamefont {P.}~\bibnamefont {Wagle}}, \bibinfo {author} {\bibfnamefont {Y.}~\bibnamefont {Chen}}, \bibinfo {author} {\bibfnamefont {N.}~\bibnamefont {Yunes}},\ and\ \bibinfo {author} {\bibfnamefont {A.}~\bibnamefont {Zimmerman}},\ }\bibfield  {title} {\bibinfo {title} {Isospectrality breaking in the {Teukolsky} formalism},\ }\href {https://doi.org/10.1103/PhysRevD.109.104026} {\bibfield  {journal} {\bibinfo  {journal} {Physical Review D}\ }\textbf {\bibinfo {volume} {109}},\ \bibinfo {pages} {104026} (\bibinfo {year} {2024})}\BibitemShut {NoStop}%
\bibitem [{\citenamefont {Jackiw}\ and\ \citenamefont {Pi}(2003)}]{jackiw_chern-simons_2003}%
  \BibitemOpen
  \bibfield  {author} {\bibinfo {author} {\bibfnamefont {R.}~\bibnamefont {Jackiw}}\ and\ \bibinfo {author} {\bibfnamefont {S.-Y.}\ \bibnamefont {Pi}},\ }\bibfield  {title} {\bibinfo {title} {Chern-{Simons} modification of general relativity},\ }\href {https://doi.org/10.1103/PhysRevD.68.104012} {\bibfield  {journal} {\bibinfo  {journal} {Physical Review D}\ }\textbf {\bibinfo {volume} {68}},\ \bibinfo {pages} {104012} (\bibinfo {year} {2003})}\BibitemShut {NoStop}%
\bibitem [{\citenamefont {Wagle}\ \emph {et~al.}(2024)\citenamefont {Wagle}, \citenamefont {Li}, \citenamefont {Chen},\ and\ \citenamefont {Yunes}}]{wagle_perturbations_2023}%
  \BibitemOpen
  \bibfield  {author} {\bibinfo {author} {\bibfnamefont {P.}~\bibnamefont {Wagle}}, \bibinfo {author} {\bibfnamefont {D.}~\bibnamefont {Li}}, \bibinfo {author} {\bibfnamefont {Y.}~\bibnamefont {Chen}},\ and\ \bibinfo {author} {\bibfnamefont {N.}~\bibnamefont {Yunes}},\ }\bibfield  {title} {\bibinfo {title} {Perturbations of spinning black holes in dynamical chern-simons gravity: Slow rotation equations},\ }\href {https://doi.org/10.1103/PhysRevD.109.104029} {\bibfield  {journal} {\bibinfo  {journal} {Phys. Rev. D}\ }\textbf {\bibinfo {volume} {109}},\ \bibinfo {pages} {104029} (\bibinfo {year} {2024})}\BibitemShut {NoStop}%
\bibitem [{\citenamefont {Li}\ \emph {et~al.}(2025)\citenamefont {Li}, \citenamefont {Wagle}, \citenamefont {Chen},\ and\ \citenamefont {Yunes}}]{li_perturbations_2025}%
  \BibitemOpen
  \bibfield  {author} {\bibinfo {author} {\bibfnamefont {D.}~\bibnamefont {Li}}, \bibinfo {author} {\bibfnamefont {P.}~\bibnamefont {Wagle}}, \bibinfo {author} {\bibfnamefont {Y.}~\bibnamefont {Chen}},\ and\ \bibinfo {author} {\bibfnamefont {N.}~\bibnamefont {Yunes}},\ }\href {https://doi.org/10.48550/arXiv.2503.15606} {\bibinfo {title} {Perturbations of spinning black holes in dynamical {Chern}-{Simons} gravity: {Slow} rotation quasinormal modes}} (\bibinfo {year} {2025}),\ \bibinfo {note} {arXiv:2503.15606 [gr-qc]}\BibitemShut {NoStop}%
\bibitem [{\citenamefont {Cano}\ \emph {et~al.}(2024)\citenamefont {Cano}, \citenamefont {Capuano}, \citenamefont {Franchini}, \citenamefont {Maenaut},\ and\ \citenamefont {V\"olkel}}]{cano_higher-derivative_2024}%
  \BibitemOpen
  \bibfield  {author} {\bibinfo {author} {\bibfnamefont {P.~A.}\ \bibnamefont {Cano}}, \bibinfo {author} {\bibfnamefont {L.}~\bibnamefont {Capuano}}, \bibinfo {author} {\bibfnamefont {N.}~\bibnamefont {Franchini}}, \bibinfo {author} {\bibfnamefont {S.}~\bibnamefont {Maenaut}},\ and\ \bibinfo {author} {\bibfnamefont {S.~H.}\ \bibnamefont {V\"olkel}},\ }\bibfield  {title} {\bibinfo {title} {Higher-derivative corrections to the kerr quasinormal mode spectrum},\ }\href {https://doi.org/10.1103/PhysRevD.110.124057} {\bibfield  {journal} {\bibinfo  {journal} {Phys. Rev. D}\ }\textbf {\bibinfo {volume} {110}},\ \bibinfo {pages} {124057} (\bibinfo {year} {2024})}\BibitemShut {NoStop}%
\bibitem [{\citenamefont {Cano}\ \emph {et~al.}(2023)\citenamefont {Cano}, \citenamefont {Fransen}, \citenamefont {Hertog},\ and\ \citenamefont {Maenaut}}]{cano_universal_2023}%
  \BibitemOpen
  \bibfield  {author} {\bibinfo {author} {\bibfnamefont {P.~A.}\ \bibnamefont {Cano}}, \bibinfo {author} {\bibfnamefont {K.}~\bibnamefont {Fransen}}, \bibinfo {author} {\bibfnamefont {T.}~\bibnamefont {Hertog}},\ and\ \bibinfo {author} {\bibfnamefont {S.}~\bibnamefont {Maenaut}},\ }\bibfield  {title} {\bibinfo {title} {The universal {Teukolsky} equations and black hole perturbations in higher-derivative gravity},\ }\href {https://doi.org/10.1103/PhysRevD.108.024040} {\bibfield  {journal} {\bibinfo  {journal} {Physical Review D}\ }\textbf {\bibinfo {volume} {108}},\ \bibinfo {pages} {024040} (\bibinfo {year} {2023})},\ \bibinfo {note} {arXiv:2304.02663 [gr-qc]}\BibitemShut {NoStop}%
\bibitem [{\citenamefont {Weller}\ \emph {et~al.}(2025)\citenamefont {Weller}, \citenamefont {Li},\ and\ \citenamefont {Chen}}]{weller_spectroscopy_2025}%
  \BibitemOpen
  \bibfield  {author} {\bibinfo {author} {\bibfnamefont {C.}~\bibnamefont {Weller}}, \bibinfo {author} {\bibfnamefont {D.}~\bibnamefont {Li}},\ and\ \bibinfo {author} {\bibfnamefont {Y.}~\bibnamefont {Chen}},\ }\bibfield  {title} {\bibinfo {title} {Spectroscopy of bumpy {BHs}: {The} nonrotating case},\ }\href {https://doi.org/10.1103/PhysRevD.111.024064} {\bibfield  {journal} {\bibinfo  {journal} {Physical Review D}\ }\textbf {\bibinfo {volume} {111}},\ \bibinfo {pages} {024064} (\bibinfo {year} {2025})}\BibitemShut {NoStop}%
\bibitem [{\citenamefont {Chung}\ and\ \citenamefont {Yunes}(2024{\natexlab{a}})}]{chung_quasi-normal_2024}%
  \BibitemOpen
  \bibfield  {author} {\bibinfo {author} {\bibfnamefont {A.~K.-W.}\ \bibnamefont {Chung}}\ and\ \bibinfo {author} {\bibfnamefont {N.}~\bibnamefont {Yunes}},\ }\bibfield  {title} {\bibinfo {title} {Quasi-normal mode frequencies and gravitational perturbations of black holes with any subextremal spin in modified gravity through {METRICS}: the scalar-{Gauss}-{Bonnet} gravity case},\ }\href {https://doi.org/10.1103/PhysRevD.110.064019} {\bibfield  {journal} {\bibinfo  {journal} {Physical Review D}\ }\textbf {\bibinfo {volume} {110}},\ \bibinfo {pages} {064019} (\bibinfo {year} {2024}{\natexlab{a}})},\ \bibinfo {note} {arXiv:2406.11986 [gr-qc]}\BibitemShut {NoStop}%
\bibitem [{\citenamefont {Chung}\ \emph {et~al.}(2024)\citenamefont {Chung}, \citenamefont {Wagle},\ and\ \citenamefont {Yunes}}]{chung_spectral_2024}%
  \BibitemOpen
  \bibfield  {author} {\bibinfo {author} {\bibfnamefont {A.~K.-W.}\ \bibnamefont {Chung}}, \bibinfo {author} {\bibfnamefont {P.}~\bibnamefont {Wagle}},\ and\ \bibinfo {author} {\bibfnamefont {N.}~\bibnamefont {Yunes}},\ }\bibfield  {title} {\bibinfo {title} {Spectral method for metric perturbations of black holes: {Kerr} background case in general relativity},\ }\href {https://doi.org/10.1103/PhysRevD.109.044072} {\bibfield  {journal} {\bibinfo  {journal} {Physical Review D}\ }\textbf {\bibinfo {volume} {109}},\ \bibinfo {pages} {044072} (\bibinfo {year} {2024})},\ \bibinfo {note} {arXiv:2312.08435 [gr-qc]}\BibitemShut {NoStop}%
\bibitem [{\citenamefont {Chung}\ and\ \citenamefont {Yunes}(2024{\natexlab{b}})}]{chung_ringing_2024}%
  \BibitemOpen
  \bibfield  {author} {\bibinfo {author} {\bibfnamefont {A.~K.-W.}\ \bibnamefont {Chung}}\ and\ \bibinfo {author} {\bibfnamefont {N.}~\bibnamefont {Yunes}},\ }\bibfield  {title} {\bibinfo {title} {Ringing out {General} {Relativity}: {Quasi}-normal mode frequencies for black holes of any spin in modified gravity},\ }\href {https://doi.org/10.1103/PhysRevLett.133.181401} {\bibfield  {journal} {\bibinfo  {journal} {Physical Review Letters}\ }\textbf {\bibinfo {volume} {133}},\ \bibinfo {pages} {181401} (\bibinfo {year} {2024}{\natexlab{b}})},\ \bibinfo {note} {arXiv:2405.12280 [gr-qc]}\BibitemShut {NoStop}%
\bibitem [{\citenamefont {Lam}\ \emph {et~al.}(2025{\natexlab{a}})\citenamefont {Lam}, \citenamefont {Chung},\ and\ \citenamefont {Yunes}}]{lam_near-extremal_2025}%
  \BibitemOpen
  \bibfield  {author} {\bibinfo {author} {\bibfnamefont {K.~K.-H.}\ \bibnamefont {Lam}}, \bibinfo {author} {\bibfnamefont {A.~K.-W.}\ \bibnamefont {Chung}},\ and\ \bibinfo {author} {\bibfnamefont {N.}~\bibnamefont {Yunes}},\ }\href {https://doi.org/10.48550/arXiv.2509.07061} {\bibinfo {title} {Near-{Extremal} {Black} {Holes} in {Modified} {Gravity} via {Spectral} {Methods}}} (\bibinfo {year} {2025}{\natexlab{a}}),\ \bibinfo {note} {arXiv:2509.07061 [gr-qc]}\BibitemShut {NoStop}%
\bibitem [{\citenamefont {Chung}\ \emph {et~al.}(2025)\citenamefont {Chung}, \citenamefont {Lam},\ and\ \citenamefont {Yunes}}]{chung_quasinormal_2025}%
  \BibitemOpen
  \bibfield  {author} {\bibinfo {author} {\bibfnamefont {A.~K.-W.}\ \bibnamefont {Chung}}, \bibinfo {author} {\bibfnamefont {K.~K.-H.}\ \bibnamefont {Lam}},\ and\ \bibinfo {author} {\bibfnamefont {N.}~\bibnamefont {Yunes}},\ }\bibfield  {title} {\bibinfo {title} {Quasinormal mode frequencies and gravitational perturbations of spinning black holes in modified gravity through {METRICS}: {The} dynamical {Chern}-{Simons} gravity case},\ }\href {https://doi.org/10.1103/g83n-rrlj} {\bibfield  {journal} {\bibinfo  {journal} {Physical Review D}\ }\textbf {\bibinfo {volume} {111}},\ \bibinfo {pages} {124052} (\bibinfo {year} {2025})},\ \bibinfo {note} {arXiv:2503.11759 [gr-qc]}\BibitemShut {NoStop}%
\bibitem [{\citenamefont {Lam}\ \emph {et~al.}(2025{\natexlab{b}})\citenamefont {Lam}, \citenamefont {Chung},\ and\ \citenamefont {Yunes}}]{lam_analytic_2025}%
  \BibitemOpen
  \bibfield  {author} {\bibinfo {author} {\bibfnamefont {K.~K.-H.}\ \bibnamefont {Lam}}, \bibinfo {author} {\bibfnamefont {A.~K.-W.}\ \bibnamefont {Chung}},\ and\ \bibinfo {author} {\bibfnamefont {N.}~\bibnamefont {Yunes}},\ }\href {https://doi.org/10.48550/arXiv.2510.05208} {\bibinfo {title} {Analytic and accurate approximate metrics for black holes with arbitrary rotation in beyond-{Einstein} gravity using spectral methods}} (\bibinfo {year} {2025}{\natexlab{b}}),\ \bibinfo {note} {arXiv:2510.05208 [gr-qc]}\BibitemShut {NoStop}%
\bibitem [{\citenamefont {Johannsen}\ and\ \citenamefont {Psaltis}(2011)}]{johannsen_metric_2011}%
  \BibitemOpen
  \bibfield  {author} {\bibinfo {author} {\bibfnamefont {T.}~\bibnamefont {Johannsen}}\ and\ \bibinfo {author} {\bibfnamefont {D.}~\bibnamefont {Psaltis}},\ }\bibfield  {title} {\bibinfo {title} {Metric for rapidly spinning black holes suitable for strong-field tests of the no-hair theorem},\ }\href {https://doi.org/10.1103/PhysRevD.83.124015} {\bibfield  {journal} {\bibinfo  {journal} {Physical Review D}\ }\textbf {\bibinfo {volume} {83}},\ \bibinfo {pages} {124015} (\bibinfo {year} {2011})}\BibitemShut {NoStop}%
\bibitem [{\citenamefont {Johannsen}(2013)}]{johannsen_inner_2013}%
  \BibitemOpen
  \bibfield  {author} {\bibinfo {author} {\bibfnamefont {T.}~\bibnamefont {Johannsen}},\ }\bibfield  {title} {\bibinfo {title} {Inner {Accretion} {Disk} {Edges} in a {Kerr}-{Like} {Spacetime}},\ }\href {https://doi.org/10.1103/PhysRevD.87.124010} {\bibfield  {journal} {\bibinfo  {journal} {Physical Review D}\ }\textbf {\bibinfo {volume} {87}},\ \bibinfo {pages} {124010} (\bibinfo {year} {2013})},\ \bibinfo {note} {arXiv:1304.8106 [gr-qc]}\BibitemShut {NoStop}%
\bibitem [{\citenamefont {Wang}\ \emph {et~al.}(2025)\citenamefont {Wang}, \citenamefont {Zhao}, \citenamefont {Zeng},\ and\ \citenamefont {Wang}}]{wang_revisiting_2025}%
  \BibitemOpen
  \bibfield  {author} {\bibinfo {author} {\bibfnamefont {X.}~\bibnamefont {Wang}}, \bibinfo {author} {\bibfnamefont {Z.}~\bibnamefont {Zhao}}, \bibinfo {author} {\bibfnamefont {X.-X.}\ \bibnamefont {Zeng}},\ and\ \bibinfo {author} {\bibfnamefont {X.-Y.}\ \bibnamefont {Wang}},\ }\bibfield  {title} {\bibinfo {title} {Revisiting the shadow of {Johannsen}-{Psaltis} black holes},\ }\href {https://doi.org/10.1103/PhysRevD.111.084054} {\bibfield  {journal} {\bibinfo  {journal} {Physical Review D}\ }\textbf {\bibinfo {volume} {111}},\ \bibinfo {pages} {084054} (\bibinfo {year} {2025})}\BibitemShut {NoStop}%
\bibitem [{\citenamefont {Kong}\ \emph {et~al.}(2014)\citenamefont {Kong}, \citenamefont {Li},\ and\ \citenamefont {Bambi}}]{kong_constraints_2014}%
  \BibitemOpen
  \bibfield  {author} {\bibinfo {author} {\bibfnamefont {L.}~\bibnamefont {Kong}}, \bibinfo {author} {\bibfnamefont {Z.}~\bibnamefont {Li}},\ and\ \bibinfo {author} {\bibfnamefont {C.}~\bibnamefont {Bambi}},\ }\bibfield  {title} {\bibinfo {title} {Constraints on the spacetime geometry around 10 stellar-mass black hole candidates from the disk's thermal spectrum},\ }\href {https://doi.org/10.1088/0004-637X/797/2/78} {\bibfield  {journal} {\bibinfo  {journal} {The Astrophysical Journal}\ }\textbf {\bibinfo {volume} {797}},\ \bibinfo {pages} {78} (\bibinfo {year} {2014})}\BibitemShut {NoStop}%
\bibitem [{\citenamefont {John}\ and\ \citenamefont {Stevens}(2019)}]{john_bondi_2019}%
  \BibitemOpen
  \bibfield  {author} {\bibinfo {author} {\bibfnamefont {A.~J.}\ \bibnamefont {John}}\ and\ \bibinfo {author} {\bibfnamefont {C.~Z.}\ \bibnamefont {Stevens}},\ }\bibfield  {title} {\bibinfo {title} {Bondi accretion in the spherically symmetric {Johannsen}–{Psaltis} spacetime},\ }\href {https://doi.org/10.1140/epjc/s10052-019-7481-1} {\bibfield  {journal} {\bibinfo  {journal} {The European Physical Journal C}\ }\textbf {\bibinfo {volume} {79}},\ \bibinfo {pages} {962} (\bibinfo {year} {2019})}\BibitemShut {NoStop}%
\bibitem [{\citenamefont {Newman}\ and\ \citenamefont {Janis}(1965)}]{newman_note_1965}%
  \BibitemOpen
  \bibfield  {author} {\bibinfo {author} {\bibfnamefont {E.~T.}\ \bibnamefont {Newman}}\ and\ \bibinfo {author} {\bibfnamefont {A.~I.}\ \bibnamefont {Janis}},\ }\bibfield  {title} {\bibinfo {title} {Note on the {Kerr} {Spinning}-{Particle} {Metric}},\ }\href {https://doi.org/10.1063/1.1704350} {\bibfield  {journal} {\bibinfo  {journal} {Journal of Mathematical Physics}\ }\textbf {\bibinfo {volume} {6}},\ \bibinfo {pages} {915} (\bibinfo {year} {1965})}\BibitemShut {NoStop}%
\bibitem [{\citenamefont {Biskupek}\ \emph {et~al.}(2021)\citenamefont {Biskupek}, \citenamefont {Müller},\ and\ \citenamefont {Torre}}]{biskupek_benefit_2021}%
  \BibitemOpen
  \bibfield  {author} {\bibinfo {author} {\bibfnamefont {L.}~\bibnamefont {Biskupek}}, \bibinfo {author} {\bibfnamefont {J.}~\bibnamefont {Müller}},\ and\ \bibinfo {author} {\bibfnamefont {J.-M.}\ \bibnamefont {Torre}},\ }\bibfield  {title} {\bibinfo {title} {Benefit of {New} {High}-{Precision} {LLR} {Data} for the {Determination} of {Relativistic} {Parameters}},\ }\href {https://doi.org/10.3390/universe7020034} {\bibfield  {journal} {\bibinfo  {journal} {Universe}\ }\textbf {\bibinfo {volume} {7}},\ \bibinfo {pages} {34} (\bibinfo {year} {2021})}\BibitemShut {NoStop}%
\bibitem [{\citenamefont {Santos}\ \emph {et~al.}(2024)\citenamefont {Santos}, \citenamefont {Nunes},\ and\ \citenamefont {De~Araujo}}]{santos_testing_2024}%
  \BibitemOpen
  \bibfield  {author} {\bibinfo {author} {\bibfnamefont {R.~M.}\ \bibnamefont {Santos}}, \bibinfo {author} {\bibfnamefont {R.~C.}\ \bibnamefont {Nunes}},\ and\ \bibinfo {author} {\bibfnamefont {J.~C.~N.}\ \bibnamefont {De~Araujo}},\ }\bibfield  {title} {\bibinfo {title} {Testing beyond-{Kerr} spacetimes with {GWTC}-3},\ }\href {https://doi.org/10.1140/epjc/s10052-024-12666-0} {\bibfield  {journal} {\bibinfo  {journal} {The European Physical Journal C}\ }\textbf {\bibinfo {volume} {84}},\ \bibinfo {pages} {302} (\bibinfo {year} {2024})}\BibitemShut {NoStop}%
\bibitem [{\citenamefont {Pratten}\ \emph {et~al.}(2021)\citenamefont {Pratten} \emph {et~al.}}]{Pratten:2020ceb}%
  \BibitemOpen
  \bibfield  {author} {\bibinfo {author} {\bibfnamefont {G.}~\bibnamefont {Pratten}} \emph {et~al.},\ }\bibfield  {title} {\bibinfo {title} {{Computationally efficient models for the dominant and subdominant harmonic modes of precessing binary black holes}},\ }\href {https://doi.org/10.1103/PhysRevD.103.104056} {\bibfield  {journal} {\bibinfo  {journal} {Phys. Rev. D}\ }\textbf {\bibinfo {volume} {103}},\ \bibinfo {pages} {104056} (\bibinfo {year} {2021})},\ \Eprint {https://arxiv.org/abs/2004.06503} {arXiv:2004.06503 [gr-qc]} \BibitemShut {NoStop}%
\bibitem [{\citenamefont {Nitz}\ \emph {et~al.}(2024)\citenamefont {Nitz}, \citenamefont {Harry}, \citenamefont {Brown}, \citenamefont {Biwer}, \citenamefont {Willis} \emph {et~al.}}]{nitz_gwastropycbc_2024}%
  \BibitemOpen
  \bibfield  {author} {\bibinfo {author} {\bibfnamefont {A.}~\bibnamefont {Nitz}}, \bibinfo {author} {\bibfnamefont {I.}~\bibnamefont {Harry}}, \bibinfo {author} {\bibfnamefont {D.}~\bibnamefont {Brown}}, \bibinfo {author} {\bibfnamefont {C.~M.}\ \bibnamefont {Biwer}}, \bibinfo {author} {\bibnamefont {Willis}}, \emph {et~al.},\ }\href {https://doi.org/10.5281/zenodo.10473621} {\bibinfo {title} {gwastro/pycbc: v2.3.3 release of {PyCBC}}} (\bibinfo {year} {2024})\BibitemShut {NoStop}%
\bibitem [{\citenamefont {Ashton}\ \emph {et~al.}(2019)\citenamefont {Ashton} \emph {et~al.}}]{ashton_bilby_2019}%
  \BibitemOpen
  \bibfield  {author} {\bibinfo {author} {\bibfnamefont {G.}~\bibnamefont {Ashton}} \emph {et~al.},\ }\bibfield  {title} {\bibinfo {title} {{BILBY}: {A} user-friendly {Bayesian} inference library for gravitational-wave astronomy},\ }\href {https://doi.org/10.3847/1538-4365/ab06fc} {\bibfield  {journal} {\bibinfo  {journal} {Astrophys. J. Suppl.}\ }\textbf {\bibinfo {volume} {241}},\ \bibinfo {pages} {27} (\bibinfo {year} {2019})},\ \bibinfo {note} {arXiv:2509.08099 [gr-qc]}\BibitemShut {NoStop}%
\bibitem [{\citenamefont {Pound}\ and\ \citenamefont {Wardell}(2022)}]{pound_black_2022}%
  \BibitemOpen
  \bibfield  {author} {\bibinfo {author} {\bibfnamefont {A.}~\bibnamefont {Pound}}\ and\ \bibinfo {author} {\bibfnamefont {B.}~\bibnamefont {Wardell}},\ }\bibfield  {title} {\bibinfo {title} {Black {Hole} {Perturbation} {Theory} and {Gravitational} {Self}-{Force}},\ }in\ \href {https://doi.org/10.1007/978-981-16-4306-4_38} {\emph {\bibinfo {booktitle} {Handbook of {Gravitational} {Wave} {Astronomy}}}},\ \bibinfo {editor} {edited by\ \bibinfo {editor} {\bibfnamefont {C.}~\bibnamefont {Bambi}}, \bibinfo {editor} {\bibfnamefont {S.}~\bibnamefont {Katsanevas}},\ and\ \bibinfo {editor} {\bibfnamefont {K.~D.}\ \bibnamefont {Kokkotas}}}\ (\bibinfo  {publisher} {Springer Nature Singapore},\ \bibinfo {address} {Singapore},\ \bibinfo {year} {2022})\ pp.\ \bibinfo {pages} {1411--1529}\BibitemShut {NoStop}%
\bibitem [{\citenamefont {Glampedakis}\ \emph {et~al.}(2017)\citenamefont {Glampedakis}, \citenamefont {Pappas}, \citenamefont {Silva},\ and\ \citenamefont {Berti}}]{glampedakis_post-kerr_2017}%
  \BibitemOpen
  \bibfield  {author} {\bibinfo {author} {\bibfnamefont {K.}~\bibnamefont {Glampedakis}}, \bibinfo {author} {\bibfnamefont {G.}~\bibnamefont {Pappas}}, \bibinfo {author} {\bibfnamefont {H.~O.}\ \bibnamefont {Silva}},\ and\ \bibinfo {author} {\bibfnamefont {E.}~\bibnamefont {Berti}},\ }\bibfield  {title} {\bibinfo {title} {Post-{Kerr} black hole spectroscopy},\ }\href {https://doi.org/10.1103/PhysRevD.96.064054} {\bibfield  {journal} {\bibinfo  {journal} {Physical Review D}\ }\textbf {\bibinfo {volume} {96}},\ \bibinfo {pages} {064054} (\bibinfo {year} {2017})},\ \bibinfo {note} {arXiv:1706.07658 [gr-qc]}\BibitemShut {NoStop}%
\bibitem [{\citenamefont {Cowling}(1941)}]{cowling_non-radial_1941}%
  \BibitemOpen
  \bibfield  {author} {\bibinfo {author} {\bibfnamefont {T.~G.}\ \bibnamefont {Cowling}},\ }\bibfield  {title} {\bibinfo {title} {The {Non}-radial {Oscillations} of {Polytropic} {Stars}},\ }\href {https://doi.org/10.1093/mnras/101.8.367} {\bibfield  {journal} {\bibinfo  {journal} {Monthly Notices of the Royal Astronomical Society}\ }\textbf {\bibinfo {volume} {101}},\ \bibinfo {pages} {367} (\bibinfo {year} {1941})}\BibitemShut {NoStop}%
\bibitem [{\citenamefont {Christensen-Dalsgaard}(2003)}]{Dalsgaard:2003book}%
  \BibitemOpen
  \bibfield  {author} {\bibinfo {author} {\bibfnamefont {J.}~\bibnamefont {Christensen-Dalsgaard}},\ }\href {https://users-phys.au.dk/jcd/oscilnotes/} {\emph {\bibinfo {title} {{Lecture Notes on Stellar Oscillations}}}}\ (\bibinfo  {publisher} {Institut for Fysik og Astronomi, Aarhus Universitet},\ \bibinfo {address} {Aarhus, Denmark},\ \bibinfo {year} {2003})\BibitemShut {NoStop}%
\bibitem [{\citenamefont {Dudley}\ and\ \citenamefont {Finley}(1977)}]{DudleyI}%
  \BibitemOpen
  \bibfield  {author} {\bibinfo {author} {\bibfnamefont {A.~L.}\ \bibnamefont {Dudley}}\ and\ \bibinfo {author} {\bibfnamefont {J.~D.}\ \bibnamefont {Finley}},\ }\bibfield  {title} {\bibinfo {title} {Separation of wave equations for perturbations of general type-$d$ space-times},\ }\href {https://doi.org/10.1103/PhysRevLett.38.1505} {\bibfield  {journal} {\bibinfo  {journal} {Phys. Rev. Lett.}\ }\textbf {\bibinfo {volume} {38}},\ \bibinfo {pages} {1505} (\bibinfo {year} {1977})}\BibitemShut {NoStop}%
\bibitem [{\citenamefont {Dudley}\ and\ \citenamefont {Finley}(1979)}]{DudleyII}%
  \BibitemOpen
  \bibfield  {author} {\bibinfo {author} {\bibfnamefont {A.~L.}\ \bibnamefont {Dudley}}\ and\ \bibinfo {author} {\bibfnamefont {I.}~\bibnamefont {Finley}, \bibfnamefont {J.~D.}},\ }\bibfield  {title} {\bibinfo {title} {Covariant perturbed wave equations in arbitrary type‐d backgrounds},\ }\href {https://doi.org/10.1063/1.524064} {\bibfield  {journal} {\bibinfo  {journal} {Journal of Mathematical Physics}\ }\textbf {\bibinfo {volume} {20}},\ \bibinfo {pages} {311} (\bibinfo {year} {1979})}\BibitemShut {NoStop}%
\bibitem [{\citenamefont {Saha}\ and\ \citenamefont {Silva}(2026)}]{SahaDudleyFinley}%
  \BibitemOpen
  \bibfield  {author} {\bibinfo {author} {\bibfnamefont {S.}~\bibnamefont {Saha}}\ and\ \bibinfo {author} {\bibfnamefont {H.~O.}\ \bibnamefont {Silva}},\ }\bibfield  {title} {\bibinfo {title} {Quasinormal modes of {Kerr-Newman} black holes: Revisiting the {Dudley-Finley} approximation},\ }\href {https://doi.org/10.1103/2wc1-yntl} {\bibfield  {journal} {\bibinfo  {journal} {Phys. Rev. D}\ }\textbf {\bibinfo {volume} {113}},\ \bibinfo {pages} {064009} (\bibinfo {year} {2026})}\BibitemShut {NoStop}%
\bibitem [{\citenamefont {Teukolsky}(1973)}]{teukolsky_perturbations_1973}%
  \BibitemOpen
  \bibfield  {author} {\bibinfo {author} {\bibfnamefont {S.~A.}\ \bibnamefont {Teukolsky}},\ }\bibfield  {title} {\bibinfo {title} {Perturbations of a {Rotating} {Black} {Hole}. {I}. {Fundamental} {Equations} for {Gravitational}, {Electromagnetic}, and {Neutrino}-{Field} {Perturbations}},\ }\href {https://doi.org/10.1086/152444} {\bibfield  {journal} {\bibinfo  {journal} {The Astrophysical Journal}\ }\textbf {\bibinfo {volume} {185}},\ \bibinfo {pages} {635} (\bibinfo {year} {1973})}\BibitemShut {NoStop}%
\bibitem [{\citenamefont {Chrzanowski}(1975)}]{chrzanowski_vector_1975}%
  \BibitemOpen
  \bibfield  {author} {\bibinfo {author} {\bibfnamefont {P.~L.}\ \bibnamefont {Chrzanowski}},\ }\bibfield  {title} {\bibinfo {title} {Vector potential and metric perturbations of a rotating black hole},\ }\href {https://doi.org/10.1103/PhysRevD.11.2042} {\bibfield  {journal} {\bibinfo  {journal} {Physical Review D}\ }\textbf {\bibinfo {volume} {11}},\ \bibinfo {pages} {2042} (\bibinfo {year} {1975})}\BibitemShut {NoStop}%
\bibitem [{\citenamefont {Kegeles}\ and\ \citenamefont {Cohen}(1979)}]{kegeles_constructive_1979}%
  \BibitemOpen
  \bibfield  {author} {\bibinfo {author} {\bibfnamefont {L.~S.}\ \bibnamefont {Kegeles}}\ and\ \bibinfo {author} {\bibfnamefont {J.~M.}\ \bibnamefont {Cohen}},\ }\bibfield  {title} {\bibinfo {title} {Constructive procedure for perturbations of spacetimes},\ }\href {https://doi.org/10.1103/PhysRevD.19.1641} {\bibfield  {journal} {\bibinfo  {journal} {Physical Review D}\ }\textbf {\bibinfo {volume} {19}},\ \bibinfo {pages} {1641} (\bibinfo {year} {1979})}\BibitemShut {NoStop}%
\bibitem [{\citenamefont {Ori}(2003)}]{ori_reconstruction_2003}%
  \BibitemOpen
  \bibfield  {author} {\bibinfo {author} {\bibfnamefont {A.}~\bibnamefont {Ori}},\ }\bibfield  {title} {\bibinfo {title} {Reconstruction of inhomogeneous metric perturbations and electromagnetic four-potential in {Kerr} spacetime},\ }\href {https://doi.org/10.1103/PhysRevD.67.124010} {\bibfield  {journal} {\bibinfo  {journal} {Physical Review D}\ }\textbf {\bibinfo {volume} {67}},\ \bibinfo {pages} {124010} (\bibinfo {year} {2003})},\ \bibinfo {note} {arXiv:gr-qc/0207045}\BibitemShut {NoStop}%
\bibitem [{\citenamefont {Keidl}\ \emph {et~al.}(2007)\citenamefont {Keidl}, \citenamefont {Friedman},\ and\ \citenamefont {Wiseman}}]{keidl_finding_2007}%
  \BibitemOpen
  \bibfield  {author} {\bibinfo {author} {\bibfnamefont {T.~S.}\ \bibnamefont {Keidl}}, \bibinfo {author} {\bibfnamefont {J.~L.}\ \bibnamefont {Friedman}},\ and\ \bibinfo {author} {\bibfnamefont {A.~G.}\ \bibnamefont {Wiseman}},\ }\bibfield  {title} {\bibinfo {title} {Finding fields and self-force in a gauge appropriate to separable wave equations},\ }\href {https://doi.org/10.1103/PhysRevD.75.124009} {\bibfield  {journal} {\bibinfo  {journal} {Physical Review D}\ }\textbf {\bibinfo {volume} {75}},\ \bibinfo {pages} {124009} (\bibinfo {year} {2007})}\BibitemShut {NoStop}%
\bibitem [{\citenamefont {Keidl}\ \emph {et~al.}(2010)\citenamefont {Keidl}, \citenamefont {Shah}, \citenamefont {Friedman}, \citenamefont {Kim},\ and\ \citenamefont {Price}}]{keidl_gravitational_2010}%
  \BibitemOpen
  \bibfield  {author} {\bibinfo {author} {\bibfnamefont {T.~S.}\ \bibnamefont {Keidl}}, \bibinfo {author} {\bibfnamefont {A.~G.}\ \bibnamefont {Shah}}, \bibinfo {author} {\bibfnamefont {J.~L.}\ \bibnamefont {Friedman}}, \bibinfo {author} {\bibfnamefont {D.-H.}\ \bibnamefont {Kim}},\ and\ \bibinfo {author} {\bibfnamefont {L.~R.}\ \bibnamefont {Price}},\ }\bibfield  {title} {\bibinfo {title} {Gravitational self-force in a radiation gauge},\ }\bibfield  {journal} {\bibinfo  {journal} {Physical Review D}\ }\textbf {\bibinfo {volume} {82}},\ \href {https://doi.org/10.1103/physrevd.82.124012} {10.1103/physrevd.82.124012} (\bibinfo {year} {2010}),\ \bibinfo {note} {publisher: American Physical Society (APS)}\BibitemShut {NoStop}%
\bibitem [{\citenamefont {Nichols}\ \emph {et~al.}(2012)\citenamefont {Nichols}, \citenamefont {Zimmerman}, \citenamefont {Chen}, \citenamefont {Lovelace}, \citenamefont {Matthews}, \citenamefont {Owen}, \citenamefont {Zhang},\ and\ \citenamefont {Thorne}}]{nichols_visualizing_2012}%
  \BibitemOpen
  \bibfield  {author} {\bibinfo {author} {\bibfnamefont {D.~A.}\ \bibnamefont {Nichols}}, \bibinfo {author} {\bibfnamefont {A.}~\bibnamefont {Zimmerman}}, \bibinfo {author} {\bibfnamefont {Y.}~\bibnamefont {Chen}}, \bibinfo {author} {\bibfnamefont {G.}~\bibnamefont {Lovelace}}, \bibinfo {author} {\bibfnamefont {K.~D.}\ \bibnamefont {Matthews}}, \bibinfo {author} {\bibfnamefont {R.}~\bibnamefont {Owen}}, \bibinfo {author} {\bibfnamefont {F.}~\bibnamefont {Zhang}},\ and\ \bibinfo {author} {\bibfnamefont {K.~S.}\ \bibnamefont {Thorne}},\ }\bibfield  {title} {\bibinfo {title} {Visualizing spacetime curvature via frame-drag vortexes and tidal tendexes. {III}. {Quasinormal} pulsations of {Schwarzschild} and {Kerr} black holes},\ }\href {https://doi.org/10.1103/PhysRevD.86.104028} {\bibfield  {journal} {\bibinfo  {journal} {Physical Review D}\ }\textbf {\bibinfo {volume} {86}},\ \bibinfo {pages} {104028} (\bibinfo {year} {2012})}\BibitemShut {NoStop}%
\bibitem [{\citenamefont {Chandrasekhar}(1983)}]{chandrasekhar_mathematical_1983}%
  \BibitemOpen
  \bibfield  {author} {\bibinfo {author} {\bibfnamefont {S.}~\bibnamefont {Chandrasekhar}},\ }\href@noop {} {\emph {\bibinfo {title} {Mathematical {Theory} of {Black} {Holes}}}}\ (\bibinfo  {publisher} {Oxford University Press},\ \bibinfo {address} {New York},\ \bibinfo {year} {1983})\BibitemShut {NoStop}%
\bibitem [{\citenamefont {Teukolsky}\ and\ \citenamefont {Press}(1974)}]{teukolsky_perturbations_1974}%
  \BibitemOpen
  \bibfield  {author} {\bibinfo {author} {\bibfnamefont {S.~A.}\ \bibnamefont {Teukolsky}}\ and\ \bibinfo {author} {\bibfnamefont {W.~H.}\ \bibnamefont {Press}},\ }\bibfield  {title} {\bibinfo {title} {Perturbations of a rotating black hole. {III} - {Interaction} of the hole with gravitational and electromagnetic radiation},\ }\href {https://doi.org/10.1086/153180} {\bibfield  {journal} {\bibinfo  {journal} {The Astrophysical Journal}\ }\textbf {\bibinfo {volume} {193}},\ \bibinfo {pages} {443} (\bibinfo {year} {1974})}\BibitemShut {NoStop}%
\bibitem [{\citenamefont {Starobinskii}(1973)}]{starobinskii_amplification_1973}%
  \BibitemOpen
  \bibfield  {author} {\bibinfo {author} {\bibfnamefont {A.~A.}\ \bibnamefont {Starobinskii}},\ }\bibfield  {title} {\bibinfo {title} {Amplification of waves during reflection from a rotating ”black hole”},\ }\href@noop {} {\bibfield  {journal} {\bibinfo  {journal} {Sov. Phys. JETP}\ }\textbf {\bibinfo {volume} {37}},\ \bibinfo {pages} {28} (\bibinfo {year} {1973})}\BibitemShut {NoStop}%
\bibitem [{\citenamefont {Starobinskii}\ and\ \citenamefont {Churilov}(1974)}]{starobinskil_amplification_1974}%
  \BibitemOpen
  \bibfield  {author} {\bibinfo {author} {\bibfnamefont {A.~A.}\ \bibnamefont {Starobinskii}}\ and\ \bibinfo {author} {\bibfnamefont {S.~M.}\ \bibnamefont {Churilov}},\ }\bibfield  {title} {\bibinfo {title} {Amplification of electromagnetic and gravitational waves scattered by a rotating ”black hole”},\ }\href@noop {} {\bibfield  {journal} {\bibinfo  {journal} {Sov. Phys. JETP}\ }\textbf {\bibinfo {volume} {65}},\ \bibinfo {pages} {1} (\bibinfo {year} {1974})}\BibitemShut {NoStop}%
\bibitem [{\citenamefont {Berens}\ \emph {et~al.}(2024)\citenamefont {Berens}, \citenamefont {Gravely},\ and\ \citenamefont {Lupsasca}}]{Berens_2024}%
  \BibitemOpen
  \bibfield  {author} {\bibinfo {author} {\bibfnamefont {R.}~\bibnamefont {Berens}}, \bibinfo {author} {\bibfnamefont {T.}~\bibnamefont {Gravely}},\ and\ \bibinfo {author} {\bibfnamefont {A.}~\bibnamefont {Lupsasca}},\ }\bibfield  {title} {\bibinfo {title} {Gravitational waves on kerr black holes: I. reconstruction of linearized metric perturbations},\ }\href {https://doi.org/10.1088/1361-6382/ad6c9c} {\bibfield  {journal} {\bibinfo  {journal} {Classical and Quantum Gravity}\ }\textbf {\bibinfo {volume} {41}},\ \bibinfo {pages} {195004} (\bibinfo {year} {2024})}\BibitemShut {NoStop}%
\bibitem [{\citenamefont {Zimmerman}\ \emph {et~al.}(2015)\citenamefont {Zimmerman}, \citenamefont {Yang}, \citenamefont {Mark}, \citenamefont {Chen},\ and\ \citenamefont {Lehner}}]{Zimmerman:2014aha}%
  \BibitemOpen
  \bibfield  {author} {\bibinfo {author} {\bibfnamefont {A.}~\bibnamefont {Zimmerman}}, \bibinfo {author} {\bibfnamefont {H.}~\bibnamefont {Yang}}, \bibinfo {author} {\bibfnamefont {Z.}~\bibnamefont {Mark}}, \bibinfo {author} {\bibfnamefont {Y.}~\bibnamefont {Chen}},\ and\ \bibinfo {author} {\bibfnamefont {L.}~\bibnamefont {Lehner}},\ }\bibfield  {title} {\bibinfo {title} {{Quasinormal Modes Beyond Kerr}},\ }\href {https://doi.org/10.1007/978-3-319-10488-1_19} {\bibfield  {journal} {\bibinfo  {journal} {Astrophys. Space Sci. Proc.}\ }\textbf {\bibinfo {volume} {40}},\ \bibinfo {pages} {217} (\bibinfo {year} {2015})},\ \Eprint {https://arxiv.org/abs/1406.4206} {arXiv:1406.4206 [gr-qc]} \BibitemShut {NoStop}%
\bibitem [{\citenamefont {Cook}\ and\ \citenamefont {Zalutskiy}(2014)}]{cook_gravitational_2014}%
  \BibitemOpen
  \bibfield  {author} {\bibinfo {author} {\bibfnamefont {G.~B.}\ \bibnamefont {Cook}}\ and\ \bibinfo {author} {\bibfnamefont {M.}~\bibnamefont {Zalutskiy}},\ }\bibfield  {title} {\bibinfo {title} {Gravitational perturbations of the {Kerr} geometry: {High}-accuracy study},\ }\href {https://doi.org/10.1103/PhysRevD.90.124021} {\bibfield  {journal} {\bibinfo  {journal} {Physical Review D}\ }\textbf {\bibinfo {volume} {90}},\ \bibinfo {pages} {124021} (\bibinfo {year} {2014})}\BibitemShut {NoStop}%
\bibitem [{\citenamefont {Leaver}(1986)}]{leaver_spectral_1986}%
  \BibitemOpen
  \bibfield  {author} {\bibinfo {author} {\bibfnamefont {E.~W.}\ \bibnamefont {Leaver}},\ }\bibfield  {title} {\bibinfo {title} {Spectral decomposition of the perturbation response of the {Schwarzschild} geometry},\ }\href {https://doi.org/10.1103/PhysRevD.34.384} {\bibfield  {journal} {\bibinfo  {journal} {Physical Review D}\ }\textbf {\bibinfo {volume} {34}},\ \bibinfo {pages} {384} (\bibinfo {year} {1986})}\BibitemShut {NoStop}%
\bibitem [{\citenamefont {Wolfram~Research}()}]{wolfram_research_inc_mathematica_nodate}%
  \BibitemOpen
  \bibfield  {author} {\bibinfo {author} {\bibfnamefont {I.}~\bibnamefont {Wolfram~Research}},\ }\href {https://www.wolfram.com/mathematica} {\bibinfo {title} {Mathematica, {Version} 14.3}},\ \bibinfo {note} {{C}hampaign, IL, 2025}\BibitemShut {NoStop}%
\bibitem [{\citenamefont {Schutz}\ and\ \citenamefont {Will}(1985)}]{schutz_black_1985}%
  \BibitemOpen
  \bibfield  {author} {\bibinfo {author} {\bibfnamefont {B.~F.}\ \bibnamefont {Schutz}}\ and\ \bibinfo {author} {\bibfnamefont {C.~M.}\ \bibnamefont {Will}},\ }\bibfield  {title} {\bibinfo {title} {Black hole normal modes - {A} semianalytic approach},\ }\href {https://doi.org/10.1086/184453} {\bibfield  {journal} {\bibinfo  {journal} {The Astrophysical Journal}\ }\textbf {\bibinfo {volume} {291}},\ \bibinfo {pages} {L33} (\bibinfo {year} {1985})}\BibitemShut {NoStop}%
\bibitem [{\citenamefont {Iyer}\ and\ \citenamefont {Will}(1987)}]{iyer_black-hole_1987}%
  \BibitemOpen
  \bibfield  {author} {\bibinfo {author} {\bibfnamefont {S.}~\bibnamefont {Iyer}}\ and\ \bibinfo {author} {\bibfnamefont {C.~M.}\ \bibnamefont {Will}},\ }\bibfield  {title} {\bibinfo {title} {Black-hole normal modes: {A} {WKB} approach. {I}. {Foundations} and application of a higher-order {WKB} analysis of potential-barrier scattering},\ }\href {https://doi.org/10.1103/PhysRevD.35.3621} {\bibfield  {journal} {\bibinfo  {journal} {Physical Review D}\ }\textbf {\bibinfo {volume} {35}},\ \bibinfo {pages} {3621} (\bibinfo {year} {1987})}\BibitemShut {NoStop}%
\bibitem [{\citenamefont {Seidel}\ and\ \citenamefont {Iyer}(1990)}]{seidel_black-hole_1990}%
  \BibitemOpen
  \bibfield  {author} {\bibinfo {author} {\bibfnamefont {E.}~\bibnamefont {Seidel}}\ and\ \bibinfo {author} {\bibfnamefont {S.}~\bibnamefont {Iyer}},\ }\bibfield  {title} {\bibinfo {title} {Black-hole normal modes: {A} {WKB} approach. {IV}. {Kerr} black holes},\ }\href {https://doi.org/10.1103/PhysRevD.41.374} {\bibfield  {journal} {\bibinfo  {journal} {Physical Review D}\ }\textbf {\bibinfo {volume} {41}},\ \bibinfo {pages} {374} (\bibinfo {year} {1990})}\BibitemShut {NoStop}%
\bibitem [{\citenamefont {Yang}\ \emph {et~al.}(2012)\citenamefont {Yang}, \citenamefont {Nichols}, \citenamefont {Zhang}, \citenamefont {Zimmerman}, \citenamefont {Zhang},\ and\ \citenamefont {Chen}}]{yang_quasinormal-mode_2012}%
  \BibitemOpen
  \bibfield  {author} {\bibinfo {author} {\bibfnamefont {H.}~\bibnamefont {Yang}}, \bibinfo {author} {\bibfnamefont {D.~A.}\ \bibnamefont {Nichols}}, \bibinfo {author} {\bibfnamefont {F.}~\bibnamefont {Zhang}}, \bibinfo {author} {\bibfnamefont {A.}~\bibnamefont {Zimmerman}}, \bibinfo {author} {\bibfnamefont {Z.}~\bibnamefont {Zhang}},\ and\ \bibinfo {author} {\bibfnamefont {Y.}~\bibnamefont {Chen}},\ }\bibfield  {title} {\bibinfo {title} {Quasinormal-mode spectrum of {Kerr} black holes and its geometric interpretation},\ }\href {https://doi.org/10.1103/PhysRevD.86.104006} {\bibfield  {journal} {\bibinfo  {journal} {Physical Review D}\ }\textbf {\bibinfo {volume} {86}},\ \bibinfo {pages} {104006} (\bibinfo {year} {2012})},\ \bibinfo {note} {arXiv:1207.4253 [gr-qc]}\BibitemShut {NoStop}%
\bibitem [{\citenamefont {Konoplya}\ and\ \citenamefont {Zhidenko}(2013)}]{konoplya_radiation_2013}%
  \BibitemOpen
  \bibfield  {author} {\bibinfo {author} {\bibfnamefont {R.~A.}\ \bibnamefont {Konoplya}}\ and\ \bibinfo {author} {\bibfnamefont {A.}~\bibnamefont {Zhidenko}},\ }\bibfield  {title} {\bibinfo {title} {Radiation processes in the vicinity of non-{Schwarzschild} and non-{Kerr} black holes},\ }\href {https://doi.org/10.1103/PhysRevD.87.024044} {\bibfield  {journal} {\bibinfo  {journal} {Physical Review D}\ }\textbf {\bibinfo {volume} {87}},\ \bibinfo {pages} {024044} (\bibinfo {year} {2013})},\ \bibinfo {note} {arXiv:1210.8430 [gr-qc]}\BibitemShut {NoStop}%
\bibitem [{\citenamefont {Moncrief}(1974)}]{moncrief_gravitational_1974}%
  \BibitemOpen
  \bibfield  {author} {\bibinfo {author} {\bibfnamefont {V.}~\bibnamefont {Moncrief}},\ }\bibfield  {title} {\bibinfo {title} {Gravitational perturbations of spherically symmetric systems. {I}. {The} exterior problem},\ }\href {https://doi.org/https://doi.org/10.1016/0003-4916(74)90173-0} {\bibfield  {journal} {\bibinfo  {journal} {Annals of Physics}\ }\textbf {\bibinfo {volume} {88}},\ \bibinfo {pages} {323} (\bibinfo {year} {1974})}\BibitemShut {NoStop}%
\bibitem [{\citenamefont {Press}\ and\ \citenamefont {Teukolsky}(1973)}]{press_perturbations_1973}%
  \BibitemOpen
  \bibfield  {author} {\bibinfo {author} {\bibfnamefont {W.~H.}\ \bibnamefont {Press}}\ and\ \bibinfo {author} {\bibfnamefont {S.~A.}\ \bibnamefont {Teukolsky}},\ }\bibfield  {title} {\bibinfo {title} {Perturbations of a {Rotating} {Black} {Hole}. {II}. {Dynamical} {Stability} of the {Kerr} {Metric}},\ }\href {https://doi.org/10.1086/152445} {\bibfield  {journal} {\bibinfo  {journal} {The Astrophysical Journal}\ }\textbf {\bibinfo {volume} {185}},\ \bibinfo {pages} {649} (\bibinfo {year} {1973})}\BibitemShut {NoStop}%
\bibitem [{\citenamefont {Kinnersley}(1969)}]{kinnersley_type_1969}%
  \BibitemOpen
  \bibfield  {author} {\bibinfo {author} {\bibfnamefont {W.}~\bibnamefont {Kinnersley}},\ }\bibfield  {title} {\bibinfo {title} {Type {D} {Vacuum} {Metrics}},\ }\href {https://doi.org/10.1063/1.1664958} {\bibfield  {journal} {\bibinfo  {journal} {Journal of Mathematical Physics}\ }\textbf {\bibinfo {volume} {10}},\ \bibinfo {pages} {1195} (\bibinfo {year} {1969})}\BibitemShut {NoStop}%
\bibitem [{\citenamefont {Wu}(2026)}]{wu2026beyondgrqnms}%
  \BibitemOpen
  \bibfield  {author} {\bibinfo {author} {\bibfnamefont {D.~G.}\ \bibnamefont {Wu}},\ }\bibfield  {title} {\bibinfo {title} {Davidwu421/beyondgrqnms: Beyond-gr quasinormal mode shifts v1.0.0},\ }\href {https://doi.org/10.5281/zenodo.18434613} {10.5281/zenodo.18434613} (\bibinfo {year} {2026}),\ \bibinfo {note} {[Data set]}\BibitemShut {NoStop}%
\bibitem [{\citenamefont {Ferrari}\ and\ \citenamefont {Mashhoon}(1984)}]{ferrari_new_1984}%
  \BibitemOpen
  \bibfield  {author} {\bibinfo {author} {\bibfnamefont {V.}~\bibnamefont {Ferrari}}\ and\ \bibinfo {author} {\bibfnamefont {B.}~\bibnamefont {Mashhoon}},\ }\bibfield  {title} {\bibinfo {title} {New approach to the quasinormal modes of a black hole},\ }\href {https://doi.org/10.1103/PhysRevD.30.295} {\bibfield  {journal} {\bibinfo  {journal} {Physical Review D}\ }\textbf {\bibinfo {volume} {30}},\ \bibinfo {pages} {295} (\bibinfo {year} {1984})}\BibitemShut {NoStop}%
\bibitem [{\citenamefont {Bryant}\ \emph {et~al.}(2021)\citenamefont {Bryant}, \citenamefont {Silva}, \citenamefont {Yagi},\ and\ \citenamefont {Glampedakis}}]{bryantEikonalQuasinormalModes2021}%
  \BibitemOpen
  \bibfield  {author} {\bibinfo {author} {\bibfnamefont {A.}~\bibnamefont {Bryant}}, \bibinfo {author} {\bibfnamefont {H.~O.}\ \bibnamefont {Silva}}, \bibinfo {author} {\bibfnamefont {K.}~\bibnamefont {Yagi}},\ and\ \bibinfo {author} {\bibfnamefont {K.}~\bibnamefont {Glampedakis}},\ }\bibfield  {title} {\bibinfo {title} {Eikonal quasinormal modes of black holes beyond general relativity {III}: scalar {Gauss}-{Bonnet} gravity},\ }\href {https://doi.org/10.1103/PhysRevD.104.044051} {\bibfield  {journal} {\bibinfo  {journal} {Physical Review D}\ }\textbf {\bibinfo {volume} {104}},\ \bibinfo {pages} {044051} (\bibinfo {year} {2021})},\ \bibinfo {note} {arXiv:2106.09657 [gr-qc]}\BibitemShut {NoStop}%
\bibitem [{\citenamefont {Cano}\ and\ \citenamefont {David}(2024)}]{canoIsospectralityEffectiveField2024}%
  \BibitemOpen
  \bibfield  {author} {\bibinfo {author} {\bibfnamefont {P.~A.}\ \bibnamefont {Cano}}\ and\ \bibinfo {author} {\bibfnamefont {M.}~\bibnamefont {David}},\ }\href {https://doi.org/10.48550/arXiv.2407.12080} {\bibinfo {title} {Isospectrality in {Effective} {Field} {Theory} {Extensions} of {General} {Relativity}}} (\bibinfo {year} {2024}),\ \bibinfo {note} {arXiv:2407.12080 [hep-th]}\BibitemShut {NoStop}%
\bibitem [{\citenamefont {Pacilio}\ and\ \citenamefont {Bhagwat}(2023)}]{pacilio_identifying_2023}%
  \BibitemOpen
  \bibfield  {author} {\bibinfo {author} {\bibfnamefont {C.}~\bibnamefont {Pacilio}}\ and\ \bibinfo {author} {\bibfnamefont {S.}~\bibnamefont {Bhagwat}},\ }\bibfield  {title} {\bibinfo {title} {Identifying modified theories of gravity using binary black-hole ringdowns},\ }\href {https://doi.org/10.1103/PhysRevD.107.083021} {\bibfield  {journal} {\bibinfo  {journal} {Physical Review D}\ }\textbf {\bibinfo {volume} {107}},\ \bibinfo {pages} {083021} (\bibinfo {year} {2023})}\BibitemShut {NoStop}%
\bibitem [{\citenamefont {Collins}\ and\ \citenamefont {Hughes}(2004)}]{collins_towards_2004}%
  \BibitemOpen
  \bibfield  {author} {\bibinfo {author} {\bibfnamefont {N.~A.}\ \bibnamefont {Collins}}\ and\ \bibinfo {author} {\bibfnamefont {S.~A.}\ \bibnamefont {Hughes}},\ }\bibfield  {title} {\bibinfo {title} {Towards a formalism for mapping the spacetimes of massive compact objects: {Bumpy} black holes and their orbits},\ }\href {https://doi.org/10.1103/PhysRevD.69.124022} {\bibfield  {journal} {\bibinfo  {journal} {Physical Review D}\ }\textbf {\bibinfo {volume} {69}},\ \bibinfo {pages} {124022} (\bibinfo {year} {2004})}\BibitemShut {NoStop}%
\bibitem [{\citenamefont {Carullo}\ \emph {et~al.}(2019)\citenamefont {Carullo}, \citenamefont {Del~Pozzo},\ and\ \citenamefont {Veitch}}]{Carullo:2019flw}%
  \BibitemOpen
  \bibfield  {author} {\bibinfo {author} {\bibfnamefont {G.}~\bibnamefont {Carullo}}, \bibinfo {author} {\bibfnamefont {W.}~\bibnamefont {Del~Pozzo}},\ and\ \bibinfo {author} {\bibfnamefont {J.}~\bibnamefont {Veitch}},\ }\bibfield  {title} {\bibinfo {title} {{Observational Black Hole Spectroscopy: A time-domain multimode analysis of GW150914}},\ }\href {https://doi.org/10.1103/PhysRevD.99.123029} {\bibfield  {journal} {\bibinfo  {journal} {Phys. Rev. D}\ }\textbf {\bibinfo {volume} {99}},\ \bibinfo {pages} {123029} (\bibinfo {year} {2019})},\ \bibinfo {note} {[Erratum: Phys.Rev.D 100, 089903 (2019)]},\ \Eprint {https://arxiv.org/abs/1902.07527} {arXiv:1902.07527 [gr-qc]} \BibitemShut {NoStop}%
\bibitem [{\citenamefont {Isi}\ and\ \citenamefont {Farr}(2021)}]{isi_analyzing_2021}%
  \BibitemOpen
  \bibfield  {author} {\bibinfo {author} {\bibfnamefont {M.}~\bibnamefont {Isi}}\ and\ \bibinfo {author} {\bibfnamefont {W.~M.}\ \bibnamefont {Farr}},\ }\href {https://doi.org/10.48550/arXiv.2107.05609} {\bibinfo {title} {Analyzing black-hole ringdowns}} (\bibinfo {year} {2021}),\ \bibinfo {note} {arXiv:2107.05609 [gr-qc]}\BibitemShut {NoStop}%
\bibitem [{\citenamefont {Capano}\ \emph {et~al.}(2023)\citenamefont {Capano}, \citenamefont {Cabero}, \citenamefont {Westerweck}, \citenamefont {Abedi}, \citenamefont {Kastha}, \citenamefont {Nitz}, \citenamefont {Wang}, \citenamefont {Nielsen},\ and\ \citenamefont {Krishnan}}]{Capano:2021etf}%
  \BibitemOpen
  \bibfield  {author} {\bibinfo {author} {\bibfnamefont {C.~D.}\ \bibnamefont {Capano}}, \bibinfo {author} {\bibfnamefont {M.}~\bibnamefont {Cabero}}, \bibinfo {author} {\bibfnamefont {J.}~\bibnamefont {Westerweck}}, \bibinfo {author} {\bibfnamefont {J.}~\bibnamefont {Abedi}}, \bibinfo {author} {\bibfnamefont {S.}~\bibnamefont {Kastha}}, \bibinfo {author} {\bibfnamefont {A.~H.}\ \bibnamefont {Nitz}}, \bibinfo {author} {\bibfnamefont {Y.-F.}\ \bibnamefont {Wang}}, \bibinfo {author} {\bibfnamefont {A.~B.}\ \bibnamefont {Nielsen}},\ and\ \bibinfo {author} {\bibfnamefont {B.}~\bibnamefont {Krishnan}},\ }\bibfield  {title} {\bibinfo {title} {{Multimode Quasinormal Spectrum from a Perturbed Black Hole}},\ }\href {https://doi.org/10.1103/PhysRevLett.131.221402} {\bibfield  {journal} {\bibinfo  {journal} {Phys. Rev. Lett.}\ }\textbf {\bibinfo {volume} {131}},\ \bibinfo {pages} {221402} (\bibinfo {year} {2023})},\ \Eprint {https://arxiv.org/abs/2105.05238} {arXiv:2105.05238 [gr-qc]} \BibitemShut {NoStop}%
\bibitem [{\citenamefont {Ma}\ \emph {et~al.}(2023)\citenamefont {Ma}, \citenamefont {Sun},\ and\ \citenamefont {Chen}}]{Ma:2023cwe}%
  \BibitemOpen
  \bibfield  {author} {\bibinfo {author} {\bibfnamefont {S.}~\bibnamefont {Ma}}, \bibinfo {author} {\bibfnamefont {L.}~\bibnamefont {Sun}},\ and\ \bibinfo {author} {\bibfnamefont {Y.}~\bibnamefont {Chen}},\ }\bibfield  {title} {\bibinfo {title} {{Black Hole Spectroscopy by Mode Cleaning}},\ }\href {https://doi.org/10.1103/PhysRevLett.130.141401} {\bibfield  {journal} {\bibinfo  {journal} {Phys. Rev. Lett.}\ }\textbf {\bibinfo {volume} {130}},\ \bibinfo {pages} {141401} (\bibinfo {year} {2023})},\ \Eprint {https://arxiv.org/abs/2301.06705} {arXiv:2301.06705 [gr-qc]} \BibitemShut {NoStop}%
\bibitem [{\citenamefont {Leaver}(1985)}]{leaver_analytic_1985}%
  \BibitemOpen
  \bibfield  {author} {\bibinfo {author} {\bibfnamefont {E.~W.}\ \bibnamefont {Leaver}},\ }\bibfield  {title} {\bibinfo {title} {An analytic representation for the quasi-normal modes of {Kerr} black holes},\ }\href {https://doi.org/10.1098/rspa.1985.0119} {\bibfield  {journal} {\bibinfo  {journal} {Proceedings of the Royal Society of London. A. Mathematical and Physical Sciences}\ }\textbf {\bibinfo {volume} {402}},\ \bibinfo {pages} {285} (\bibinfo {year} {1985})}\BibitemShut {NoStop}%
\end{thebibliography}%

\end{document}